\documentclass[a4paper,11pt]{article}

\usepackage{jheppub} 

\usepackage{lscape}
\usepackage{array}
\usepackage{natbib}
\usepackage{amsfonts}
\usepackage{amssymb}
\usepackage{bbm} 
\usepackage{amsthm} 
\usepackage{graphicx}
\usepackage{imakeidx}
\usepackage[utf8]{inputenc}
\usepackage[T1]{fontenc}
\usepackage{tikz}
\usetikzlibrary{arrows.meta, positioning, decorations.pathreplacing, calligraphy}

\makeindex

\renewcommand{\i}{{\rm i}}

\newcommand\bi{\begin{itemize}}
\newcommand\ei{\end{itemize}}

\newcommand\bspl{\begin{split}}
\newcommand\espl{\end{split}}

\newcommand{\susyL}[1]{\underset{\text{SUSY Locus}}{\longrightarrow}}
\newcommand{\be}{\begin{equation}}
\newcommand{\ee}{\end{equation}}
\newcommand{\bea}{\begin{eqnarray}}
\newcommand{\eea}{\end{eqnarray}}

\newtheorem{theorem}{Observation}[section]

\theoremstyle{definition}

\theoremstyle{definition}

\begin{document}

\title{\boldmath The Schwarzian from gauge theories}

\author{Alejandro Cabo-Bizet\,$^a$}

\affiliation[a]{Università del Salento, Dipartimento di Matematica e Fisica Ennio De Giorgi, and I.N.F.N. - sezione di Lecce, Via Arnesano, I-73100 Lecce, Italy}

\emailAdd{acbizet@gmail.com}

\abstract{  
The continuum of holographic dual gravitational charges is recovered out of the discrete spectrum of~$U(N)$ $\mathcal{N}=4$ SYM on~$\mathbb{R}\times S^3\,$. In such a limit, the free energy of the free gauge theory is computed up to logarithmic contributions and exponentially suppressed contributions. Assuming the supergravity dual prediction to correctly capture strong-coupling results in field theory, the answer is bound to encode a complete low-temperature expansion of the Gibbons-Hawking gravitational on-shell action, valid well beyond the vicinity of supersymmetric black hole solutions. The formula recovers the long awaited Schwarzian contribution at low enough temperatures for certain choices of flows to the continuum. One such choice identifies the chemical potentials and thermodynamic charges in field theory with the chemical potentials and  thermodynamic charges of the dual black holes. For such a flow the computed mass-gap kinematically matches the conjectured strong-coupling result obtained by Boruch, Heydeman, Iliesiu, and Turiaci in supergravity, including small $1/\lambda$-corrections. The {emergent} reparameterizations, broken by the selection of the Schwarzian, correspond to redefinitions of the relevant cutoff scale. Observations are made regarding the existence of~$\frac{1}{8}$-BPS black holes and how this is in tension with
BPS inequalities. The RG-flow procedure leading to these results opens a way to understanding the emergence of chaos in gauge theories and its relation to non-extremal and non-supersymmetric black hole physics.
}



\maketitle


\section{Introduction}
\label{sec1}

In a quantum system with a semi-positive Hamiltonian~$\Delta\geq 0$ and a discrete spectrum with degeneracies~$d(\Delta)$ the Taylor expansion of the partition function
\begin{equation}
Z[\beta]=\text{Tr} e^{-\beta\Delta}=e^{-\mathcal{F}[\beta]}
\end{equation}
at zero temperature~$\beta=\infty$ is trivially equal to the number of ground states. \footnote{This is because every finite-temperature correction to the partition function beyond the vacuum degeneracy~$d(0)$, $Z = d(0) + d(\Delta_{\text{min}}) e^{-\beta \Delta_{\text{min}}} + \ldots$ is exponentially suppressed.} Equivalently, the Taylor expansion of the free energy~$\mathcal{F}$ at zero temperature~$\beta=\infty\,$ is trivially equal to minus the logarithm of the number of ground states.~\footnote{ $\mathcal{F}``="-\log d(0)\,$.}
If the discrete system has a holographic dual description~\cite{tHooft:1973alw,Maldacena:1997re,Witten:1998qj,Gubser:1998bc}\cite{Polchinski:1995mt} such that in an RG flow procedure~\footnote{Very much in the mathematical spirit of~\cite{Wilson:1974mb,Wilson:1983xri}\cite{Gell-Mann:1954yli,Gross:1973id}.} -- denoted as~$\Lambda\to\infty\,$ in this introduction --~\footnote{To make contact with supergravity ~$N\gg1$ is also required.~The scale~$\Lambda$, whose meaning we will explain below, is a conceptually different scale than~$N$ (first of all it is dimensionful). Whenever we connect with the theory obtained in the RG-expansion $\Lambda\to\infty$ to supergravity it is implicitly assumed that~$\Lambda R \gg N\,\gg\,1\,$. That said, the parameter leading the expansions to the continuum is $\Lambda$, for us~$N$ will always be fixed and large $N\,\gg\, 1\,$ in such an expansion. Indeed, in the last few years we have come to understand that this is the kind of expansions that pick up semiclassical black hole physics, not a naive large-$N$ expansion. In a naive large-$N$ expansion where $\Lambda$ is kept fixed while $N$ is expanded around infinity, the physics of a gas of multigravitons in pure AdS is recovered not a black hole.} it reduces to a semiclassical theory of gravity with black hole solutions at arbitrary Bekenstein-Hawking temperature~$1/\beta$~\cite{Bekenstein:1972tm,Hawking:1974rv,Strominger:1996sh,Callan:1996dv,Horowitz:1996fn,Breckenridge:1996sn,Horowitz:1996ay,Witten:1998zw}\cite{Susskind:1994vu,Sen:1995in}, then one reaches a contradiction, as the gravitational free energy~$\mathcal{F}_{g}$
\begin{equation}
\mathcal{F}\underset{\Lambda\to\infty}{\to} \mathcal{F}_{g}
\end{equation}
of any such black hole is bound to have a non-trivial perturbative expansion around~$\beta=\infty\,$. For example, the Gibbons-Hawking gravitational on-shell action~\cite{Gibbons:1976ue} of rotating and electrically charged black holes in~$AdS_5\,$~\cite{Chen:2005zj,Kunduri:2005zg}, and in particular its low temperature expansion, has been recently studied, with varied motivations, for example, in~\cite{Cabo-Bizet:2018ehj,Larsen:2019oll,Goldstein:2019gpz,Iliesiu:2020qvm,Heydeman:2020hhw,Boruch:2022tno,Turiaci:2023wrh}. The first goal of this paper is to solve this apparent contradiction.

That~$\mathcal{F}[\beta]\,$, the free energy of the fundamental theory has a trivial Taylor expansion at zero temperature while $\mathcal{F}_{g}[\beta]\,$, the free energy of the infrared effective theory has a non-trivial one, strongly suggests that the RG-flow procedure~$\Lambda\to\infty\,$, which must be applied before expanding in low-temperatures~$\beta\to\infty\,$,~\footnote{As it corresponds to a semiclassical gravitational limit with black hole solutions at any temperature.} should map the discrete spectrum of the fundamental theory (and its stringy dual formulation) to a continuum spectrum in the infrared. 

In the grand-canonical ensemble (with fixed chemical potentials), we will find that the operation \(\Lambda \rightarrow \infty\) corresponds to localizing the free energy \(\mathcal{F}\) of the fundamental system to a \(\frac{1}{\Lambda}\)-vicinity of its singular locus. By localization, we mean discarding exponentially suppressed contributions at large values of the scaling parameter \(\Lambda\), in the spirit of~\cite{Beccaria:2023hip}. 

In a generic gauge theory there are several such singularities,~\footnote{For example, already at~$\beta=\infty$, for the superconformal index of~$\mathcal{N}=4$ SYM, this is known to be the case~\cite{Beccaria:2023hip,Cabo-Bizet:2021plf}~\cite{ArabiArdehali:2021nsx,Ardehali:2021irq,Jejjala:2021hlt}.} and disconnected families of them are expected to correspond to inequivalent \(\Lambda \rightarrow \infty\) RG-flow procedures. It is also expected that they correspond to different saddle points~\cite{Choi:2018hmj,Benini:2018ywd,Honda:2019cio,ArabiArdehali:2019tdm,Kim:2019yrz,Cabo-Bizet:2019osg,Cabo-Bizet:2019eaf,ArabiArdehali:2019orz,Cabo-Bizet:2020nkr,Benini:2020gjh,GonzalezLezcano:2020yeb,Copetti:2020dil,Cabo-Bizet:2020ewf,Amariti:2021ubd,Cassani:2021fyv,Aharony:2021zkr,Choi:2021rxi,Mamroud:2022msu,Choi:2023tiq,Aharony:2024ntg}\cite{Agarwal:2020zwm,Murthy:2020rbd} which may or may not be intersected by the complex contour one wishes to integrate over in order to work at fixed and large charges. In this paper, we will focus on one such leading singularity, that is, one that corresponds to the potentially most dominant saddle point(s).~\footnote{\label{ftn:DiscreteGroupOrbits}These dominating saddle points come in representations of discrete groups~\cite{GonzalezLezcano:2020yeb,Amariti:2021ubd}\cite{Cabo-Bizet:2021jar}.} The test theory will be $U(N)$ $\mathcal{N}=4$ SYM on~$\mathbb{R}\,\times\,S^3$ where there is convincing evidence that such a leading saddle point determines the analytic part of the asymptotic expansion of the free energy in the~$\Lambda\to\infty$ RG flow procedure~\underline{at zero temperature~$\beta=\infty$}. This is, of the free energy of the superconformal index~\cite{Beccaria:2023hip}. We will further confirm that a smooth deformation of it continues to dominate the perturbative expansion in the continuum $\Lambda\to\infty\,$ beyond the superconformal locus.

This leading localization procedure may seem abstract at first. It has a clear physical meaning though. In the microcanonical ensemble it amounts to ignoring charge eigenvalues $E>0$ that are larger than a certain large energy scale~$E_{max}=\Lambda^{n+1} R^n\,$, where~$R$ is the radius of the~$S^3\,$.~{Then requiring $\Lambda$ to be much larger than~$\frac{1}{R}$, implies that the spacing among contiguous eigenvalues~$\delta E \,=\,\frac{1}{R}$ becomes infinitesimally small compared to the hierarchy of eigenvalues explored~$\frac{\delta E}{E_{max}} \sim \frac{1}{(\Lambda R)^{n+1}}\,\ll\,1\,$. \emph{Judiciously ignoring the small spacing}, $\frac{\delta E}{E_{max}}$, {in dimensionless variable~$0<\frac{E}{E_{max}}<1$}, one reaches a continuum spectrum that unlike the original spectrum of the fundamental theory, remains to be continuum after coming back to the dimensionful variable~$0\,<\,E\,< \,E_{max}\,$. The integer power~$n$ may be conveniently selected depending on the landing point of the selected $\Lambda R\to\infty$ limit. The effective theory associated with any such a continuum spectrum will be called~\emph{infrared theory}.~\footnote{Of course a theory contains more data than its spectrum, but for the purposes of this work we will assume that other more refined data of the effective theory in the continuum, such as correlators, can be recovered from $\mathcal{N}=4$ SYM by the $\Lambda$-dependent deformation of chemical potentials to be mentioned below. }

{More details of this microcanonical version of the procedure will be reported elsewhere, but let us make an obvervation. The just described expansion to the continuum is not unique. The process of ignoring the small spacing among eigenvalues in dimensionless variables\,, i.e. $\frac{\delta E}{E_{max}}\,$, amounts to dropping certain choices of Fourier modes in the Fourier expansion of the eigenvalue density of the UV complete theory, e.g. of~$\mathcal{N}=4$ SYM,. In this way the Dirac delta-functions accounting for the discrete spectrum of the UV theory get smoothed out. Obviously, the selection of truncated Fourier modes is ambiguous and  in grand-canonical ensemble this ambiguity is naturally related to the ambiguity in selecting different large-$\Lambda$ expansions.~\footnote{This ambiguity can be also understood to be related to a renormalization of chemical potentials in grand-canonical ensemble or to a chemical potential dependent redefinition of the cutoff~$\Lambda\,$. }  } 

One important observation to make regarding this procedure is that \emph{any possible infinitesimally small mass-gap remainning in an expansion to the continuum $\Lambda R\to\infty$, can not be equated with the mass gap of the discrete spectrum of the UV theory, i.e. free $\mathcal{N}=4$ SYM, which is of order $\mathcal{O}(\frac{1}{R})=\mathcal{O}(\frac{\Lambda ^0}{R})\,$. Namely, the latter is non-infinitesimal in the expansion $\Lambda \to\infty$ because it is by construction $\Lambda$-independent.}  We will comeback in due time to elaborate on this.~\footnote{Please read around the emphasized comment below equation~\eqref{eq:EmphComment}.}

Using the grand-canonical description of the expansion(s) to the continuum, the complete analytic part of the asymptotic expansion of $\mathcal{F}$ around a generic reference $\frac{1}{16}$-BPS locus located at $\alpha_0=0\,$~\footnote{Sometimes it will be called the \emph{perturbative part}.} will be computed at any value of~$N$.~\footnote{We will not pay attention to chemical potential independent~$\mathcal{O}((\Lambda R)^0)$ contributions, such as those coming from counting the number of equally contributing saddle points like the ones mentioned in footnote~\ref{ftn:DiscreteGroupOrbits}.} The answer takes the form ($n=2$)
\be\label{eq:FInfinityIntro}
\mathcal{F}_{\infty}\,=\,\beta_0\,\mathcal{E}_0\,+\,\sum_{p=-1}^{n}\sum_{q=0}^4\sum^{\infty}_{r\,=\,0}(\Lambda R)^p\, L_{p+1;q,r}[\varphi_v,\varphi_w, \underline{u}]\,\frac{{F_{p;q;r} \alpha_0^q}}{(\beta_0)^{r} \omega_{1,0}\omega_{2,0}}\,,
\ee
~\footnote{The sum over powers of temperature~$\sum_{r=0}^{\infty}$ can be solved analytically as a rational function of~$\beta_0$. We will not report such expressions here because they are too large, but their complete form can be found in the shared Mathematica notebook. Many of the function coefficients denoted as $F_{p;q;r}$ vanish trivially, for example, $F_{p;0;r\geq 1}=F_{p;q\geq 1;0}=0\,$. Also, as it will be elaborated upon in due time there is implicit dependence on~$\Lambda R$ in the \emph{auxiliary  chemical potentials}, $\{\beta_0,\alpha_0,\omega_{1,0},\omega_{2,0}\}\,$. This ambiguity in implicit dependence is precisely the ambiguity in selection of large-$\Lambda$ expansions of $\mathcal{N}=4$ SYM.} where $\beta_0 = \beta \Lambda$. More details on this formula will be given in section~\ref{sec:3}, equation~\eqref{eq:FInfinity}, here we just notice that it includes terms that range from low-temperature corrections including Casimir-energy like contributions of order~$\mathcal{O}(\beta)$, and arbitrarily high positive powers of temperature~$\frac{1}{\beta}\,$.~Sometimes, in an abuse of notation,~\eqref{eq:FInfinityIntro} will be called the free energy of the holographic infrared theory or simply \emph{the infrared free energy}. However,~\eqref{eq:FInfinityIntro} really serves as a master formula encoding the infrared free energy of infinitelly many possible large-$\Lambda$ expansions to the continuum.  Assuming supergravity predictions are correctly capturing strong-coupling results in field theory and selecting the appropriate expansion to the continuum, then in virtue of analyticity,~\eqref{eq:FInfinityIntro} is bound to recover the complete low-temperature expansion of the Gibbons-Hawking onshell action of the black holes of~\cite{Chong:2005hr}, even well beyond the vicinity of supersymmetric black hole solutions.~\footnote{Using their $\frac{1}{16}$-BPS solutions~\cite{Gutowski:2004ez,Gutowski:2004yv} as reference point for the expansion~\cite{Cabo-Bizet:2018ehj,Cassani:2019mms,Bobev:2021qxx,Cassani:2022lrk}.}

By low-temperature expansion we mean an expansion of the master formula~\eqref{eq:FInfinityIntro} that takes us to zero-temperature and that precisely at zero-temperature, it approaches the supersymmetric locus~$\alpha_0=0\,$.~\footnote{This $\beta\to\infty$ limits of the continuum reductions studied in this paper are not counting ground states but BPS states in $\mathcal{N}=4$ SYM.} There are as many families of such expansions as independent superconformal indices (sectors) in $\mathcal{N}=4$ SYM and each such family is defined by a set of~\emph{boundary conditions} or precisely linear constraints upon chemical potentials.~$\frac{1}{4}$,~$\frac{1}{8}$ and~$\frac{1}{16}$-BPS boundary conditions and their corresponding expansions are all encoded in~\eqref{eq:FInfinityIntro}. 

After picking up a superconformal locus ($\alpha=0$) to expand around, there are infinitely many ways to reach it as $\Lambda R\to \infty\,$. This is precisely the ambiguity of limits that we mentioned above.
The reparameterization group invariance that the Schwarzian breaks~\cite{Maldacena:2016upp,Jensen:2016pah,Kitaev:2017awl,Stanford:2017thb}\cite{Almheiri:2014cka} is one that generates motion within such family of limits to the continuum. It corresponds, as well, to a set of chemical potential-dependent redefinitions of the cutoff~$\Lambda\mapsto \Lambda^\prime(\Lambda)\,$ such that~$\frac{\Lambda}{\Lambda^\prime (\Lambda)}\to1$ as $\Lambda R\to \infty\,$. For this reason sometimes we will also call it renormalization ambiguity. Such renormalizations would change the form of the free energy and thus the new thermodynamic charges as well --which are gradients of the deformed free energy with respect to the new chemical potentials --. The deformed charges become non-linearly related to the original ones. Such non-linear relation determines the relation among density of states via the chain-rule. Schematically, $\rho_{def}(E_{def})=\frac{d E}{d E_{def}}\,\rho(E)\,$. This makes contact with the previously mentioned microcanonical perspective on the ambiguity of limits to the continuum.

To show that the master formula~\eqref{eq:FInfinityIntro} captures sensible near-horizon physics after selecting the appropriate flow to the continuum, we will explicitly recover the first contributions in an infrared expansion of the free energy of super Schwarzian theories. For this we will select to work in a near~$\frac{1}{8}$-BPS sector, and in a near-$\frac{1}{16}$-BPS sector. Once this is done, it will then be essentially obvious that a limit to the continuum exists that takes~\eqref{eq:FInfinityIntro} to a complete quantum Schwarzian description, not just at the semiclassical level (See the discussion at the end of section~\ref{sec:QUantumSchwarzian}.).

The first Schwarzian contribution that we will study, the one localized around a $\frac{1}{8}$-BPS sector, and infinitely many other higher low-temperature corrections upon it,~\footnote{These expansions around $\beta=\infty\,$, and the corresponding corrections, are taken over equation~\eqref{eq:FInfinityIntro} not over the complete free energy of $\mathcal{N}=4\,$ which as we explained before it has only exponentially suppressed contributions in such expansion. From now on when we refer to small-temperature expansion or small-temperature corrections we always refer to expansions of obervables in the infrared effective theory where the spectrum is already continuum. } will be shown to be protected by superconformal symmetry. Namely, it does not receive corrections in $\lambda\,$. The second Schwarzian contributions that we will study, the ones localized around a $\frac{1}{16}$-BPS sector, are not protected by superconformal symmetry and they must receive corrections in~$\lambda\,$.

Interestingly, even for the \emph{unprotected} Schwarzian contributions the corresponding mass gap is matched by a zero-gauge coupling computation with the appropriate selection of limit to the continuum. This will trivially follow from protectedness of the reference superconformal index $\alpha_0=0$ and analyticity. The explanation will be given in~\ref{sec:ArgumentFreeTheoryStrongCoupling}. The complementary renormalization~\footnote{..., or equivalently, the particular choice of limit to the continuum (we will use both interpretations indistinctly, see~\ref{subsec:31}.),...} that is needed to match the answer obtained at 't Hooft coupling $\lambda=0$ with the answer obtained at $\lambda=\infty$ (as reported in~\cite{Boruch:2022tno}) is kinematically fixed by the identification of chemical potentials and thermodynamic charges among the boundary theory (precisely, its infrared effective description) and the gravitational solution in the bulk. This specific limit to the continuum will be studied in section~\ref{sec:BoundaryBulkIdentification}.

As mentioned, the protected Schwarzian contributions are expected to compute exact finite-temperature corrections of the gravitational onshell action of~$AdS_5$ solutions~\cite{Chong:2005hr,Chong:2005da,Wu:2011gq} about their $\frac{1}{8}$-BPS locus.~\footnote{The solutions of~\cite{Chong:2005hr,Chong:2005da,Wu:2011gq} include $\frac{1}{16}$-BPS black holes~\cite{Gutowski:2004ez,Gutowski:2004yv}~\cite{Cabo-Bizet:2018ehj,Choi:2018hmj,Benini:2018ywd} but also horizon-less solutions such as BPS solitons~\cite{Chong:2005hr,Chong:2005da}. So, even in the case~$\frac{1}{8}$-BPS black holes do not exist, this point may have a physical meaning, i.e., a dual horizonless geometry. $\frac{1}{16}$-BPS rotating solitons have been already found in, e.g.,~\cite{Chong:2005hr,Cvetic:2005zi}. Examples of $\frac{1}{8}$-BPS solitons with electric charge and no rotation have been already found in~\cite{Gava:2006pu}. Other horizonless geometries, such as LLM geometries~\cite{Lin:2004nb} have been shown to correspond to the $\frac{1}{2}$-BPS sector~\cite{Maoz:2005nk,Chang:2024zqi,Deddo:2024liu}. So, more general horizonless geometries may very well correspond to the $\frac{1}{4}$- and $\frac{1}{8}$-BPS sectors.} From this protected infrared limit we will learn something about near~$\frac{1}{8}$-BPS black holes. For near-$\frac{1}{8}$-BPS black holes with at least two almost-equal electric charges~$Q_1\approx Q_3$ (but unequal), it will be shown that semipositivity bounds of the fundamental theory imply that the mixed-ensemble free energy of the infrared theory should come solely from the semiclassical Schwarzian contribution. Thus, it scales bilinearly with temperature~$T$ and a chemical potential~$(\alpha-\frac{1}{2})\,$.~\footnote{This is the chemical potential controlling the limit to the supersymmetric locus~\cite{Cabo-Bizet:2018ehj}.} This result strongly suggests that these black hole solutions have a near-vanishing horizon area, a conclusion that resonates with the results of \cite{Goldstein:2019gpz} (in gravity) and of~\cite{Chang:2024zqi,Chang:2023ywj,Eleftheriou:2022kkv} (in field theory).

The Schwarzian contributions which are not protected against coupling corrections are the ones corresponding -- at strong coupling -- to the breaking of re-parameterizations of the time coordinate in the near-horizon~$AdS_2$ geometry of black holes~\cite{Chong:2005hr,Chong:2005da,Wu:2011gq} that are near-$\frac{1}{16}$-BPS, but \underline{not near~$\frac{1}{8}$-BPS}. These include the ones studied in~\cite{Boruch:2022tno} in the context of minimally gauged five-dimensional supergravity. 

As announced before, for an appropriate flow to the continuum, the free-field theory computation ($\lambda=0$) of the Schwarzian mass gap matches the gravitational result ($\lambda=\infty$) obtained in reference~\cite{Boruch:2022tno}. {{In this near-$\frac{1}{16}$-BPS case} logarithmic divergencies will be understood to come from the reference superconformal index ($\alpha_0=0$) confirming the absence of~$\log T/T_{breakdown}$ corrections, as expected from the supergravity perspective~\cite{Boruch:2022tno}\cite{Turiaci:2023wrh}.}~\footnote{For the near-$\frac{1}{8}$-BPS case there may be subtleties that spoil this last conclusion. We leave for the future to study that case.}

It should be said that the main observation of this paper goes far beyond the near-BPS expansion and it is summarized in section~\ref{sec:ArgumentFreeTheoryStrongCoupling}, equation~\eqref{eq:TheProposal}:

\vspace{.2cm}

{\emph{The free field theory analysis of the partition function of $\mathcal{N}=4$ SYM, complemented with an appropriate choice of limit to the continuum,~\footnote{Or equivalently, with an appropriate renormalization of chemical potentials. } captures exact holographic finite-temperature corrections to the free-energy of the dual black holes, including terms that will be identified as coming from an $\mathcal{N}=2$ Schwarzian theory~\cite{Stanford:2017thb}\cite{Heydeman:2020hhw,Boruch:2022tno}. Furthermore, the limit to the continuum can be always refined in order to incorporate the quantum corrections of the corresponding super-Schwarzian. }}
\vspace{.2cm}

The RG-flow procedure leading to the previous conclusions opens a way to realize, analytically, how chaos may emerge~\cite{Cotler:2016fpe,Saad:2018bqo,Turiaci:2023jfa} within higher dimensional non-averaged systems, such as four-dimensional $\mathcal{N}=4$ SYM~\cite{Chen:2022hbi,Chen:2023mbc}\cite{Choi:2022asl}. 

This paper focuses on~$\mathcal{N}=4$ SYM on $\mathbb{R}\times S^3$, but the proposed RG-flow procedure can be applied to any other example of superconformal gauge theories with a known discrete spectrum. Thinking not only in the context of gauge/gravity dualities, these ways of flowing from the discretuum to the continuum can be applied to any gauge theory with a known discrete spectrum, which can be enforced by placing the theory in a box with appropriate boundary conditions, for example. It can also be applied to systems that are already known to be realized in nature.

The content of this paper is organized as follows. Section~\ref{sec:2} revisits the computation of the partition function of four-dimensional~$U(N)$~$\mathcal{N}=4$ SYM on~$\mathbb{R}\times S^3$ at zero gauge coupling and sets up conventions. Special emphasis is put on illustrating the constraints (boundary conditions) that reduce the partition function to~$\frac{1}{16}$-BPS indices and~$\frac{1}{8}$-BPS indices. Section~\ref{sec:3} introduces the RG flow procedures that map the discrete field theory spectrum to the continuum of charges to eventually match with supergravity. It also summarizes the derivation of the master-formula~\eqref{eq:FInfinityIntro}, and explains why it captures the Gibbons-Hawking on-shell action well beyond supersymmetry and extremality. Section~\ref{sec:4} proceeds to compute a Schwarzian contribution around the~$\frac{1}{8}$-BPS locus and its mass gap based solely on requiring reality conditions of physical charges and entropy, without using the holographic dictionary. Section~\ref{sec:5} proceeds to show how~\eqref{eq:FInfinityIntro} recovers the Schwarzian contribution around the generic $\frac{1}{16}$-BPS
locus and its mass gap by identifying chemical potentials and continuum charges in the boundary with those in the bulk. It also compares the obtained results with the conjectured supergravity duals. Section~\ref{sec:6} ends the main body of the article with a summary of results, open questions, and observations. Some technical details and results are relegated to the Appendices.~\footnote{Our analysis concerns $AdS_5/CFT_4$ but we expect an analogous RG-flow reduction will be implemented in~$AdS_4/CFT_3\,$, eventually. Relevant observations in that context have been put forward recently in~\cite{Benini:2022bwa}.}

\subsection{On some of the results in this paper}
Before moving on to the bulk of the paper we pause to clearly state three important results ahead. In the diagram below, these results are summarized as three causally connected stages or steps. The first two steps are results obtained solely within the gauge theory side without using any gravitational side input.

\begin{figure}[ht]
  \centering
\includegraphics[width=1.0\textwidth]{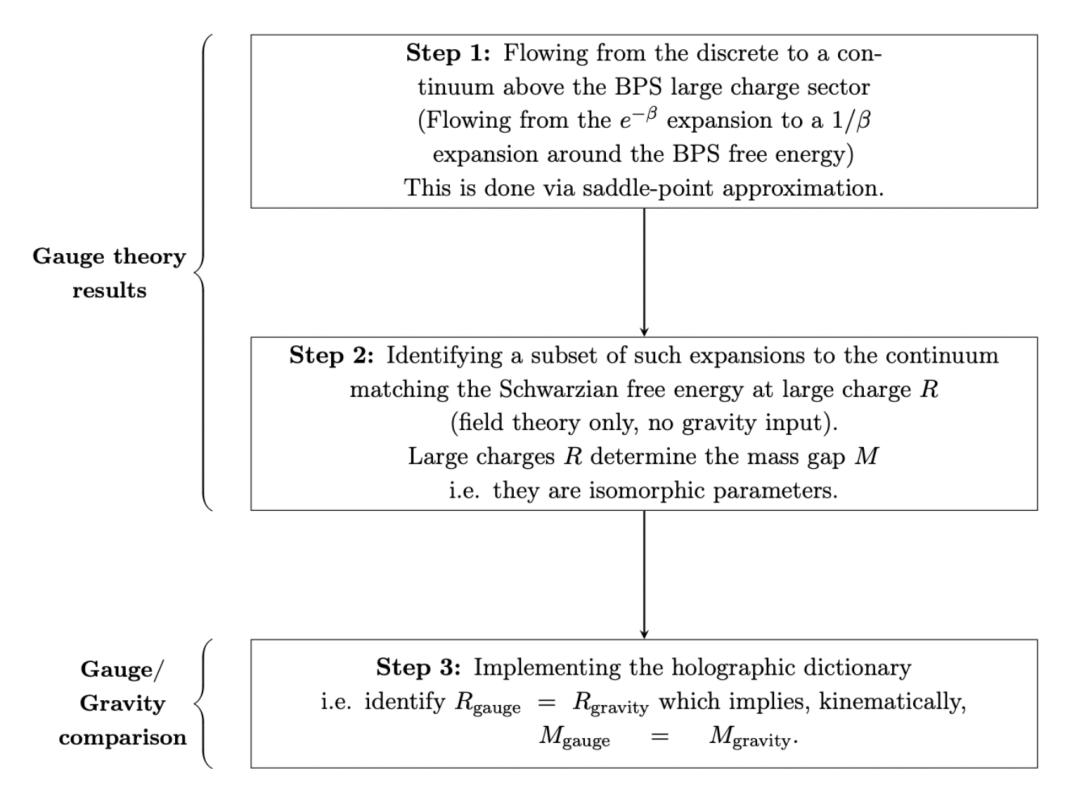}
  \label{fig:example}
\end{figure}

\paragraph{Explanation of the match with gravity} 

How can a mass gap computed in gravity be matched by a zero-coupling computation in the gauge theory after simply using the holographic dictionary? This may sound too good to be true, it is however, tautologically true once Step 2 has been reached in field theory.

In gauge theory (resp. gravity) the Schwarzian emerges in the presence of an expectation value for the R-charge $R$ (resp. electric charge). Let us denote such expectation value as $R_{\text{SUSY}}\,$. 

This is the expectation value of $R$ within the supersymmetric set of states
\be
R_{\text{SUSY}}\,=\, R\Big|_{\text{SUSY operators}}\,.
\ee

The expectation value $R_{SUSY}$ is a linear function of the temperature $T\,$, 
\be
R_{\text{SUSY}}\,=\,R_{\text{SUSY}}(T)\,=\,R_{BPS}+R_1 T\,.
\ee

On the other hand the (super)Schwarzian theory is defined by its mass gap $M$ (aside from its intrinsic temperature $T$ and possibly other chemical potentials); which happens to be a susceptibility in temperature $T$ at $T=0$ of the R-charge average value $R_{SUSY}=R_{SUSY}(T)\,$. 
\be
M \,\propto\, \frac{1}{R_1}\,.
\ee
Please refer to equations \eqref{eq:DefSuscept}-\eqref{eq:MassGapSusceptibility} for the details on this isomorphism.~\footnote{In particular the proportionality constant between the mass gap and the average charge $R$ is pure imaginary. We will comment on the physical interpretation of this fact.}

This explains why identifying background R-charges in both sides of the duality, automatically identifies the mass gaps. 

{\theorem{}{Of course, just matching charges on both sides, i.e., imposing the holographic dictionary most probably will not be enough to match all further higher temperature corrections beyond the Schwarzian one (which is linear in temperature) in both sides of the duality. The fact that all the dynamics of the Schwarzian is encoded in fluctuations of expectation values of global charges suggests that the family of large charge expansions leading to Step 2, may be understood as a hydrodynamic expansion of $\mathcal{N}=4$ SYM. It would be very interesting to make this observation more precise.}}

\section{The partition function}
\label{sec:2}

The space of states of $\mathcal{N}=4$ SYM on~$S^3$ can be constructed with a set of 16 raising and lowering operators and an auxiliary vacuum state~$|0\rangle$~\cite{Kinney:2005ej}. These operators can be divided into 8 bosons and 8 fermions.
The bosons, which we will denote as~$a^{\pm}$, $a_{\pm}$, $b^{\pm}$,~$b_{\pm}$  form an 8-dimensional spinoral representation of the conformal group in four-dimensions~$SO(2,4)$.
The fermions, which we will denote as~$\mathfrak{f}^{1,2,3,4}$\,,\, $\mathfrak{f}_{1,2,3,4}\,$, form an 8-dimensional spinorial representation of the R-symmetry group~$SO(6)\,$. These operators obey canonical commutation rules,
\begin{equation}\label{eq:CommRel}
\begin{split}
\left[a^{\eta},a_{\gamma}\right] &=\delta^{\eta}_{\gamma}\,,\,\qquad \left[b^{\dot{\eta}},b_{\dot{\gamma}}\right]\,=\,\delta^{\dot{\eta}}_{\dot{\gamma}}\,,\qquad \{\mathfrak{f}^{n},\mathfrak{f}_{m}\}\,=\,\delta^{n}_m\,,\\\eta,\, \dot{\eta}\,,\,\gamma\,,&\dot{\gamma}\,=\,-,+\,\qquad n,m=1,2,3,4\,.
\end{split}
\end{equation}
The Fock vacuum is defined by the conditions
\begin{equation}
a^{\pm}\,|0\rangle\,=\,0\,\,,\, b^{\pm}\,|0\rangle\,=\,0\,\,,\,\mathfrak{f}^{1}\,|0\rangle\,=\,0\,,\,\mathfrak{f}^{2}\,|0\rangle\,=\,0\,,\,\mathfrak{f}_{3}\,|0\rangle\,=\,0\,,\,\mathfrak{f}_{4}\,|0\rangle\,=\,0\,.
\end{equation}
The~$\mathfrak{f}^{1,2}$ are lowering operators and the~$\mathfrak{f}^{3,4}$ are rising operators. Operators with supraindices are complex conjugated to operators with subindices . The scalar single-states operators in the theory $X_{1}$, $X_2$, $X_3$,~$\overline{X}_1$, $\overline{X}_2$, and~$\overline{X}_3$ are isomorphic to the states~\footnote{For more details of this construction please refer to~\cite{Alday:2005kq} and~\cite{Kinney:2005ej}. In this section, we will use conventions which are closely related to the ones in Appendix A of~\cite{Kinney:2005ej}. }
\be
\begin{split}
X_1&\,\leftrightarrow\,\mathfrak{f}_1\mathfrak{f}_2\mathfrak{f}^3\mathfrak{f}^4|0\rangle\,,\, \qquad\,\,\,\,\overline{X}_1\,\leftrightarrow\, |0\rangle\,,\,\\ X_2&\,\leftrightarrow\,\mathfrak{f}_1 \mathfrak{f}^4|0\rangle\,,\qquad \qquad \,\,\overline{X}_2\,\leftrightarrow\,\mathfrak{f}_2 \mathfrak{f}^3|0\rangle\,,\, \\{X}_3&\,\leftrightarrow\,\mathfrak{f}_2 \mathfrak{f}^4|0\rangle\,,\,\qquad\qquad\,\overline{X}_3\,\leftrightarrow\,\mathfrak{f}_1 \mathfrak{f}^3|0\rangle\,. 
\end{split}
\ee

The symmetry generators in this theory, e.g. 
dilations~$E$,~ the two independent angular momenta~$J_{1}$ and~$J_2$ and R-charges~$R_{1}$,~$R_2$ and~$R_3$
\be\label{eq:BosSymm}
\{E, J_1^3, J_2^3, R_1, R_2, R_3\}
\ee
can be expressed as quadratic combinations of oscillators~$a$, $b$ and~$\mathfrak{f}\,$.~\footnote{We will use the definitions of charges given in Appendix A of~\cite{Kinney:2005ej}, equation~(A.2), with the relations~$\widetilde{\mathfrak{f}}^n=\alpha^{n}_{\text{there}}\,$,~$(a^{\eta})_{here}\,=\,(a_{\eta})_{there}\,$, ~$(b^{\dot{\eta}})_{here}\,=\,(b_{\dot{\eta}})_{there}$. Moreover,~$(R_{a})_{\text{here}}=(-R_{a})_{\text{there}}\,$.}
In particular, the 32 supercharges of the theory are
\be
\mathcal{Q}^{n,\pm}=\widetilde{\mathfrak{f}}^{n}a_{\pm},\, \mathcal{S}_{n,\pm}=\widetilde{\mathfrak{f}}_n a^{\pm}, \, \overline{\mathcal{Q}}^{n,\pm}\,=\,\widetilde{\mathfrak{f}}_{n} b_{\pm}\,,\, \overline{\mathcal{S}}_{n,\pm}\,=\, \widetilde{\mathfrak{f}}^{n} b^{\pm}\,.
\ee
where
\be
\begin{split}
\widetilde{\mathfrak{f}}^n &= \mathfrak{f}_n \qquad \text{if} \qquad n\,=\,1\,,\,2\,,\\
\widetilde{\mathfrak{f}}^n &= \mathfrak{f}^n \qquad \text{if} \qquad n\,=\,3\,,\,4\,.
\end{split}
\ee
The~$\dagger$-operation raises/lowers indices of single oscillators~$\mathfrak{f}$,~$a$ and~$b\,$. Thus~$\mathcal{S}_{n,\pm}=(\mathcal{Q}^{n,\pm})^\dagger\,$, $\overline{\mathcal{S}}_{n,\pm}=(\overline{\mathcal{Q}}^{n,\pm})^\dagger\,$. In our conventions the R-charge generators are
\be
\begin{split}
R_{1}\,:=\,\widetilde{\mathfrak{f}}_2\widetilde{\mathfrak{f}}^2 -\widetilde{\mathfrak{f}}_1 \widetilde{\mathfrak{f}}^1 \,,\\
R_{2}\,:=\,\widetilde{\mathfrak{f}}_3\widetilde{\mathfrak{f}}^3 -\widetilde{\mathfrak{f}}_2 \widetilde{\mathfrak{f}}^2 \,,\\
R_{3}\,:=\,\widetilde{\mathfrak{f}}_4 \widetilde{\mathfrak{f}}^4 -\widetilde{\mathfrak{f}}_3 \widetilde{\mathfrak{f}}^3\,.
\end{split}
\ee
Using these expressions one obtains the R-charges of the scalar single states, which are summarized in Table~\ref{table:Tabla0}.
\begin{table}[h!]
    \centering
    \begin{tabular}{cccc}
      Scalars    & $R_1$ & $R_2$ & $R_3$ \\
        $X_1$ & $0$  & $-1$ & $\,\,\,0$\\
        $X_2$ & $-1$ & $+1$ & $-1$ \\
        $X_3$ & $+1$ & $\,\,\,0$  & $-1$ \\
    \end{tabular}
    \caption{R-charges of the scalar single-states~$X_{1,2,3}$. Barred scalars have opposite charges.}
    \label{table:Tabla0}
\end{table}

We will be interested in two sets of complex-conjugated supercharges. The first couple
\begin{equation}
\mathcal{Q}:=\mathcal{Q}^{4,-}\,,\qquad \mathcal{S}:=\mathcal{Q}^\dagger=\mathcal{S}_{4,-}\,,
\end{equation}
whose anti-commutation relation, following from~\eqref{eq:CommRel}, is
\be\label{eq:SemipositivityDelta}
\Delta\,=\,2\{\mathcal{Q},\mathcal{S}\}\,=\,H\,-\,2 J^3_1\,-\,2\,\sum_{k=1}^{3} \frac{k}{4}\, R_k\,\geq\,0\,.
\ee
To obtain this commutation relation one must use the fact that the central element of the oscillator algebra
\be
\mathcal{C}\,:=\, b^+b_+\,+\,b^-b_- \,-\,a^+ a_+\,-\,a^-a_- -\widetilde{\mathfrak{f}}_n \widetilde{\mathfrak{f}}^n \,=\, -\,Z_1\,-\,B_1 \,-\,2\,=\,-2\,,
\ee
equals~$-2\,$, and that the~$U(1)_J$ and~$U(1)_B$ charges~\cite{Alday:2005kq}
\be
-Z_1\,=\,N_b \,-\, N_a\,,\qquad\,\,\,\,\,\, -B_1\,=\,-\widetilde{\mathfrak{f}}_n\widetilde{\mathfrak{f}}^n\,+\,2\,=\,N_\beta \,-\,N_\alpha\,,
\ee
vanish on the physical states of the theory.~\footnote{For example, these contraints imply that at the level of eigenvalues~$\widetilde{\mathfrak{f}}_4\widetilde{\mathfrak{f}}^4 \,=\,-\sum_{n=1}^3 \widetilde{\mathfrak{f}}_n\widetilde{\mathfrak{f}}^n\,+\,2\,$. The oscilators~$\widetilde{\mathfrak{f}}_n$ are the ones denoted as~$\alpha_n$ in reference~\cite{Kinney:2005ej}.
} In our conventions, the number operators are~\cite{Alday:2005kq}
\be
\begin{split}
N_b&\,=\, b_+ b^+\,+\, b_- b^-\,, \quad\,\,\,\,\, N_a\,=\,  a_+ a^+\,+\, a_- a^-\,,\\
N_{\beta}&\,=\,\mathfrak{f}^3\mathfrak{f}_3 \,+\,\mathfrak{f}^4\mathfrak{f}_4\,,\qquad\quad N_{\alpha}\,=\,  \mathfrak{f}_1 \mathfrak{f}^1 +\mathfrak{f}_2\mathfrak{f}^2\,.
\end{split}
\ee
$Q$ and~$S$ will be the supercharges preserved by the BPS sector around which we want to compute low-temperature corrections. The other two supercharges we will work with are, either
\begin{equation}
\overline{\mathcal{Q}}^{2,+}\,, \qquad \overline{\mathcal{S}}_{2,+}= (\overline{\mathcal{Q}}^{2,+})^\dagger\,,
\end{equation}
or
\be
\overline{\mathcal{Q}}^{2,-}\,, \qquad \overline{\mathcal{S}}_{2,-}= (\overline{\mathcal{Q}}^{2,-})^\dagger\,,
\ee
and their anticommutation relations are
\be\label{eq:Q2pm}
\begin{split}
\,2\{\overline{\mathcal{Q}}^{2,+},\overline{\mathcal{S}}_{2,+}\}&\,=\,H\,+\,2 J^3_2\,+\,\frac{R_1}{2}-{R_2}-\frac{R_3}{2}\\
&\,=\Delta_+\,:=\,\Delta \,+\,\Delta^{(2)}_{+}\,\geq\,0\\
\,2\{\overline{\mathcal{Q}}^{2,-},\overline{\mathcal{S}}_{2,-}\}&\,=\,H\,-\,2 J^3_2\,+\,\frac{R_1}{2}-{R_2}-\frac{R_3}{2}\\
&\,=\,\Delta_-\,:=\,\Delta \,+\,\Delta^{(2)}_{-}\,\geq\,0
\end{split}
\ee
where
\be
\Delta^{(2)}_{\pm}\,:=\, 2\bigl({J}^3_1\,\pm\,{J}^3_2\bigr)\,+\, {R_1+R_3}\,.
\ee
The weights of these supercharges under the action of the bosonic symmetries~\eqref{eq:BosSymm} are
\be\label{eq:ChargesofTheSupercharge}\begin{split}
\mathcal{Q}^{4,-}\,\to\,\bigl\{\frac{1}{2},-\frac{1}{2},0,0,1\bigr\}\,,\qquad\overline{\mathcal{Q}}^{2,\pm}\,\to\,\bigl\{\frac{1}{2},0,\pm\frac{1}{2},-1,+1,0\bigr\}\,.
\end{split}
\ee
For latter reference we note that the set of bosonic charges that commute simultaneously with~$\mathcal{Q}^{4,-}$,~$\mathcal{S}_{4,-}$,~$\overline{\mathcal{Q}}^{2,+}$, and~$\overline{\mathcal{S}}_{2,+}$ is generated by
\be\label{eq:MutuallyCommutingCharges}
\Delta\,,\, R_1+R_2\,,\, H\,+\,J^3_1\,-\,R_2/2\,, \,+\, J^3_{2}\,-\, R_2/2\,.
\ee
The set of those that commute simultaneously with~$\mathcal{Q}^{4,-}$,~$\mathcal{S}_{4,-}$,~$\overline{\mathcal{Q}}^{2,-}$, and~$\overline{\mathcal{S}}_{2,-}$ is generated instead by
\be
\Delta\,,\, R_1+R_2\,,\, H\,+\,J^3_1\,-\,R_2/2\,, \,-\, J^3_{2}\,-\, R_2/2\,.
\ee

\subsection{The matrix integral at zero gauge coupling}

\label{sec:2p1}

In the language of the~$\mathcal{N}=1$ superconformal symmetry corresponding to the~$U(1)$ R-charge generator 
\be
R_3
\ee
the fundamental field content of the theory organizes in a vector, three chiral and three anti-chiral multiplets. The vector multiplet is composed of a vector field with~$R_3=0$ and~\emph{flavour} charges~$R_1=R_2=0$, and a gaugino with a chiral (and antichiral) components with~$R_3=+1$ (and~$-1$) and flavour charges $R_1=R_2=0\,$. The R-charges of the chiral multiplets can be found in table~\ref{table:Tabla1}. The R-charges of the anti-chiral multiplets take the opposite values.
\begin{table}[h!]
    \centering
    \begin{tabular}{cccc}
      chiral mutliplet $I$    & $R_1$ & $R_2$ & $R_3$ \\
        1 & $0$  & $-1$ & $\,\,\,0$\\
        2 & $-1$ & $+1$ & $-1$ \\
        3 & $+1$ & $\,\,\,0$  & $-1$ \\
    \end{tabular}
    \caption{R-charges of the chiral multiplets. The same ones as their corresponding scalar components quoted in table~\ref{table:Tabla0}. The antichiral multiplets have opposite charges. The scalars in the chiral(antichiral)-multiplets have the same R-charges as their multiplets. The fermions in the chiral(resp. antichiral)-multiplets have the same charges as the scalars under~$R_{1}$ and~$R_2$. Their charge under~$R_3$ increases by~$+1$ (resp.$-1$) with respect to the R-charge of the scalar in the same multiplet. }
    \label{table:Tabla1}
\end{table}

For later use we note that
\be\label{eq:MinusOneF}
(-1)^F \,=\, e^{\pi \i (R_1+R_3)}\,.
\ee
After a straightforward computation the partition function~$Z$ of the free theory~\cite{Aharony:2003sx}
\be\label{eq:PartitionFunction}
Z=e^{-\mathcal{F}}=Z[x,u,v,w,t,y]:=\text{Tr}_{\mathcal{H}}x^\Delta u^{-R_1-R_3} v^{-R_1} w^{-R_2} t^{2 (H+J^3_1)} y^{2 J^3_2}
\ee
reduces to 
\be\label{eq:PartitionFunction2}
Z\,=\, e^{-\mathcal{F}}\,=\,\int [DU] e^{-\mathcal{F}^{\infty}_{sl}[x,u,v,w,t,y; U]}
\ee
where
\be\label{eq:FSL}
\begin{split}
-\mathcal{F}^{\Lambda}_{sl}[x,u,v,w,t,y; U]&\,:=\,\sum_{j=1}^{\Lambda^{n+1}} \frac{1}{j} \left(f_{Bos}[x^j,u^j,v^j,w^j,t^j,y^j] \right. \\ &\left.\qquad +\, (-1)^{j+1} f_{Fer}[x^j,u^j,v^j,w^j,t^j,y^j]\right)\text{Tr}U^j \text{Tr}U^{\dagger j}\,,
\end{split}
\ee
and
\be\label{eq:BosTotalSL}
\begin{split}
f_{Bos}&:= f^{(V)}_{Bos}+\sum_{I=1}^3f^{(I)}_{Bos} \\
&\,=\,-\frac{t^6 y^2 \left(u^2 x^4
   \left(v \left(u^2 v+w^2\right)+w\right)+x^2 \left(u^2 v (v w+1)+w^2\right)\right)}{u^2 v w \left(t^3-y\right) \left(t^3
   y-1\right) \left(t x^2-y\right) \left(t x^2 y-1\right)}
    \\
 &+\frac{t^5 y \left(2 t^3 x^4 y-2 t^2 x^2 \left(y^2+1\right)+t y-2 x^4
   \left(y^2+1\right)\right)}{\left(t^3-y\right) \left(t^3 y-1\right) \left(t
   x^2-y\right) \left(t x^2 y-1\right)}
    \\
 &+ \frac{t^4 u^2 v w x^2
   \left(y^2+1\right)^2+t^2 y^2 \left(u^4 v^2 x^2+u^2 \left(v^2 w+v w x^2
   \left(w+x^2\right)+v+w x^2\right)+w^2\right)}{u^2 v w \left(t^3-y\right) \left(t^3
   y-1\right) \left(t x^2-y\right) \left(t x^2 y-1\right)}\,,
   \end{split}
\ee
\be\label{eq:FerTotalSL}
\begin{split}
f_{Fer}&:= f^{(V)}_{Fer}+\sum_{I=1}^3f^{(I)}_{Fer}\\
&= \frac{t^3 y \left(x^2 \left(u^2 v \left(t^2 (-v)+v w+1\right)+w \left(w-t^2 (v
   w+1)\right)\right)+v w\right)}{u v w \left(t^3-y\right) \left(t^3 y-1\right) \left(t
   x^2-y\right) \left(t x^2 y-1\right)} \\
   &+\frac{t^2 y \left(t^4 (-v) y \left(u^2 x^2+w\right)-t^3 u^2 v w x^4+t^2 y \left(v
   \left(u^2 v+w^2\right)+w\right)+u^2 v^2 x^2 y\right)}{u v w \left(t^3-y\right)
   \left(t^3 y-1\right) \left(t x^2-y\right) \left(t x^2 y-1\right)} \\
\end{split}
\ee
\be\label{eq:FBosFFer}
\begin{split}
\qquad\qquad   &+\frac{t^2 x^2 y^2 \left(-\left(t^4 \left(u^2 v^2+w\right)\right)+t^2 \left(u^2-1\right)
   v+v w+1\right)}{u v \left(t^3-y\right) \left(t^3 y-1\right) \left(t x^2-y\right)
   \left(t x^2 y-1\right)}\\&+\frac{t^2 y^2 \left(-t^2 x^4 \left(u^2 v (v w+1)+w^2\right)+t v y \left(u^2
   x^2+w\right)+u^2 v w x^4\right)}{u v w \left(t^3-y\right) \left(t^3 y-1\right)
   \left(t x^2-y\right) \left(t x^2 y-1\right)}\\&-\frac{t^3 x^2 y^3 \left(t^2 \left(v \left(u^2 \left(v+w
   x^2\right)+w^2\right)+w\right)-w \left(u^2 v^2+w\right)\right)}{u v w
   \left(t^3-y\right) \left(t^3 y-1\right) \left(t x^2-y\right) \left(t x^2 y-1\right)}\,.
\end{split}
\ee
Let us give some details on how this expressions above were derived. The contributions coming from vector multiplets are
\be
\begin{split}
f^{(V)}_{Bos}&=\sum_{j=1}^{\infty}\sum_{j^3_1=-\frac{j+1}{2}}^{\frac{j+1}{2}}\sum_{j^3_2=-\frac{j-1}{2}}^{\frac{j-1}{2}} (x^{\Delta_+}t^{2(\epsilon^{(1)}_j+j^3_1)}y^{2 j^3_2} u^{{-R_1}-{R_3}} v^{-R_1} w^{{-R_2}})  \\&\,+\,\sum_{j=1}^{\infty}\sum_{j^3_1=-\frac{j-1}{2}}^{\frac{j-1}{2}}\sum_{j^3_2=-\frac{j+1}{2}}^{\frac{j+1}{2}} (x^{\Delta_-}t^{2(\epsilon^{(1)}_j-j^3_1)}y^{-2 j^3_2} u^{{+R_1}+{R_3}} v^{+{R_1}} w^{+{R_2}})
\end{split}
\ee
and
\be
\begin{split}
f^{(V)}_{Fer}&\,=\,\sum_{j=1}^{\infty}\sum_{j^3_1=-\frac{j}{2}}^{\frac{j}{2}}\sum_{j^3_2=-\frac{j-1}{2}}^{\frac{j-1}{2}} x^{\Delta^+}t^{2(\epsilon_j^{(\frac{1}{2})}+j^3_1)}y^{2 j^3_2}u^{{-R_1}-{R_3}}v^{{-R_1}}w^{{-R_2}} \\&\,+\,\sum_{j=1}^{\infty}\sum_{j^3_1=-\frac{j-1}{2}}^{\frac{j-1}{2}}\sum_{j^3_2=-\frac{j}{2}}^{\frac{j}{2}} x^{\Delta^-}t^{2(\epsilon_j^{(\frac{1}{2})}-j^3_1)}y^{-2 j^3_2}u^{{+R_1}+{R_3}} v^{+{R_1}}w^{+{R_2}}\,.
\end{split}
\ee
The contributions coming from chiral+antichiral multiplets are
\be
\begin{split}
f^{(I)}_{Bos}&\,=\,\sum_{j=0}^{\infty}\sum_{j^3_1=-\frac{j}{2}}^{\frac{j}{2}}\sum_{j^3_2=-\frac{j}{2}}^{\frac{j}{2}} x^{\Delta^+}t^{2(\epsilon_j^{(0)}+j^3_1)}y^{2 j^3_2}u^{{-R_1}-{R_3}} v^{{-R_1}}w^{{-R_2}} \\&\,+\,\sum_{j=0}^{\infty}\sum_{j^3_1=-\frac{j}{2}}^{\frac{j}{2}}\sum_{j^3_2=-\frac{j}{2}}^{\frac{j}{2}} x^{\Delta^-}t^{2(\epsilon_j^{(0)}-j^3_1)}y^{-2 j^3_2}u^{{+R_1}+{R_3}} v^{+{R_1}}w^{+{R_2}}
\end{split}
\ee
and
\be
\begin{split}
f^{(I)}_{Fer}&\,=\,\sum_{j=1}^{\infty}\sum_{j^3_1=-\frac{j}{2}}^{\frac{j}{2}}\sum_{j^3_2=-\frac{j-1}{2}}^{\frac{j-1}{2}} x^{\Delta^+}t^{2(\epsilon_j^{(\frac{1}{2})}+j^3_1)}y^{2 j^3_2}u^{{-R_1}-{R_3}}v^{{-R_1}}w^{{-R_2}} \\&\,+\,\sum_{j=1}^{\infty}\sum_{j^3_1=-\frac{j-1}{2}}^{\frac{j-1}{2}}\sum_{j^3_2=-\frac{j}{2}}^{\frac{j}{2}} x^{\Delta^-}t^{2(\epsilon_j^{(\frac{1}{2})}-j^3_1)}y^{-2 j^3_2}u^{{+R_1}+{R_3}}v^{+{R_1}}w^{+{R_2}}\,.
\end{split}
\ee
In these expressions we have used the following definitions
\be
\Delta^{\pm}=\epsilon_j\mp 2 j^3_1\mp\frac{1}{2} {R_1} \mp {R_2}\mp \frac{3}{2}{R_3}\,,\,
\epsilon_j^{(1)}\,=\,j+1\,,\, \epsilon^{(\frac{1}{2})}_j\,=\,j+\frac{1}{2}\,,\,\epsilon^{(0)}_j\,=\,j+1\,,
\ee
where~$j^3_1$ and~$j^3_2$ are the eigenvalues of~$J^3_1$ and~$J^3_2$ respectively.

After resumming the series in these contributions above, and summing the result over the R-charge values of the vector multiplet components, we obtain
\be
\begin{split}
f^{(V)}_{Bos}&\,=\,\frac{2 t^8 x^4 y^2-2 t^7 x^2 y \left(y^2+1\right)+t^6 y^2-2 t^5 x^4 y \left(y^2+1\right)+t^4 x^2
   \left(y^2+1\right)^2+t^2 x^4 y^2}{\left(t^3-y\right) \left(t^3 y-1\right) \left(t x^2-y\right) \left(t x^2
   y-1\right)} \\
f^{(V)}_{Fer}   &\,=\,\frac{t^2 y \left(-t^4 y-t^3 u^2 x^4 \left(y^2+1\right)+t^2 \left(u^2-1\right) x^2 y+t \left(y^2+1\right)+u^2 x^4
   y\right)}{u \left(t^3-y\right) \left(t^3 y-1\right) \left(t x^2-y\right) \left(t x^2 y-1\right)}\,.
\end{split}
\ee
After summing 
\begin{equation}
\begin{split}
f^{(I)}_{Bos}&\,=\,\frac{t^2 y^2 u^{{R_1}+{R_3}} v^{{R_1}} w^{{R_2}} x^{\frac{{R_1}}{2}+{R_2}+\frac{3 {R_3}}{2}+1} \left(1-t^4 x^2\right)
   \left(u^{-2 \left({R_1}+ {R_3}\right)} v^{-{2 R_1}} w^{- { 2 R_2}} x^{{-R_1}- { 2 R_2}-3 {R_3}}+1\right)}{\left(t^3-y\right) \left(t^3
   y-1\right) \left(t x^2-y\right) \left(t x^2 y-1\right)} \\
 f^{(I)}_{Fer}  &\,=\, \frac{t^2 y u^{{-R_1}- {R_3}} v^{{-R_1}} w^{{-R_2}} x^{\frac{1}{2} \left({-R_1}- { 2 R_2}-3 {R_3}+1\right)} \left(-t^3 x^2
   \left(y^2+1\right)+t^2 y+x^2 y\right)}{\left(t^3-y\right) \left(t^3 y-1\right) \left(t x^2-y\right) \left(t x^2
   y-1\right)} \\
& \qquad\qquad  +\frac{t^3 y u^{+{R_1}+{R_3}} v^{+{R_1}} w^{+{R_2}} x^{\frac{1}{2} \left(+{R_1}+ { 2 R_2}+{3R_3}+3\right)} \left(y \left(-t^3-t
   x^2+y\right)+1\right)}{\left(t^3-y\right) \left(t^3 y-1\right) \left(t x^2-y\right) \left(t x^2 y-1\right)}\,,
   \end{split}
\end{equation}
over the R-charges~$I=1,2,3$ reported in the table~\ref{table:Tabla1} and summing over contributions coming from bosons and fermions in the vector multiplets, we obtain the total contributions from bosons~\eqref{eq:BosTotalSL} and fermions~\eqref{eq:FerTotalSL} to~$\mathcal{F}_{sl}^\Lambda\,$.

For completeness, we report the translation to the writing of the partition function given in equation~(2.5) of~\cite{Choi:2018hmj},
\be
\begin{split}
e^{-\beta_{\text{there}}} &\,=\, x\,t^2\,,\, e^{-\omega_{1\text{there}}}\,=\,\frac{t}{y x}\,,\, e^{-\omega_{2\text{there}}}\,=\,\frac{t y}{x}\,,\\ \, e^{\Delta_{1\text{there}}}&\,=\,{x w}\,,\,e^{\Delta_{2\text{there}}}\,=\,x \frac{ u^2 v}{w}\,,\,e^{\Delta_{3\text{there}}}\,=\,\frac{x}{v}\,,\\ \,e^{2\Delta_{\text{there}}}&\,=\,e^{\Delta_{1\text{there}}+\Delta_{2\text{there}}+\Delta_{3\text{there}}}\,=\,{x^{3} u^2}\,.
\end{split}
\ee
Eventually, we will use the following definitions of rapidities in terms of chemical potentials
\be\label{eq:DefChemicalPotentials}
\widetilde{x}^2\,:=\, \frac{t x^2 }{y}\,,\,x\,=\, {e^{-\beta}}\,,\, \frac{t^3}{y}\,=\, e^{-\omega_1}\, ,\,{t^3}{y}\,=\, e^{-\omega_2} \,,\, {t^2}{v}\,=\,e^{-\varphi_v}\,,\,\frac{w}{t^2}\,=\,e^{\varphi_w}
\ee
which can be equivalently written as follows
\be\label{eq:RelsRapiditiesChemicalPotentials}
t\,=\,e^{-\frac{1}{6} (\omega_1+\omega_2)}\,,\,y\,=\,e^{\frac{\omega_1-\omega_2}{2}}\,,\,v\,=\,e^{\frac{1}{3} (\omega_1+\omega_2-3 \varphi_v)}\,,\,w\,=\,e^{-\frac{1}{3} (\omega_1+\omega_2-3 \varphi_w)}\,.
\ee

\paragraph{The BPS locus $\alpha=\pm\frac{1}{2}$}
Note that from~\eqref{eq:MinusOneF} it follows that at
\be\label{2.35}
u:= e^{2\pi\text{i}\alpha}= e^{\pm \pi\i}
\ee
cancellations happen and the dependence on~$x$ dissapears in
\be
f_{Bos}[x^j,\ldots]+ (-1)^{j+1}f_{Fer}[x^j,\ldots] \,=\,\mathcal{I}^{4,-}_{sl}[v^j,w^j,t^j,y^j]
\ee
where
\be\label{eq:Index4mSl}
\mathcal{I}^{4,-}_{sl}[v,w,t,y]\,:=\, 1\,-\,\frac{\left({t^2}{v}-1\right) \left(\frac{t^2}{w}-1\right) \left(\frac{t^2
   w}{v}-1\right)}{\left(\frac{t^3}{y}-1\right) \left(1-t^3 y\right)}\,.
\ee
This implies that at the value of chemical potential~\eqref{2.35} the partition function equals
\be
Z[x,u=-1,v,w,t,y]\,=\,e^{-\mathcal{F}[x,u=-1,v,w,t,y]}\,=\,\mathcal{I}^{4,-}(v,w,t,y)\,,
\ee
where
\be\label{eq:TraceIndex4m}
\begin{split}
\mathcal{I}^{4,-}&:= \int [DU] e^{\sum_{j=1}^\infty \frac{1}{j} \,\mathcal{I}^{4,-}_{sl}(v^j,w^j,t^j,y^j)\, \text{Tr}U^j \text{Tr}U^{\dagger j}}\\ &=\text{Tr}_{\mathcal{H}}(-1)^F\,x^{2\{Q,S\}} v^{-R_1} w^{-R_2} t^{2 (H+J^3_1)} y^{2 J^3_2} \,=:\, \mathcal{I}_1\,,
\end{split}
\ee
is the $\frac{1}{16}$-BPS superconformal index counting states in the cohomologies of~$Q=\mathcal{Q}^{4,-}$ and~$S=\mathcal{S}_{4,-}=Q^\dagger\,$, for which~$\Delta=0\,$~\cite{Romelsberger:2005eg,Kinney:2005ej}\cite{Dolan:2007rq,Dolan:2008qi,Nawata:2011un}. States which are not in such cohomology do not contribute to this index and thus~$\mathcal{I}^{4,-}$ does not depend on~$x\,$, as it has been explicitly shown in~\eqref{eq:Index4mSl}. We will also denote this index as~$\mathcal{I}_1$.

Further imposing
\be\label{eq:EnhancementSUSY}
w\,=\, v{ t y}\qquad  \text{or}\qquad  \varphi_w \,=\,-\varphi_v +\omega_1+\omega_2\,,
\ee
on~\eqref{eq:Index4mSl} we find that
\be
\mathcal{I}_{sl}^{4,-;2,+}[v,t,y]\,=\,1-\frac{\left(1-{t^2}{v}\right) \left(1-\frac{t }{v y}\right)}{1-\frac{t^3}{y}}\,.
\ee
Indeed,~\eqref{eq:EnhancementSUSY} implies, more generally, that
\be\label{eq:IdentitySuSYEnhancement}
Z[x,u\,=\,-1,v,w\,=\,{v}{t y},t,y]\,=\,\mathcal{I}^{4,-;2,+}[v,t,y]\,.
\ee
where
\be\label{eq:TraceIndex4m2p}
\begin{split}
\mathcal{I}^{4,-;2,+}&:= \int [DU] e^{\sum_{j=1}^\infty \frac{1}{j} \,\mathcal{I}^{4,-;2,+}_{sl}(v^j,t^j,y^j)\, \text{Tr}U^j \text{Tr}U^{\dagger j}} \\ &=\text{Tr}_{\mathcal{H}}(-1)^F\,x^{2\{Q,S\}} v^{-R_1-R_2} t^{2 (H+J^3_1)-R_2} y^{2 J^3_2-R_2}
\end{split}
\ee
is the $\frac{1}{8}$-BPS superconformal index counting states in the cohomologies of~$\mathcal{Q}=\mathcal{Q}^{4,-}$,~$\mathcal{S}=\mathcal{S}_{4,-}\,$,~$\overline{\mathcal{Q}}^{2,+}$ and~$\overline{\mathcal{S}}_{2,+}\,$: the Macdonald index~\cite{Gadde:2011uv} associated to the latter four supercharges. 
\subsection{An index to compute higher order thermal corrections at strong coupling}
\label{sec:2p2}

To compute physical thermal corrections around the point at which supersymmetric cancellations occur
\be
\alpha=\frac{1}{2}
\ee
one can define the following restriction of the fully refined partition function~\eqref{eq:PartitionFunction2}
\be\label{eq:Index2}
\begin{split}
Z[x,u=e^{2\pi\i\alpha},e^{-2\pi\i(\alpha-1/2)}\widetilde{v},w={\widetilde{v}}{t y}\,,\,t,\,y]&\,=\,\mathcal{I}_2[x,e^{2\pi\i\alpha},\widetilde{v},t,y]\,=\, e^{-\mathcal{F}_{\frac{1}{16}\text{near}\frac{1}{8}}}\,.
\end{split}
\ee
Equation~\eqref{eq:IdentitySuSYEnhancement} implies that at~$\alpha=\frac{1}{2}$ the partition function~$\mathcal{I}_2$ reduces to a 1/8-BPS index
\be
\mathcal{I}_2[x,-1,\widetilde{v},t,y]\,=\,\mathcal{I}^{4,-;2,+}[\widetilde{v},t,y]\,.
\ee
Remarkably, for any~$\alpha\,\neq\, \frac{1}{2}\,$, the restricted partition function~$\mathcal{I}_2\,$ remains a superconformal index.~\footnote{That is why we have attached the subindex~$\frac{1}{16}near\frac{1}{8}$ to the free energy in equation~\eqref{eq:Index2}.} That follows from the fact that the Taylor coefficients of~\eqref{eq:Index2} at~$\alpha=\frac{1}{2}$
\be
\text{Tr}_{\mathcal{H}} (R_3)^n (-1)^F x^{\Delta}\widetilde{v}^{-R_1-R_2} t^{2 (H+J^3_1)-R_2} y^{2 J^3_2-R_2}\,
\ee
are protected observables under the supercharges~$\overline{\mathcal{Q}}^{2,+}$ and~$\overline{\mathcal{S}}_{2,+}\,$.~\footnote{Not under~$\mathcal{Q}$ and~$\mathcal{S}\,$ because they do not commute with~$R_3\,$ (See~\eqref{eq:ChargesofTheSupercharge}). } Indeed, starting from the fully refined partition function~\eqref{eq:PartitionFunction} and using the definition~\eqref{eq:Index2}, a straightforward computation gives us
\be
\mathcal{I}_2\,=\,\int [DU] e^{\sum_{j=1}^\infty \frac{1}{j} \,\mathcal{I}_{2,sl}(x^j,e^{2\pi\i j\alpha},\widetilde{v}^j,t^j,y^j)\, \text{Tr}U^j \text{Tr}U^{\dagger j}}
\ee
where
\be\label{eq:Indices}
\mathcal{I}_{2,sl}(x,e^{2\pi\i \alpha},\widetilde{v},t,y)\,=\,1\,+\,\frac{\left(1+e^{-2 i \pi  \alpha } t^2 \widetilde{v}\right) \left(\frac{t}{\widetilde{v} y}-1\right)
   \left(1+{e^{2 i \pi  \alpha } \widetilde{x}^2}\right)}{\left(\frac{t^3}{y}-1\right)
   \left(\widetilde{x}^2-1\right)}\,.
\ee
Note that~\eqref{eq:Indices} is the known expression for the single-letter maximally refined superconformal index. The interesting feature of~\eqref{eq:Indices} is that its rapidites depend explicitly on the physical temperature~$\frac{1}{\beta}$ of the system. For latter convenience we recall that
\be
\widetilde{v}\,:=\,e^{2\pi\i(\alpha-1/2)}v\,=\,\frac{e^{-\varphi_{\widetilde{v}}}}{t^2}\,, \qquad \widetilde{x}^2\,:=\, \frac{t x^2 }{y}\,.
\ee

\subsection{Protected near-{1}/{8}-BPS low-temperature corrections}
\label{sec:2p3}

Let us explain what has just been found. If we define inverse temperature~$\beta$ as the chemical potential dual to the twisted Hamiltonian obtained from the anticommutation of~$\mathcal{Q}_1=\mathcal{Q}$ and its complex conjugated supercharge~$\mathcal{S}_1=\mathcal{S}\,$, then states in the cohomology of~$\mathcal{Q}_1$ and~$\mathcal{S}_1$ are \emph{zero-temperature states} in the sense that their contribution to the partition function does not depend on temperature.  Indeed, the restricted partition function receiving contribution only from states in the~$\mathcal{Q}_1$ and $\mathcal{S}_1$-cohomology, which happens to be the superconformal index $\mathcal{I}_1=\mathcal{I}^{4,-}$, does not depend on~$\beta\,$.

On the other hand, states in the cohomology of~$\mathcal{Q}_2=\overline{\mathcal{Q}}^{2,+}$ and its complex conjugated supercharge~$\mathcal{S}_2=\overline{\mathcal{S}}_{2,+}$, which are not in the cohomology of~$\mathcal{Q}_1$ and~$\mathcal{S}_1$, are \emph{finite-temperature states} in the sense that their contribution to the physical partition function depends explicitly on temperature. The latter finite-temperature corrections are protected by~$\mathcal{Q}_2$,$\,\mathcal{S}_2$-supersymmetry and thus are computed by another superconformal index~$\mathcal{I}_2$ counting states in the cohomologies of~$\mathcal{Q}_2\,$ and~$\mathcal{S}_2\,$. Consequently, they do not receive corrections in the gauge coupling and can be computed, exactly, at zero gauge coupling.

In virtue of AdS/CFT conjecture, this predicts that~$\mathcal{I}_2$ encodes perturbatively low-temperature corrections of the gravitational on-shell action of the solutions of~\cite{Wu:2011gq}, when the latter family of solutions is expanded around its~$\frac{1}{8}$-BPS locus ($\Delta=\Delta_+=0$).

\section{The {holographic} low-temperature expansion}
\label{sec:3}

As explained in the introduction, to obtain perturbative corrections in \(\frac{1}{\mathfrak{b}}=\frac{1}{\beta R}\)~\footnote{Meaning by $\mathfrak{b}$ the dimensionful inverse temperature obtained by substituting~$\beta\to \frac{\mathfrak{b}}{R}$ in the previous equations.} consistent with the dual gravitational picture, it is necessary to implement an RG flow mechanism by which the discrete spectrum of the gauge theory effectively becomes dense. Otherwise, the Taylor expansion of \(\mathcal{F}\) at $\mathfrak{b}=\infty$ trivializes (as one can explicitly check from~\eqref{eq:PartitionFunction2}). 

This section defines such RG flow procedure to the continuum. It also derives the infrared free energy emerging after such flow, and it explains why it is bound to encode the low-temperature expansion of the Gibbons-Hawking onshell action of the black holes of~\cite{Chong:2005hr,Chong:2005da} even well beyond their BPS locus~\cite{Gutowski:2004ez,Gutowski:2004yv}.

\vspace{.1cm}

\subsection{The expansion to the continuum} \label{subsec:31}  The discreteness of the spectrum of $\mathcal{N}=4$ SYM is controlled by the radius of the~$S^3$,~$R\,$. In the discussion above we have fixed~$R=1\,$. The dependence in~$R$ can recovered by substituting
\be\label{eq:ContinuumDefinition}
\beta\,=\, \frac{\mathfrak{b}}{R}\,,\qquad \omega_{1}\,=\, \frac{\mathfrak{w}_1}{R}\,,\qquad \omega_{2} \,=\,\frac{\mathfrak{w}_2}{R}\,,\qquad \alpha\,-\,\frac{1}{2}\,=\,\frac{\mathfrak{a}}{R}\,,
\ee
in the partition function~$Z\,$ in~\eqref{eq:PartitionFunction2} (after using the relations~\eqref{eq:RelsRapiditiesChemicalPotentials}).

The naive limit to the continuum,~$R\to\infty\,$, with the dimensionless ratios in the left-hand sides of~\eqref{eq:ContinuumDefinition} fixed and finite is not the limit we are looking for.

For reasons that were explained in the introduction and that we will comeback to discuss below, we need another length scale, say~$1/\Lambda\,$. Then we define
\be\label{eq:Def}
\mathfrak{b}\,=:\, \frac{\beta_{0}[\Lambda R]}{\Lambda}\,,\qquad \, \mathfrak{a}\,=:\,\frac{{\alpha}_{0}[\Lambda R]}{\Lambda}\, \,,\qquad \mathfrak{w}_{1}\,=:\, \frac{\omega_{1,0}[\Lambda R]}{\Lambda}\,,\qquad \mathfrak{w}_{2}\,=:\, \frac{\omega_{2,0}[\Lambda R]}{\Lambda}\,\,.
\ee
The parameter functions in the numerators, from now on called~\emph{auxiliary potentials},
\be\label{eq:AuxiliaryChemicalPotentials}
\underline{\mu}\,:=\,\{\mu_i\}\,=\,\{\beta_0\,,\,\alpha_0\,,\,\omega_{1,0}\,,\,\omega_{2,0}\}
\ee
are dimensionless, and around~$\Lambda R= \infty\,$ are assumed to behave as follows
\be\label{eq:RedefScales}
\mu[\Lambda R]\,=\, \mu_i^{(0)} \Bigl(1\,+\,\sum_{p=1}^{\infty} \frac{\mu_i^{({p})}}{(\Lambda R)^{p}}\Bigr)\,.
\ee
The functions
\be\label{eq:MuiP}
\mu_i^{(p)}= \mu_i^{(p)}[\underline{\mu}^{(0)}]\,,\qquad p\,\geq\, 1\,,
\ee
are meromorphic functions of the \emph{leading behaviors} $\mu_i^{(0)}$, 
\be
\mu_i^{(p)}\,=\,\mu_i^{(p)}[ \underline{\mu}^{(0)}]\,.
\ee
Different choices of functions~$\mu_i^{(p>0)}$ represent different ways to RG-flow towards the continuum. Sometimes we will call these~$\mu_i^{(p>0)}\,$, \emph{moduli of the space of limits} or \emph{of the space of RG-flows}. 

For example, let us assume two choices of moduli
\be\label{eq:TransModuli}
\mu^{(p>0)}_i\,\to\,\mu^{\prime(p>0)}_i\implies \underline{\mu}\to\underline{\mu}^\prime\,:=\, \underline{f}(\Lambda R,\underline{\mu})
\ee
and define the invariant object
\be\label{eq:Invariant}
\mathcal{F}\,=\,{F}[\mathfrak{b},\mathfrak{a},\mathfrak{w}_1,\mathfrak{w}_2]\,.
\ee
Let us compute the expansions of the latter with both choices of moduli. From~\eqref{eq:Def},~\eqref{eq:TransModuli}, and~\eqref{eq:Invariant} one obtains
\be\label{eq:DefFInfty}
\begin{split}
\mathcal{F}_\infty&\,=\,{F}_{
\Lambda=\infty}[\frac{\underline{\mu}}{\Lambda}] \,,
\\
\mathcal{F}^\prime_\infty&\,=\,{F}_{
\Lambda=\infty}[\frac{\underline{\mu}^\prime}{\Lambda}]\,=\,{F}_{
\Lambda=\infty}[\frac{f(\Lambda,\underline{\mu})}{\Lambda}] \,,
\end{split}
\ee
where \underline{the subindex $\Lambda=\infty$ means asymptotic expansion around $\Lambda=\infty$} of the indexed quantity \underline{ignoring subleading non-meromorphic corrections} in the chemical potentials.~\footnote{This is, ignoring subleading logarithms and exponentially suppressed contributions.} Then, if~$\underline{f}(\Lambda,\underline{\mu})$ does not generate isometries of~$F$~\footnote{The problem of systematically classifying isometries of $F$ or even simpler, of $F_{\Lambda=\infty}$, will be left for future work. }
\be
\mathcal{F}_\infty \,\neq\, \mathcal{F}^\prime_\infty\,.
\ee
Different choices of~\eqref{eq:MuiP} can generate the same form of infrared free energy. There is redundancy in their choice. For example, all possible changes of~$\mathcal{F}_{\infty}$ can be generated by redefinitions of one out of the two angular velocities,~$i=3$ or~$i=4\,$. Namely, in complex redefinitions of either
\be\label{eq:TransfSingleVar}
 \omega^{(p)}_{1,0 }[\underline{\mu}^{(0)}]\,\qquad \text{or }\qquad \omega^{(p)}_{2,0}[\underline{\mu}^{(0)}].
\ee
~\footnote{\label{ftn:Scale} In particular, there are choices of these functions that generate the same form of $\mathcal{F}_\infty$ as the $\mathcal{O}(\Lambda^0)$ scalings of~$\mu_2 :=\alpha_0\to C \alpha_0\,$. For instance, the change~$\omega_{a,0}\to\omega_{a,0}(1+C_1 \frac{\alpha_0}{\Lambda})\,$ with $C_1$ being an~$\mathcal{O}((\Lambda R)^0)\,$ meromorphic function of the physical chemical potentials~$\{\beta,\alpha,\omega_1,\omega_2\}\,$, generates the same change in~$\mathcal{F}_\infty$ as the change induced by keeping fixed the $\omega_{a,0}$ and appropriately scaling~$\alpha_0\,$. We will also use the latter kind of reparameterization, which is not of the kind~\eqref{eq:TransModuli}, without explicitly invoking the former one, which induces it, and it is of the kind~\eqref{eq:TransModuli}.} Alternatively, they can be generated by reparameterizations of the cutoff scale
\be\label{eq:ReparametrizationsLambda}
\Lambda \to \Lambda \Bigl(1 \,+\, \sum^\infty_{p=1} \frac{\Lambda^{(p)}[\underline{\mu}^{(0)}]}{\Lambda R}\Bigr)\,.
\ee
at fixed~$\underline{\mu}^{(p)}\,$, i.e.\,, by redefinitions of the~$\Lambda^{(p)}[\underline{\mu}^{(0)}]$'s, which are generic meromorphic functions of the~$\underline{\mu}^{(0)}\,$.
The physical conditions that fix the Schwarzian action, e.g. the reality condition on BPS charges and the non-linear constraint among BPS charges~\cite{Cabo-Bizet:2018ehj}, will happen to break part of these complex reparameterizations, loosely speaking their ``imaginary" part. We will comeback to illustrate this in section~\ref{subsec:RealityConditions}.

From now on and \underline{until the end of this section} the implicit dependence of the~$\mu_{i}$'s, on $\Lambda R\,$ will be ignored. Thus, by \emph{expanding at large-$\Lambda R$} it will be meant expanding in every other dependence on~$\Lambda R$, which is not the one implicit in the auxiliary potentials~$\mu$'s.
An important role in our discussion will be played by the following infinitesimal vicinities (at large enough~$\Lambda$)
\be\label{eq:ContinuumSUSY}
\mathfrak{b}\,\sim\, \frac{\beta_{0}}{\Lambda}\,,\qquad \, \mathfrak{a}\,\sim\,\frac{{\alpha}_{0}}{\Lambda}\, \,,\qquad \mathfrak{w}_{1}\,\sim\, \frac{\omega_{1,0}}{\Lambda}\,,\qquad \mathfrak{w}_{2}\,\sim\, \frac{\omega_{2,0}}{\Lambda}\,.
\ee
They correspond, in the sense explained in the introduction, to a leading RG-flow to the continuum. 

\label{Steps14}
\subsection{The RG flow procedure: the infrared free energy}

\textbf{Comment on convention}: In order to save some notation and ease the reading, usually we will use the same notation, e.g. $\mathcal{F}$ or $\mathcal{F}_{\infty}$, for the free energy of the gauge theory, before and after extremizing over gauge chemical potentials $u$ (i.e. before and after imposing the Gauss constraint). In instances where it is relevant to highlight the difference we will add a subindex $sl$ to denote the free energy before imposing Gauss constraint i.e. $\mathcal{F}_{\infty,sl}$ together with an extra input variable corresponding to the gauge potentials $\underline{u}\,$.~\footnote{Concretely, this abuse of notation is sort of irrelevant because in the analysis in this paper regarding the large charge expansions $\Lambda R\to \infty$ extremizing with respect to the gauge potentials $u$ amounts to substituting the latter by $0\,$. }

Let us move on to compute the holographic low-temperature expansion of the free energy~$\mathcal{F}\,$. To do so we follow the RG-flow procedure below:

\begin{itemize}
\item[\underline{Step 1}] Truncate $\mathcal{F}$ at a power~$LR$ which eventually we will assume to be~$\Lambda^{{n+1}}R^{n+1}\,$, e.g.,~$n=2\,$, as follows
\be\label{eq:Truncation}
\begin{split}
\mathcal{F}&\,\to\, \mathcal{F}_{\Lambda}\,=\,\mathcal{F}_{\Lambda}[\beta,\alpha,\omega_1,\omega_2,\varphi_v,\varphi_w,\underline{u}]\\&:=\, \mathcal{F}^{\Lambda=(LR)^{\frac{1}{n+1}}}_{sl}[x,u,v,w,t,y;e^{2\pi \text{i} u_i}]\,+\,\sum_{j=1}^{LR}\,\sum_{\rho\,\neq\, 0} \frac{e^{2\pi \text{i} j\rho(u)}}{j} \,.
\end{split}
\ee
~$\mathcal{F}^{\Lambda}_{sl}$ was defined in~\eqref{eq:FSL}. The truncation~\eqref{eq:Truncation} is justified because we are interested in probing the physics of states with charges below the energy scale $\mathcal{O}(1)\Lambda^{n+1}R^n\,$, as $\Lambda R\to\infty\,$, forgetting about \emph{heavier} states. We will comeback below to further illustrate the physical relevance of this truncation.~\footnote{In concrete expansions it is convenient to think of the cutoff~$LR$ as independent of~$\Lambda R$ while expanding the summand at large-$\Lambda R\,$. Then one obtains an effective potential for the $u_i\,$'s. Then one can extemize such potential at $LR\gg 1$ (at this stage it is already safe to take~$LR=\infty$) and find all the perturbative corrections to the saddle-point $u^\star$ in the $\frac{1}{\Lambda R}$-expansion. Corrections coming from the dependence of $LR$ on $\Lambda R$ will be exponentially suppressed, not perturbatively suppressed, and thus for the purposes of this paper they are not essential. }

The~$u_i$'s, with $i=1,\dots, N$, are the~$N$ gauge potentials. They are related to the~$N$ eigenvalues~$U_i$ of the unitary matrix~$U$ as follows~$U_i=e^{2\pi\text{i} \text{u}_i}$. The term
\be\label{eq:VandermondeDetCont}
\sum_{j=1}^{LR}\,\sum_{\rho\,\neq\, 0} \frac{e^{2\pi \text{i} j\rho(u)}}{j}
\ee
is the truncated contribution coming from the Vandermonde determinant. $\rho \neq 0$ denotes the non-vanishing adjoint weights of~$U(N)$, namely $\sum_{\rho\neq0}=\sum_{i\neq j=1}^N\,$ and~$\rho(u)=u_i-u_j\,$. The goal is to compute the effective off-shell potential for the~$u_i$'s at large-$\Lambda R\,$. The next step is to extremize the effective potential with respect to the~$u_i$'s and find its leading saddle point.  We already know that at any order in the low-temperature expansion
\be\label{eq:LeadingSaddle}
\begin{split}
u_i\,=\,u^\star_i& \,=\, \mathcal{O}\Bigl(\frac{1}{\Lambda R}\Bigr)\,.
\end{split}
\ee
This is because the effective potential for the~$u_i$'s equals the one of the superconformal index at very leading order in the $\frac{1}{\Lambda R}$-expansion and all~$\beta\,$, by definition. That is, at~$\Lambda R=\infty\,$, $\alpha=\frac{1}{2}\,$ and the partition function truncates to the index which does not depend on~$\beta\,$.

~\eqref{eq:LeadingSaddle} is enough to obtain the saddle-point approximation to~$\mathcal{F}$ at order $\mathcal{O}(\Lambda^n R^{n-1})$ and $\mathcal{O}(\Lambda^{n-
1}R^{n-2})\,$. To compute the exact saddle point approximation to~$\mathcal{F}$ at order~$\mathcal{O}(\Lambda^{n-2}R^{n-3})$ it is necessary to compute the~$\mathcal{O}(\frac{1}{\Lambda R})\,$. We will develop all necessary tools to compute all such perturbative corrections to the gauge saddle point. That said, ~$\mathcal{O}(\Lambda^{n-2}R^{n-3})$ corrections to~$\mathcal{F}$ (and all its subleading analytic corrections), depend on the choice of moduli~$\mu^{(p)}_i\,$, e.g., we can always choose a moduli representative for which these corrections vanish. Thus, in order to compare to semiclassical gravity the question we need to answer is whether there exists a limit that is isomorphic to the one studied in supergravity~\cite{Boruch:2022tno}. To address that question,~\eqref{eq:LeadingSaddle} is enough. To find features of quantum gravity (beyond higher derivative corrections) within the gauge theory, subleading analytic and non-analytic corrections to $u^\star$ at large~$\Lambda R\,$ are needed. Computing those lies beyond the scope of this paper. We plan to address this question in forthcoming work. For the present exposition it will be enough to work with~\eqref{eq:LeadingSaddle}.

\item[\underline{Step 2}] Substitute the relation between chemical potentials and the scale~$\Lambda$~\eqref{eq:Def} in $\mathcal{F}_\Lambda\,$.

\item[\underline{Step 3}] Expand the summand around
\be
(\Lambda R) \,=\,\infty \,.
\ee
(ignoring the implicit dependence on $\Lambda R$ in the auxiliary potentials) and then expand each term in the precedent expansion around
\be
\beta_0\,=\,\infty\,,
\ee
keeping the other dimensionless quantities finite.

The result of this two-step expansion has the form~\footnote{\label{ftn:ExactFormula}Each individual (series) term in this expansion that may potentially diverge in the limit~$\Lambda\to\infty$ we compute, whenever it is possible, in the region of its arguments $\{\varphi_v,\varphi_w, u_i\}$ where its limit converges, and then we analytically extend such finite answer to the complex domain. Some singularities, like the power-like or the~$\log(\Lambda)$ ones will be physical and of course, it will not be possible to remove them using the previous regularization trick. There is a more elegant and general way to compute this asymptotic expansion, allowing to obtain also exact expressions at finite~$\Lambda$, e.g., in the spirit of~\cite{Garoufalidis:2018qds}. In this paper we will not try to compute exponentially suppressed D-instantonic contributions, for us it will be enough to use the pragmatic method above enunciated which leads to the analytic tail and allows to conclude that the reminder is a combination of either logarithmic singularities (independent of temperature) and exponentially suppressed terms. The finite-$\Lambda$ completion of~$\mathcal{F}-\mathcal{F}_{\infty}\,$, will be addressed in future work.}
\be\label{eq:SchematicExpansion}
\begin{split}
\mathcal{O}((\Lambda R) ^n)+\ldots+\mathcal{O}((\Lambda R)^0) &\\ + \,&\mathcal{O}((\Lambda R)^{-1})\\
&\,+\,\text{Li}^\Lambda_{1}\Bigl(1+\mathcal{O}(\frac{1}{\Lambda R})\Bigr)-\text{contributions}\\
&\qquad\,+\,\text{Li}^\Lambda_{p\,<\,0}\Bigl(1+\mathcal{O}(\frac{1}{\Lambda R})\Bigr)-\text{contributions}\\&\qquad \qquad \,+\, \text{Li}^\Lambda_{p\,<\,0}\Bigl(x\Bigr)-\text{contributions}\,.
\end{split}
\ee
where
\be\label{eq:TruncatedDilogMain}
\text{Li}^{\Lambda}_{p}(z):=\sum_{j=1}^{LR} \frac{z^j}{j^p}\,.
\ee

The monomials with powers of order
\be
\begin{split}
\mathcal{O}((\Lambda R)^{n})\,,\,\ldots\,,\mathcal{O}((\Lambda R)^0) 
\end{split}
\ee
entering in this expansion we denote as~\emph{Type I} contributions. The monomials with powers
\be
\begin{split}
 &\mathcal{O}((\Lambda R)^{-1})\,.
\end{split}
\ee
we denote as~\emph{Type II} contributions. Type I and II contributions to~$\mathcal{F}$ can be further organized in powers of~$\alpha_0\,$. Their powers up to order~$\mathcal{O}(\alpha_0^4)$ we collect in
\be
\mathcal{F}_{\infty}\,:=\,\mathcal{O}((\Lambda R)^n)\,+\,\ldots \,+\, \mathcal{O}((\Lambda R)^{-1})\,.
\ee
In particular, the powers~$\mathcal{O}(\alpha_0^3)$ are of order
\be
\mathcal{O}((\Lambda R)^0)\,,\,\mathcal{O}((\Lambda R)^{-1})\,.
\ee
The powers~$\mathcal{O}(\alpha_0^4)$ are only of Type II
\be
\mathcal{O}((\Lambda R)^{-1})\,.
\ee
In this expansion of~$\mathcal{F}$, there are no powers of order higher or equal than $\mathcal{O}(\alpha_0^5)$ entering in Type I and Type II contributions. All such contributions turn out to be exponentially suppressed around~$\Lambda R=\infty\,$. Indeed, we advance that all perturbative contributions to $\mathcal{F}$ beyond Type I and Type II vanish in this expansion. This is, all possible corrections to~$\mathcal{F}$ beyond Type I and Type II are either logarithmic (i.e. non meromorphic) or exponentially suppressed at~$\Lambda R=\infty\,$, as we will proceed to explain below.  The conclusion will be that in the $\Lambda R\to\infty$ expansion
\be
\mathcal{F}\,=\, \mathcal{F}_{\infty} \,+\, \underbrace{\text{log-corrections} + \text{exp-suppresed corrections}}_{\text{subleading non-meromorphic}}\,.
\ee

Before moving on to explain this, let us note that there are Casimir energy-like terms, i.e. contributions that grow as $\beta_0$ around $\beta_0=\infty\,$ in this large $\Lambda R$ expansion. We collect them in a term denoted as
\be\label{eq:CasimirFactor}
\beta_0\,\mathcal{E}_{0}\,=\,\beta \mathcal{E}\,, \qquad \mathcal{E}:\,=\,\Lambda \mathcal{E}_0\,.
\ee
These contributions enclose Type I contributions of order~$\mathcal{O}(\alpha_{0}\Lambda^0)$, and Type II contributions of order~$\mathcal{O}(\alpha_{0}^2\Lambda^{-1})\,$. There are no other contributions emerging at powers higher that~$\beta_0$ around~$\beta_0=\infty\,$ in this large-$\Lambda R$ expansion.~\footnote{What happens is that all of them are exponentially suppressed, and are of the type IV and V to be defined below. }

Summarizing, all Type I and Type II contributions can be organized in the form ($n=2$)
\be\label{eq:FInfinity}
\mathcal{F}_{\infty,sl}\,=\,\beta_0\,\mathcal{E}_0\,+\,\sum_{p=-1}^{n}\sum_{q=0}^4\sum^{\infty}_{r\,=\,0}(\Lambda R)^p\, L_{p+1;q,r}[\varphi_v,\varphi_w, \underline{u}]\,\frac{{F_{p;q;r} \alpha_0^q}}{(\beta_0)^{r} \omega_{1,0}\omega_{2,0}}
\ee
where the~$F_{p;q;r}$ are $N$-dependent homogenous polynomials of order~$(-p-q+r+2)\,\geq\,0$ in the variables~$\omega_{1,0}$ and~$\omega_{2,0}\,$. Some subsets of them vanish, e.g., $F_{2;3;r}=F_{1;3;r}=F_{2;4;r}=F_{1;4;r}=F_{0;4;r}=0$ as it was noted in the previous paragraph. Also~$F_{p;0;r\geq 1}=F_{p;q\geq 1;0}=0\,$.
The $L_{p+1;q;r}$ denotes a linear and finite combination of regularized polyLogs $Li_{p+1}^\Lambda$ of order~$p+1\,$ combined symmetrically into periodic Bernoulli polynomials of the same order.~\footnote{{For the superconformal index, $r=q=0$, these truncation of the perturbative expansion in the large-$\Lambda$-expansion up to logarithmic and exponentially suppressed contributions is implicit in previous results and expectations, at least in the simplest case $\omega_1=\omega_2$~\cite{GonzalezLezcano:2020yeb,Cassani:2021fyv,ArabiArdehali:2019tdm,ArabiArdehali:2021nsx,Cabo-Bizet:2021plf}.~\eqref{eq:FInfinity} is the generalization of tis previous observation from the index to the partition function.}}~The details of formula~\eqref{eq:FInfinity} up to order~$\mathcal{O}(\frac{1}{\beta})$ are reported in equation~\eqref{eq:FslTotal}. A Mathematica notebook testing the derivation of~\eqref{eq:FInfinity} has been shared.

To ease presentation we will drop the subindex $sl$ from $\mathcal{F}_{\infty,sl}\,$. Only when absolutely necessary we will reinstate the subscript $sl\,$.

We note that all {Type II} contributions are proportional to linear combinations of terms of the form
\be
\text{Li}^\Lambda_{0}(e^{\varphi})\,+\,\text{Li}^\Lambda_{0}(e^{-\varphi})\,,
\ee
which, after using analytic continuation, converge exponentially fast as~$\Lambda\to\infty$ to
\be
\text{Li}_{0}(e^{\varphi})\,+\,\text{Li}_{0}(e^{-\varphi})\,=\,-1\,.
\ee

Let us pause and reiterate an important partial conclusion:
\begin{itemize}
\item
Contributions which are not organized in~$\mathcal{F}_{\infty}\,$, are defined to be either~\emph{Type III},~\emph{Type IV}, or~\emph{Type V}. They are either logarithmic corrections or exponentially suppressed corrections. This is, Type I and Type II contributions encode {all the possible perturbative corrections} to $\mathcal{F}^{\Lambda}$ in the large-$\Lambda R$ expansion just enunciated.~\footnote{Recall that we are ignoring the arbitrary dependence on $\Lambda R$ implicit in the auxiliary potentials. Such dependence can generate infinitelly many subleading $\mathcal{O}(\frac{1}{\Lambda R})$ corrections but at the moment we are ignoring corrections coming in that way. They will be essential to consider in due time though.} Let us proceed to explain this. The detailed answer is given in appendix~\ref{app:EffPotentialComplete}.
\end{itemize}

The {Type III}-contributions
\be
\text{Li}^\Lambda_{1}\Bigl(1+\mathcal{O}(\frac{1}{\Lambda R})\Bigr)-\text{contributions}
\ee
\underline{come only from}~$\mathcal{O}(\alpha_0^0)\,$, i.e., from the superconformal index $\mathcal{I}_1\,$. Some of these contributions come from the~$N$ zero modes~$\rho=0$. When combined they become
proportional to a linear combination of terms of the form~\footnote{Their expressions, using a specific regularization scheme have been given in equations (3.64)+(3.65) of~\cite{Beccaria:2023hip}.}
\be
\text{Li}^{\Lambda}_{1}\bigl(1-\omega_{1,2}\bigr)\,\to\,-\log\omega_{1,2} \,,
\ee
and half of
\be
-\text{Li}^{\Lambda}_{1}\bigl(1-\omega_{1}-\omega_2\bigr)\,\to\,+\log{\Bigl(\omega_{1}+\omega_{2}\Bigr)}\,,
\ee
with $\omega_{1,2}\,=\,\frac{\omega_{1,2;0}}{\Lambda R}\,$. These are physical~$\text{log}(\Lambda)$ divergencies. They should correspond to logarithmic quantum corrections in the bulk. As they come solely from $\mathcal{O}(\alpha^0_0)$ contributions (the BPS ones) then they are independent on $\beta\,$. 
There are other logarithmic divergencies associated to middle-dimensional walls of non-analyticities~\cite{Cabo-Bizet:2021plf}. We will not study these logarithmic contributions in this paper, but we advance that they are also independent of~$\beta\,$, as expected.

The {Type IV} contributions
\be
\text{Li}^\Lambda_{p\,<\,0}\Bigl(1+\mathcal{O}(\frac{1}{\Lambda R})\Bigr)-\text{contributions}
\ee
organize in linear combinations of terms of the form ($y=\mathcal{O}(\frac{1}{\Lambda R})$)
\be
\text{Li}^{\Lambda}_{-p}(1+y)\,+\,(-1)^p\,\text{Li}^{\Lambda}_{-p}(1-y)\,, \qquad  p\,\geq\,1\,,
\ee
which, using analytic continuation, converge exponentially fast as~$\Lambda R\to\infty$ to
\be
\text{Li}_{-p}(1+y)\,+\,(-1)^p\,\text{Li}_{-p}(1-y)\,=0\,.
\ee

The {Type V} contributions organize in linear combinations of terms of the form
\be
\text{Li}^{\Lambda}_{-p}(e^{\Phi})\,+\,(-1)^p\,\text{Li}^{\Lambda}_{-p}(e^{-\Phi})\,, \qquad  p\,=\,0\,,\,1\,,\, \ldots\, 
\ee
which, using analytic continuation, converge exponentially fast as~$\Lambda R\to\infty$ to
\begin{equation}
\text{Li}_{-p}(e^{\Phi})\,+\,(-1)^p\,\text{Li}_{-p}(e^{-\Phi})\,=\,0\,,
\end{equation}
where~$\Phi$ can be one of the elements in the list~$\{\varphi_v\,,\,\varphi_w\,,\,-\varphi_v\,-\,\varphi_w\,\}\,$ added to~$\pm2\pi\text{i}u_{ij}\,$'s.

\item[\underline{Step 3}]
Substitute the dimensionless auxiliary parameters back in terms of the physical quantities
\be
\beta_0 \,\to\, \mathfrak{b} \Lambda_1\,,\quad \alpha_0\,\to\,\mathfrak{a}\Lambda_2\,, \quad \omega_{1,0}\,\to\,\mathfrak{w}_1\Lambda_3\,,\quad  \omega_{2,0}\,\to\,\mathfrak{w}_2 \Lambda_4\,.
\ee
The obtained expression for the complete perturbative asymptotic expansion of the free energy, in terms of the physical chemical potentials
\be
\mathcal{F}_{\infty}[\frac{\mathfrak{b}}{R},\frac{1}{2}+\frac{\mathfrak{a}}{R},\frac{\mathfrak{w}_1}{R}\,,\,\frac{\mathfrak{w}_2}{R}\,,\ldots]\,=\,\mathcal{F}_{\infty}[\beta\,,\,\alpha\,,\omega_{1},\omega_{2}\,,\ldots]\,,
\ee
extends naturally to the physical low-temperature region~
\be
\frac{\mathfrak{b}}{R} \,=\,\beta\,\gg\,1\,.
\ee

\item[\underline{Step 4.}] {\bf The proposal:} The holographic near BPS expansion of the Gibbons-Hawking free energy (in minimal gauged-supergravity) relates to the free energy $\mathcal{F}_{\infty}$ we have computed-- after extremizing with respect to the gauge potentials $u$, as follows~
\be\label{eq:TheProposal}
\begin{split}
\mathcal{F}_{g}[\beta_g\,,\,\alpha_g\,,\omega_{g1},\omega_{g2}] &\,\,{\equiv}\,\, \mathcal{F}_{\infty,sl}[\beta\,,\,\alpha\,,\omega_{1},\omega_{2}\,,\,\varphi_v\,,\,\varphi_w; \underline{u}^\star]\\&\,+\,\Bigl(\lambda=\infty\text{ meromorphic corrections}\Bigr)\,\\
&=\mathcal{F}_{\infty}[\beta\,,\,\alpha\,,\omega_{1},\omega_{2}\,,\,\varphi_v\,,\,\varphi_w]\\&\,+\,\Bigl(\lambda=\infty\text{ meromorphic corrections}\Bigr)\,
\end{split}
\ee
where~$\varphi_v\, =\varphi_v[\beta,\alpha,\omega_1,\omega_2]\,=\,\varphi_w\,$ is fixed by the zero R-charge condition~$R_1=R_2=0$, and $u^\star=u^\star[\beta,\alpha,\omega_1,\omega_2,\varphi_v,\varphi_w]\,$ is fixed by the Gauss-constraint. The expansion of~$\mathcal{F}_{\infty,sl}$ up to~$\mathcal{O}(\frac{1}{\beta})$ is reported in~\eqref{eq:FslTotal}.

It should be recalled that there is an arbitrary implicit dependence on $\Lambda R$ in the auxiliary potentials $\underline{\mu}$ entering in the~$\{\beta,\alpha,\omega_1,\omega_2\}\,$, as detailed in~\eqref{eq:Def}, and after imposing the natural {holographic dictionary} relation~\footnote{The symbol $\overset{\cdot}{=}$ denotes equal up to periodic relations implied by quantization of charges i.e. relations such as~$\alpha \leftrightarrow  \alpha + j $ for any integer~$j\,$.}
\be\label{eq:HolographicDictionary}
\begin{split}
\beta_g&\,\overset{\cdot}{=}\,\beta\,,\,\quad \alpha_g\,\overset{\cdot}{=}\,\alpha\,,\,\quad \omega_{g1}\overset{\cdot}{=}\omega_1\,,\,\quad \omega_{g2}
\overset{\cdot}{=}\omega_2\,.
\end{split}
\ee

In the bulk, the infinitelly many choices of functions~$\underline{\mu}=\underline{\mu}[\Lambda R]$, which are implicit in the right-hand side of~\eqref{eq:TheProposal}, correspond to different ways to reach the BPS limit~$\alpha=\frac{1}{2}\,$~\cite{Cabo-Bizet:2018ehj}. We will comeback to comment on this below.

The identification~$\equiv$ is understood as an equality up to~$\mathcal{O}((\Lambda R)^0)$ ambiguities in the choice of holographic renormalization scheme in the gravitational side~\cite{deHaro:2000vlm} (the left-hand side). 
These ambiguities include the holographic dual of the field-theoretic Casimir factor~\eqref{eq:CasimirFactor}. In the field theory side these ambiguities can be understood as deformations of the moduli~\eqref{eq:MuiP}
\be\label{eq:MuiPLast}
\mu^{(p)}_i \to \mu^{(p)}_i =\mu^{(p)}_{i}\Bigl(1+ \mathcal{O}(\frac{1}{\Lambda R})\Bigr)\,, \qquad  p\geq 1\,,
\ee
that generate $\mathcal{O}((\Lambda R)^0)$ deformations of the free energy~$\mathcal{F}_{\infty}$ (after imposing Gauss contraint on $\mathcal{F}_{\infty,sl}$). An explicit example will be given in~\eqref{eq:Choice of counterterms}. There, the order $\mathcal{O}\Bigl((\Lambda R)^0\Bigr)$ Casimir energy prefactor~\eqref{eq:CasimirFactor} will be removed by one such deformation.

Our goal in the following sections will be to compare the right hand side of this proposal against the left hand side in the near-BPS region. One important conclusion of our analysis will be that of constraining the relevance of
\be\nonumber
\Bigl(\lambda=\infty\text{ meromorphic corrections}\Bigr).
\ee
Let us move on to do so.
\end{itemize}

\subsection{Implications of cohomological cancellations and analyticity: Coupling corrections}

In this subsection we explain in more details the last stage, step 4, in the RG flow procedure above stated. The emphasis will be put on explaining how protectedness at the BPS locus $\alpha_0=0$ imposes a non trivial constraint on possible meromorphic coupling dependent corrections to the coarse grained effective action $\mathcal{F}_\infty\,$(with Gauss constraint imposed). 

We will start by constraining the possible meromorphic coupling-dependent corrections to $\mathcal{F}_\infty$ away from the BPS locus, purely using field theory reasoning, without resorting to the AdS/CFT dual picture's inputs. The only hypothesis to use will be:

\begin{itemize}
\item[1.] Protectedness at the BPS locus ($\alpha_0=0$). 
\end{itemize}

It is important to keep in mind that:
\begin{itemize}
\item The conclusions will only apply to coupling-dependent corrections that are meromorphic in the chemical potentials.
\item There may be corrections to the complete UV free energy $F$ at finite values of the couplings which are non-meromorphic though. However, by definition, the coarse grained effective action $\mathcal{F}_\infty$ captures only the leading behaviour in the IR --which is meromorphic in the chemical potentials. The ignored logarithmic corrections and exponentially suppressed contributions are subleading at large $\Lambda R\,$.
\item Consequently, as it has been already stated, our conclusions do not concern subleading non-analytic corrections in the chemical potentials.
\end{itemize}

Thus, effectively, we will use the following second hypothesis in the following:
\begin{itemize}
\item[2.] Meromorphy (in chemical potentials) of coupling-dependent corrections to $\mathcal{F}_\infty\,$.
\end{itemize}

\paragraph{Field theory analysis}
Supersymmetry implies that any coupling $\lambda=g_{YM}^2 N$ corrections and in particular, strong coupling $\lambda=\infty$ corrections, must be subleading in the~$\frac{1}{\Lambda R}$-expansion above introduced, and starting at order $\mathcal{O}(\alpha_0)\,$. This means they should grow at most as $\Lambda R$, but not faster than that (see~\eqref{eq:ConstraintCouplingFinite} below). 

Otherwise, the index~$\mathcal{I}_1$ would receive~$\lambda=\infty$ corrections. The index is a protected quantity, though, and it does not depend on the coupling $\lambda$. 
This is because every $\lambda$-dependent correction to the UV free energy $-\log I_1=\mathcal{F}$ coming from the exchange (vacuum to vacuum) of a given state in the UV theory is cancelled against corrections coming from its supersymmetric partner state (which comes with an extra $-1$).

Only states that are unpaired by SUSY, i.e. that transform into null states and are not the SUSY transformation of some other state contribute to the index.
These are precisely BPS states whose propagator at any value of $\lambda=\lambda_0$ is bound to not receive coupling corrections due to the cancellations above explained.

{Thus, one concludes that the propagators of these states are coupling independent and hence they can be computed at the free theory.}

Consequently, the vaccum free energy $\mathcal{F}_\infty$ at the BPS locus $\alpha_0=0$ is independent of the gauge coupling $\lambda\,$.

Thus, hypothesis 1 implies that corrections in~$\lambda$ must start at order $\mathcal{O}(\alpha_0)$ or higher. This is, they must have the form (assuming small $\alpha_0$, after the large-$\Lambda$ truncation is already implemented)
\be\label{eq:ConstraintCouplingFinite}
\begin{split}
\Delta_\lambda \mathcal{F}_\infty & := \mathcal{F}_{\infty}^{\lambda} -\mathcal{F}_{\infty} \\&\,=\,\sum_{m,n=0\atop m-n+1\,\geq \,d(m,n)-2}^{\infty}C_{m,n}(\lambda) \frac{\alpha_0^{m+1}}{\Lambda^{m-n+1}\beta_0^{n}}\\&\,=\,  C_{0,0}(\lambda) \frac{\alpha_0}{\Lambda} \,+\,C_{0,1}(\lambda)\frac{\alpha_0}{\beta_0}\,+\, \ldots\,,
\end{split}
\ee
where $d(m,n)$ is the degree of divergence of $C_{m,n}(\lambda)$ at large $\Lambda\,$. These $C_{m,n}$'s are rational functions of~$\omega_1$ and $\omega_2\,$. The condition
\be\label{eq:ConstrPowersCmns}
m-n+1\,\geq \,d(m,n)-2
\ee
guarantees that all constraints are of the same order or subleading with respect to the BPS free energy which grows as ${\Lambda^2}\,$ (but not higher). This constraint follows from the following observation:

{\theorem{}{In any correlated limit $\alpha_0\to0$ combined with $\Lambda\to\infty$, independently on how fast $\alpha_0\to 0$ in relation to how fast $\Lambda\to \infty\,$, all the corrections to the BPS onshell action must be subleading with respect to the BPS free-energy which grows as $\Lambda^2\,$. This implies that
\be\label{eq:EqConstraintsCmns2}
C_{m,n}(\lambda) \frac{\alpha_0^{m+1}}{\Lambda^{m-n+1}\beta_0^{n}}
\ee
must grow at most as $\Lambda^2$ at large $\Lambda$ but not faster, i.e., the constraint~\eqref{eq:ConstrPowersCmns}.
}}
\vspace{.1cm}

Note that the dependence in $\beta_0$ of these corrections has been assumed to be regular as $\beta_0\to\infty\,$. Namely, they have been assumed to be non-negative powers $n\geq 0$, $\beta_0^n\,$ in the denominators of~\eqref{eq:ConstraintCouplingFinite} and~\eqref{eq:EqConstraintsCmns2}. This contraint comes from the following observation:

{\theorem{}{This is because the contributions to the thermal partition function come always in the form $e^{-\frac{\beta_0}{\Lambda} \times \Delta }$ where $\Delta\geq 0$. Thus, for large enough $\beta_0$ at fixed $\Lambda$ only states with $\Delta=0$ contribute and dependence on $\beta_0$ must dissapear. Thus, the coarse grained free energy~$\mathcal{F}^\lambda_{\infty}$ must be constant in the limit $\beta_0\to \infty\,$ (when other chemical potentials and $\Lambda$ are kept fixed).}}

Analogously as for the free energy  $\mathcal{F}_\infty$ of the free theory-- see ~\eqref{eq:DefFInfty}, the finite coupling extension $\mathcal{F}^{\lambda}_\infty$ is defined as
\be
\mathcal{F}^\lambda_\infty\,=\,{F}^\lambda_{
\Lambda=\infty}[\frac{\underline{\mu}}{\Lambda}] \,.
\ee
Again, the subscript $\Lambda=\infty$ denotes the very same expansion around $\Lambda=\infty$ chosen to define the free theory $\mathcal{F}_\infty$  (among the infinitely many possibilities). The superscript $\lambda$ denotes the very same superscripted observable but computed at finite coupling $\lambda\,$ instead of as $\lambda=0\,$. This is, for instance, using the path integral formulation of the thermal partition function at finite coupling
(the one that will be compared with minimally gauged supergravity)
\be\label{eq:PartitionFunctionMinGGrav}
 Z^\lambda\,=\, e^{-{F^\lambda}}\,=\,\text{Tr}_{\mathcal{H}} e^{-\beta {\Delta} -\omega_1 \mathcal{J}_+ -\omega_2\mathcal{J}_--2\pi \text{i}\alpha \mathcal{R}}
\ee
where $\mathcal{H}$ is the complete space of states of $SU(N)$ $\mathcal{N}=4$ SYM and
 \be
 \begin{split}
 \mathcal{J}_{\pm}&\,:=\, {J}^3_1\pm{J}^3_2+ \frac{\mathcal{R}}{2}\,,\qquad \mathcal{R}\,=\,{\frac{R_1}{6}+\frac{R_2}{3}+R_3} \,,\\
 \Delta&\,=\,H\,-\,2 J^3_1\,-\,2\,\sum_{k=1}^{3} \frac{k}{4}\, R_k\\
 &\,=\, H\,-\,\mathcal{J}_{+}-\mathcal{J}_--3\mathcal{R}\,.
 \end{split}
 \ee
Note that as the spin and R-charges are protected, then the only coupling dependence in $F^\lambda$ comes from the anomalous dimensions in the eigenvalues of $\Delta(\lambda)\,$.

The constraints~\eqref{eq:ConstraintCouplingFinite} imply that the variation introduced by the coupling $\lambda$ in the large-$\Lambda R$ free energy $\mathcal{F}_{\infty}$, which we will enote below as $\Delta_{\lambda}\mathcal{F}_\infty\,$, can be recast as a chemical potential-dependent redefinition of the cut-off $\Lambda^\prime=\Lambda^\prime(\lambda)$ in the family~\eqref{eq:ReparametrizationsLambda}. On the other hand, the latter redefinition can be recast as a redefinition of the (auxiliary) chemical potentials $\underline{\mu}^\prime =\underline{\mu}(\lambda)$ in the family~\eqref{eq:TransModuli}. Said in equations,
\be\label{eq:RenormCouplingDep}
\begin{split}
\mathcal{F}^\lambda_\infty&\,=\,\Delta_\lambda \mathcal{F}_\infty +\mathcal{F}_\infty\,\\&=\,{F}_{
\Lambda=\infty}[\frac{\underline{\mu}}{\Lambda^\prime}] \\&={F}_{
\Lambda=\infty}[\frac{\underline{\mu}^\prime}{\Lambda}]\,.
\end{split}
\ee
To identify the functions
\be
\underline{\mu}^\prime\,=\, \underline{\mu}^\prime(\lambda) 
\ee
one would need to compute the partition function in field theory~\eqref{eq:PartitionFunctionMinGGrav} around a generic value of \(\lambda\) and then apply the coarse graining process denoted as Step 2, starting from $F^\lambda\,$. Currently, there are no tools to do this directly at $\lambda=\infty\,$ without invoking AdS/CFT. At weak coupling, though, this computation is possible.~\footnote{In this paper we have not done this computation. We leave it for future work.}

The conclusion from the relations~\eqref{eq:RenormCouplingDep} is that any possible correction in coupling constant $\lambda$ (analytic in chemical potentials) to $\mathcal{F}_{\infty}$ corresponds to motion within the moduli space of limits to the continuum (or, equivalently, to \emph{a renormalization of chemical potentials}, see~\eqref{eq:AuxiliaryChemicalPotentials})
\be
\mu^{(p)}_{i} \to \mu^{(p)}_{i,\lambda=\infty} =\mu^{(p)}_{i}\Bigl(1+ \mathcal{O}(\frac{1}{\Lambda R})\Bigr)\,,
\ee
or equivalently, to a chemical-potential dependent redefinition of the cutoff~$\Lambda\,$~\eqref{eq:TransfSingleVar}.

In particular, making explicit the dependence on physical chemical potentials (which without loss of generality can be assumed to be the ones of the free gauge theory)
\be
\frac{\underline{\mu}}{\Lambda}=\{\beta,\alpha,\omega_1,\omega_2\}
\ee
this implies that strong coupling corrections are bound to be reproducible from the free gauge-theory computation  $\mathcal{F}_{\infty}\,$,
\be
\mathcal{F}^{\lambda=\infty}_{\infty}=\mathcal{F}^{\lambda=\infty}_{\infty}[\beta,\alpha,\omega_1,\omega_2]=\mathcal{F}_{\infty}[\beta^{\lambda=\infty}\,,\,\alpha^{\lambda=\infty}\,,\omega^{\lambda=\infty}_1,\omega^{\lambda=\infty}_{2}]
\ee
after implementing a precise renormalization of chemical potentials of the form~\footnote{Or equivalently, after shifting source terms.} 
\be\label{eq:Dictionary1}
\beta^{\lambda=\infty}\,=\,\beta(1+\mathcal{O}(\frac{1}{\Lambda R}))\,,\,\,\alpha^{\lambda=\infty}\,=\,\alpha(1+\mathcal{O}(\frac{1}{\Lambda R}))\,,\,\,\omega^{\lambda=\infty}_{1,2}\,=\,\omega_{1,2}(1+\mathcal{O}(\frac{1}{\Lambda R}))\,.
\ee
It may be possible to derive Callan-Symanzik like equations for this flow starting from the path integral definition of the partition function $Z^\lambda\,$. We leave the exploration of this for future work.

\subsection{Gravitational onshell action from a free field computation: A proposal}

That the Gibbons-Hawking onshell action of the CCLP solutions in minimally gauged 5d supergravity $\mathcal{F}_g$ is by definition meromorphic in chemical potentials when expanded around a connected part of its BPS locus, and that such corrections -- at least in an infinitesimal vicinity of the BPS locus, obey the \emph{subleading-ness} constraint above quoted for field-theoretic corrections, namely they are subleading functions of the BPS chemical potentials with respect to the BPS on-shell action, imply that
\be\label{eq:RefinedRelation}
\begin{split}
\mathcal{F}_{g}[\beta_g\,,\,\alpha_g\,,\omega_{g1},\omega_{g2}]
&\,=\, \mathcal{F}_{\infty}[\beta^{\prime}_g\,,\,\alpha^{\prime}_g\,,\omega^{\prime}_{g1},\omega^{\prime}_{g2}]\,,
\end{split}
\ee
where $\mathcal{F}_\infty$ is the large-$\Lambda R$ free energy computed in the free-gauge theory, i.e. at $\lambda=0\,$, with renormalized chemical potentials 
\be\label{eq:Dictionary2}
\beta^{\prime}_g\,=\,\beta(1+\mathcal{O}(\frac{1}{\Lambda R}))\,,\,\,\alpha^{\prime}_g\,=\,\alpha(1+\mathcal{O}(\frac{1}{\Lambda R}))\,,\,\,\omega^{\prime}_{g1,g2}\,=\,\omega_{1,2}(1+\mathcal{O}(\frac{1}{\Lambda R}))\,,
\ee
for the holographic dictionary~\eqref{eq:HolographicDictionary}
\be\label{eq:HolographicDictionary2} 
\begin{split}
\beta_g&\,\overset{\cdot}{=}\,\beta\,,\,\quad \alpha_g\,\overset{\cdot}{=}\,\alpha\,,\,\quad \omega_{g1}\overset{\cdot}{=}\omega_1\,,\,\quad \omega_{g2}
\overset{\cdot}{=}\omega_2\,.
\end{split}
\ee
Relation~\eqref{eq:RefinedRelation} is the improved version of the previously introduced proposal~\eqref{eq:TheProposal}. In section~\ref{sec:5} we will test this relation at order $\mathcal{O}(\frac{1}{\beta})\,$.

In virtue of $AdS/CFT$ it is then expected the equality among renormalized chemical potentials
\be
\beta^{\prime}_g\,=\,\beta^{\lambda=\infty}\,,\,\,\alpha^{\prime}_g\,=\,\alpha^{\lambda=\infty}\,,\,\,\omega^{\prime}_{g1,g2}\,=\,\omega_{1,2}^{\lambda=\infty}\,.
\ee
Currently, there are no tools to check this in field theory at coupling $\lambda=\infty\,$ (without invoking the AdS/CFT dual).

\vspace{.2cm}

\paragraph{Strong coupling corrections from free fields and renormalization} \label{sec:ArgumentFreeTheoryStrongCoupling}

Before moving on, let us pause to comment on what we think is one of the main conclusions in this paper~\eqref{eq:RefinedRelation}:
\vspace{.2cm}

{\emph{The free field theory analysis complemented with the appropriate renormalization of chemical potentials, captures exact holographic finite-temperature effects. }}
\vspace{.2cm}

\noindent Within the gauge theory, we have mentioned two independent types of renormalizations of chemical potentials, one which is dynamically generated, coming from coupling corrections~\eqref{eq:Dictionary1} and another that comes from redefinitions of the large-$\Lambda R$ expansion~\eqref{eq:MuiP}. Remarkably, the former can be recast as a particularization --i.e. \emph{a choice of reference frame}, of the latter ambiguity in the choice-of-expansion.

Namely, coupling corrections to the near BPS free-energy are fully encoded in the free field-theory answer together with an appropriate choice of the reference frame just mentioned. 

In particular, in section~\ref{sec:5} we will recover the complete expression of the Schwarzian mass gap at $\lambda=\infty$ (the one computed in supergravity~\cite{Boruch:2022tno}) from the free field theory $\lambda=0$ computation of $\mathcal{F}_{\infty}$ and a non-trivial choice of renormalization~\eqref{eq:MuiPLast}~\footnote{Or equivalently, choice of limit to the continuum. We will use both terminologies indistinctly.} which is kinematically fixed by the holographic dictionary. Namely, it is fixed by the identification of chemical potential and/or thermodynamic charges among the boundary and the bulk, as stated before. 

This result constitutes a non-trivial check of~\eqref{eq:RefinedRelation}. Beyond near-BPS physics \eqref{eq:RefinedRelation} it implies that one can reproduce semiclassical finite-temperature black hole thermodynamics from free-field computations and appropriate renormalization of chemical potentials in the dual gauge theory. 

As for high enough corrections in $\frac{1}{\beta}$ and $(\alpha-\frac{1}{2})\,$, \underline{we do not} expect the suplementary renormalization to be kinematically fixed by the identification of chemical potentials and/or thermodynamic charges among the boundary and the bulk. 

\subsection{More on the continuum limit in microcanonical picture} \label{sec:ContrinuumLimitSection}

As previously announced, the parameter~$\Lambda$
is the energy scale controlling the extension
\be
{L}_{max}\,=\, \Lambda^{n^\prime} R^{n^\prime-1} 
\ee
of a domain of charges (states) that we are interested in doing physics at (from now on we drop the subindex $max$). When~$\Lambda R$ is large enough, then the distance among contiguous charge eigenvalues:
\be
\delta L\,=\,\frac{1}{R}
\ee
becomes, necessarily
\be\label{eq:EmphComment}
\frac{\delta L}{L}\,=(\Lambda R)^{-n^\prime} \ll 1\,
\ee
and thus in dimensionaless variables the spectrum becomes quasi-continuous. One can then judiciously substitute the quasi-continuous spectrum by a closely related continuum one, but generically there is no a unique choice for that as advanced in the Introduction. ~\footnote{This ambiguity will be studied in a forthcoming paper. For the moment we advance that it is related to the ambiguities in redefinitions of chemcial potentials, or cutoffs that will be introduced below.} 

If such a deformation from quasi-continuum to the continuum is not implemented then one simply remains with the original discretuum in the physical dimensionful variables and there is no really a limit to the continuum. Namely, such discretum is nothing but a portion of the spectrum of the discrete fundamental theory, $\mathcal{N}=4$ SYM. 

\emph{This observation clarifies why the infinitesimal mass gap of $\mathcal{N}=4$ SYM above $\Delta=0$ (in dimensionful variable), which is related to the spacing $\delta \Delta$, can not be identified with the mass gap of the continuum infrared theory above $\Delta=0$, which in dimensionful variables is always of order $1\,$, and in particular, independent of~$\Lambda\,$. }

Where is this proposed scaling of~$L$ with~$\Lambda$ coming from? In the spirit of the large-charge localization approach discussed in~\cite{Beccaria:2023hip}, we can ask for the hierarchy of charges for which the truncated free energy can be safely localized to its asymptotic behaviour within the singular vicinity
\be\label{eq:ExpansionChemicalPotentialsLargeCharge}
\beta\,\sim\, \frac{\beta_{0}}{\Lambda R}\,,\qquad \, {\alpha}\,\sim\, \frac{1}{2}\,+\,\frac{{\alpha}_{0}}{\Lambda R}\, \,,\qquad \omega_{1}\,\sim\, \frac{\omega_{1,0}}{\Lambda R}\,,\qquad \omega_{2}\,\sim\, \frac{\omega_{2,0}}{\Lambda R}\,,
\ee
defined by the double expansion~
\be\label{eq:DoubleExpansion}
\Lambda R\,\gg\,1\qquad \text{and} \qquad \beta_0 \,\gg\, 1\,,
\ee
which is consistent with the hierarchy
\be\label{eq:LargeChargeExpansion1}
\Lambda\,\gg\, \frac{\beta_0}{R}\,\gg\, \frac{|\alpha_0|}{R}\,,\, \frac{|\omega_{1,0}|}{R}\,,\,\frac{|\omega_{2,0}|}{R}\,,
\ee
and for which the following dimensionless quantities are kept finite 
\be
|\alpha_0|\,,|\omega_{1,0}|\,,|\omega_{2,0}|\, =\,\mathcal{O}(1)\,,\qquad\frac{|\alpha_0|}{|\omega_{1,0}|} \,=\,\mathcal{O}(1)\,, \qquad \frac{|\omega_{1,0}|}{|\omega_{2,0}|}\,=\,\mathcal{O}(1)\,.
\ee
In terms of the dimensionful chemical potentials~\eqref{eq:LargeChargeExpansion1} looks like
\be\label{eq:ExpansionChemicalPotentials3}
R\,\gg\, \mathfrak{b}\,\gg\,\mathfrak{w}_1\,,\,\mathfrak{w}_2\,,\,\mathfrak{a}\,.
\ee
In the expansion~\eqref{eq:LargeChargeExpansion1} we will find (in the following sections) that the free energy happens to scale as~
\be
{\Lambda^n}{R^{n-1}}
\ee
where the scaling power~$n\,=\,1$ or $2$, is determined by the theory and by how much supersymmetry is preserved by the states whose contributions to the partition function do not cancel in the limit~$\alpha\,\to\,\frac{1}{2}\,$. The charges, instead, scale as~
\be
\Lambda^{n^\prime} R^{n^\prime-1}\,
\ee
where the positive integer~$n^\prime$ is fixed by demanding that the source-term of the corresponding charge, scales as the free energy near its leading singularity. More concretely, if the charges generate isometries in the~$S_3$ then $n^\prime \,=\,n+1\,$, if not, then~$n^\prime\,=\,n\,$.~\footnote{Recall that due to twisting some R-charges can also generate rotations in~$S^3\,$.} 

For example, the~$n^\prime$'s associated to the charges
\be
\begin{split}
\sqrt{2}J_{\pm} &:=\,{J}^3_1\pm{J}^3_2+ \frac{R_1+R_3}{2}+\frac{\Delta}{3}\,=\,\frac{\Delta_{\pm}}{2}-\frac{\Delta}{6}\,,
\end{split}
\ee
are fixed as~$n^\prime =n+1$ by demanding that their source terms (in the ensemble described in table~\ref{table:updated_elements3})
\be\label{eq:DomainOFcharges}
\omega_{1} \sqrt{2}J_-  \qquad \text{and} \qquad \omega_{2} \sqrt{2}J_+\,
\ee
scale as~
\be
\,{\delta L}^{\frac{1}{n^\prime}}L^{\frac{n^\prime-1}{n^\prime}}\,\sim\,\frac{L}{\Lambda R}\,\sim\,\Lambda^n\,R^{n-1}\,,
\ee
in the continuum limit
\be
\frac{\delta L}{ L}\,\to\, 0\,.
\ee
From the scaling of chemical potentials~\eqref{eq:ExpansionChemicalPotentialsLargeCharge} and that of free energy as~$\Lambda R\to \infty\,$ (at any~$R$), there follows the definning properties of the hierarchy of charges associated to its singularity~\eqref{eq:ExpansionChemicalPotentialsLargeCharge}
\be\label{eq:AsymptoticConstraints}
\begin{split}
 \sqrt{2}J_{\pm}&\,=\,\mathcal{O}((\Lambda R)^0)\,L_{J_{\pm}}\,,\,\\ R_3&\,=\,\mathcal{O}((\Lambda R)^0)\,L_{J_{\pm}}\,,\\   R_{1,2}&\,=\, \mathcal{O}((\Lambda R)^0)\,L_{R_{1,2}}\,, \\ 
 \Delta&\,=\,\mathcal{O}((\Lambda R)^0)\,L_{\Delta} \,,
 \end{split}
\ee
where the scales determining the extension of the truncations are
\be
\begin{split}
L_{J_{\pm}} &\,=\,N^2\,\Lambda^{n+1}R^n\,,\\   L_{R_{1,2}}&\,=\, N^2\,\Lambda^{n}R^{n-1}\,,\qquad \\ 
L_{\Delta} &\,=\,N^2\,|\alpha_0|\frac{\Lambda^{n+1}R^n}{\beta^2_0} \,.
 \end{split}
\ee
These are the image domains obtained by Legendre-dualizing the singular vicinity~\eqref{eq:ExpansionChemicalPotentialsLargeCharge}. Namely, this will come from the computations presented in the forthcoming sections, following the Steps 1 to Step 4 above described.

Note that~$\Delta\,$, the semi-positive charge that measures distance from the~$\mathcal{Q}$, and $\mathcal{S}$  supersymmetric locus ($\Delta=0$),  may be also large in the expansion~\eqref{eq:LargeChargeExpansion1}, although much smaller than~$J_{\pm}\,$. This means that the expansion~\eqref{eq:LargeChargeExpansion1} also probes states that are not necessarily close to be~$\mathcal{Q}$ and~$\mathcal{S}$-supersymmetric.~\footnote{The large-charge expansion studied in~\cite{Beccaria:2023hip} only supersymmetric states contribute~$(\Delta=0)\,$.} 


An alternative understanding of this RG flow proposal using the microscopic picture just described, will be presented in upcoming work.

\paragraph{A comment on conventions}
Following the holographic dictionary we find natural to identify~\cite{Cabo-Bizet:2018ehj}
\be
R\,=\,\ell_{AdS_5}\,=\, \frac{1}{g} \,.
\ee
From now on we work in natural units and fix 
\be
R\,=\, 1 \,\implies\, \mathfrak{b}\,=\,\beta\,.
\ee

\paragraph{The many ways to reach the BPS locus (in the continuum)}

As explained before, there are infinitely many ways to deform the RG-flow limits~\eqref{eq:ExpansionChemicalPotentialsLargeCharge}. These deformations correspond to changes in the choice of functions~\eqref{eq:MuiP} $\underline{\mu}^{(p)}[\underline{\mu}{(0)}]\,$.
This is, the auxiliary chemical potentials~$\underline{\mu}$ can have arbitrary implicit dependence on~$\Lambda$, as long as the~$\underline{\mu}$ obey the boundary conditions declared in~\eqref{eq:RedefScales},
\be\label{eq:PassiveRedef}
\begin{split}
\beta_0&\,\to\,\widetilde{\beta}_0\,(1\,+\,\frac{\widetilde{\beta}^{(1)}}{\Lambda}\,+\,\ldots)\\
\alpha_0 &\,\to\, \widetilde{\alpha}_0\,\Bigl(C_{0,0}\,+\, \frac{C_{1,0}}{\widetilde{\beta}_0}\,+\,\frac{\widetilde{\alpha}_0 }{\widetilde{\beta}_0}C_{1,1}\,+\,\ldots\Bigr) \\
\omega_{a,0}& \to \widetilde{\omega}_{a,0}\Bigl(1\,+\, \frac{\widetilde{\omega}^{(1)}_a}{\Lambda}\,+\, \ldots\,\Bigr)
\end{split}
\ee
~\footnote{Even the case~$C_{0,0}\underset{\Lambda\to\infty}{{\to}} c \neq 1$ can be induced by a redefinition of the functions $\omega_{1,2}$ as explained in footnote~\ref{ftn:Scale}. } with 
\be
\begin{split}
C_{0,0}&\,=\,C_{0,0}(\omega_a)\,=\,\mathcal{O}(\Lambda^{0})\,,\qquad \\\,
C_{1,0}&\,=\,C_{1,0}(\omega_a)=\mathcal{O}(\Lambda^{-1})\,,\quad C_{1,1}\,=\,C_{1,1}(\omega_a)\,=\,\mathcal{O}(\Lambda^{-1})\,,
\end{split}
\ee
as~$\Lambda=\infty\,$. As we will illustrate below, given
\be\label{eq:ExpansionChemicalPotentialsLargeChargeTilde}
{\beta}\,=\,\frac{\widetilde{\beta}_{0}}{\Lambda}\,,\qquad \, {\alpha}\,=\, \frac{1}{2}\,+\,\frac{\widetilde{\alpha}_{0}}{\Lambda}\, \,,\qquad \omega_{1}\,=\, \frac{\widetilde{\omega}_{1,0}}{\Lambda}\,,\qquad {\omega}_{2}\,= \,\frac{ \widetilde{\omega}_{2,0}}{\Lambda}\,,
\ee
the dependence of the free energy~$\mathcal{F}$ on the physical variables
\be\label{eq:OriginalPotentials}
\beta\,,\quad \alpha\,, \quad \, \omega_1\,,\quad \omega_2\,,
\ee
in the large-$\Lambda$, large-$\widetilde{\beta}_0$ expansion at fixed~$\{\widetilde{\alpha}_0,\widetilde{\omega}_{1,0},\widetilde{\omega}_{2,0}\}$ (called Expansion~2) changes with respect to one obtained via the large-$\Lambda$, large-${\beta}_0$ expansion~\eqref{eq:LargeChargeExpansion1} at fixed~$\{\alpha_0,\omega_{1,0},\omega_{2,0}\}\,$ (called Expansion~1).

\paragraph{Active transformation trick}  The same change in the choice of limits to the continuum, in the sense of~\eqref{eq:MuiP} i.e. choices of~$\underline{\mu}^{(p)}[\underline{\mu}^{(0)}]\,$, can be implemented also by redefining chemical potentials. Take one of them, for instance~$\alpha$, and redefine it as follows
\be\label{eq:ReparametrizationAlpha}
\alpha \,=\, C\, (\widetilde{\alpha}-\frac{1}{2})\,+\,\frac{1}{2}.
\ee
Then localize the free energy which is at this point a function of the tilded chemical potentials, around the small vicinities 
\be
\widetilde{\beta}\,\sim\,\frac{{\beta}_0}{\Lambda}\,,\quad \widetilde{\alpha}\,\sim\,\frac{1}{2}+\frac{{\alpha}_0}{\Lambda}\,,\quad  \widetilde{\omega}_{1}\,\sim\, \frac{{\omega}_{1,0}}{\Lambda}\,,\qquad \widetilde{\omega}_{2}\,\sim \,\frac{ {\omega}_{2,0}}{\Lambda}\,, 
\ee
where~$ C=C(\widetilde{\alpha})$ is the origin preserving reparameterization defined as
\be\label{eq:C00}
\begin{split}
C&\,=\,C_{0,0}\,+\, \frac{C_{1,0}}{{\beta}_0}\,+\,\frac{{\alpha}_0 }{{\beta}_0}\,C_{1,1}\,+\,\ldots\, \\
    &\,=\,C_{0,0}\,+\, \frac{1}{\widetilde{\beta}}\,C_{1,0}\,+\,\frac{(\widetilde{\alpha}-\frac{1}{2})}{\widetilde{\beta}}\,C_{1,1}\,+\,\ldots\,,
\end{split}
\ee
(with the same functions $C_{i,j}$'s as in~\eqref{eq:PassiveRedef}) and write its expansion in terms of the tilded physical potentials. At last, replace back
\be
\widetilde{\alpha} \,\to\, \alpha\,, \qquad \widetilde{\beta}\to \beta\,, \qquad \ldots
\ee
(i.e. drop the tildes). The answer obtained for~$\mathcal{F}_{\infty}$ after this procedure, as a function of the variables without tildes ~$\{\alpha, \beta, \omega_1,\omega_2\}\,$,  is the same one obtained after implementing the Expansion 2 in the physical chemical potentials~$\{\alpha, \beta, \omega_1,\omega_2\}$ (i.e. without modifying the background geometry, and the backgound flavour potentials, in which the field theory is quantized).

This trick will be used to go from the results obtained with a reference large-$\Lambda$ expansion to the results obtained with another expansion. We stress, though, that the chemical potentials of the fundamental theory will remain the ones to be identified with the gravitational ones. The abstract object that changes with different choices is the way to approach the BPS locus in the continuum,~\footnote{In the gravitational dual black hole solution these different limits to the continuum correspond to different ways of approaching the semiclassical BPS solution~\cite{Cabo-Bizet:2018ehj}.} not the identification of chemical potentials between the fundamental theory and the near horizon (gravitational) one.

\paragraph{Expansions to roots-of-unity}
\label{par:OrbifoldRGflows}
Before moving on we point out there are many possible limits that we forsee may be relevant in forthcoming developments. They are limits to roots of unity
\be\label{eq:EvenMoreGeneralExpansion}
\beta\,\sim\,\frac{\beta_{0}}{\Lambda}\,+\, \mathfrak{r}_1  \,,\quad \, \alpha -\frac{1}{2}\,\sim\,\frac{{\alpha}_{0}}{\Lambda}\,+\, \mathfrak{r}_2 \,,\quad \omega_{1}\,\sim\, \frac{\omega_{1,0}}{\Lambda}\,+\, \mathfrak{r}_3\,,\quad \omega_{2}\,\sim\, \frac{\omega_{2,0}}{\Lambda}\,+\, \mathfrak{r}_4\,,
\ee
where
\be
\mathfrak{r}_1 \,,\,\mathfrak{r}_2\,,\, \mathfrak{r}_3\,,\,\mathfrak{r}_4 \,\in\, \mathbb{Q}\,,
\ee
and, again, the auxiliary chemical potentials~$\underline{\mu}$ can have arbitrary implicit dependence on~$\Lambda$, as long as it respects the boundary conditions at~$\Lambda\to\infty$ imposed by the ansatz~\eqref{eq:RedefScales}.
In those limits one also obtains Schwarzian contributions with mass gap parameter depending, generically, on~$\mathfrak{r}_{1,2,3,4}\,$. We leave the study of this for the future.

\section{The protected near-{1}/{8}-BPS Schwarzian mass gap}\label{subsec:ProtectedExpansion}
\label{sec:4}

As mentioned before the leading saddle-point of the Gauss-constraint is~$U\sim e^{2\pi \i u^\star}\,\times\,1_{N\times N}$ (details on this are presented in appendix~\ref{app:Saddle}). Thus, from now on
\be\label{eq:GaugeSaddlePoint}
\text{Tr} U^n \text{Tr} U^{-n} \,\to\, N^2\,+\, \mathcal{O}\Bigl(\frac{\alpha_0}{\Lambda},\frac{\beta_0}{\Lambda},\frac{\alpha_0}{\beta_0},\frac{\omega_{1,0}}{\beta_0},\frac{\omega_{2,0}}{\beta_0}\Bigr)\,,
\ee
where
\be\label{eq:SubleadingGaugeSaddle}
\mathcal{O}\Bigl(\frac{\alpha_0}{\Lambda},\frac{\beta_0}{\Lambda},\frac{\alpha_0}{\beta_0},\frac{\omega_{1,0}}{\beta_0},\frac{\omega_{2,0}}{\beta_0}\Bigr)
\ee
stands for corrections that are first-order in at least one of the small dimensionless parameters in the expansion~\eqref{eq:ExpansionChemicalPotentialsLargeCharge}.

To compute low-temperature corrections of the free energy~$\mathcal{F}=\mathcal{F}_{\frac{1}{16}\text{near}\frac{1}{8}}$ we focus on the asymptotic behaviour of its Taylor coefficients around the point of (extra) cancellations~$\alpha=\frac{1}{2}$ 
\be
\mathcal{F}^{(p)}\,:=\,\frac{1}{p!}\,\partial^{p}_{\alpha} \mathcal{F} \bigg|_{\alpha=\frac{1}{2}}\,,\qquad p\,=\,0\,,\,1\,,\,2\,.
\ee
In the naive zero-temperature limit~$\beta\,\to\,\infty$ all the~$\mathcal{F}_{\frac{1}{16}\text{near}\frac{1}{8}}^{(p\geq 1)}$ vanish exponentially fast. Instead, in the holographic low-temperature expansion obtained after implementing~\eqref{eq:ExpansionChemicalPotentialsLargeCharge}, or more precisely after following the Steps 1 to 4 in~\ref{Steps14}, we obtain~\footnote{These expansions can be computed directly by expanding~\eqref{eq:FslTotal}. We have chosen to explain how they can be derived from scratch. They can be also derived with the shared Mathematica file.} 
\be\label{eq:116F18}
\begin{split}
\mathcal{F}_{\frac{1}{16}\text{near}\frac{1}{8}}^{(0)}&\,=\,- N^2 \, \frac{\Lambda}{\omega_{1,0}}\,L_2(\varphi_{\widetilde{v}})
+\,\mathcal{O}(\Lambda^0)\,\sim\,- N^2 \, \frac{L_2(\varphi_{\widetilde{v}})}{\omega_{1}}\,
\\
\mathcal{F}_{\frac{1}{16}\text{near}\frac{1}{8}}^{(1)}&\,=\,\pi \text{i} N^2 \, \frac{\Lambda^2}{\beta_0\omega_{1,0}}\,\Bigl( 1\,+\,\mathcal{O}\Bigl(\frac{\omega_{1,0}}{\beta_0},\frac{\omega_{2,0}}{\beta_0}\Bigr)\Bigr)L_2(\varphi_{\widetilde{v}}) \,+\, \mathcal{O}(\Lambda) \\
&\,\sim\,\pi \text{i} N^2\,\frac{1}{\beta\,\omega_{1}}\,\Bigl( 1\,+\,\mathcal{O}\Bigl(\frac{\omega_1}{\beta},\frac{\omega_2}{\beta}\Bigr)\Bigr)\,L_2(\varphi_{\widetilde{v}})\,,\\
\mathcal{F}_{\frac{1}{16}\text{near}\frac{1}{8}}^{(2)}&\,=\, \pi^2 N^2\frac{\Lambda^2}{\beta_0\omega_{1,0}}\Bigl(1+\mathcal{O}\Bigl(\frac{\omega_{1,0}}{\beta_0}, \frac{\omega_{2,0}}{\beta_0}\Bigr)\Bigr)\,L_1(\varphi_{\widetilde{v}})\,+\, \mathcal{O}(\Lambda) \\
&\,\sim\,\pi^2 N^2\frac{1}{\beta\omega_{1}}\Bigl(1+\mathcal{O}\Bigl(\frac{\omega_{1}}{\beta}, \frac{\omega_{2}}{\beta}\Bigr)\Bigr)\,L_1(\varphi_{\widetilde{v}})\,
\end{split}
\ee
where
\be
\begin{split} 
L_1(\varphi_{\widetilde{v}})&:=\,-\,\text{Li}^{\Lambda}_1\left(e^{\varphi_{\widetilde{v}}}\right)\,+\,\text{Li}^\Lambda_1\left(e^{-\varphi_{\widetilde{v}}}\right)\,,
\\
L_2(\varphi_{\widetilde{v}})&:=-\,2  \,\text{Li}^\Lambda_2(1)\,+\,\text{Li}^\Lambda_2\left(e^{-\varphi_{\widetilde{v}}}\right)\,+\,\text{Li}^\Lambda_2\left(e^{\varphi_{\widetilde{v}}}\right)\,.
\end{split}
\ee
The equivalence relation~$\sim$ indicates that the quantities in the left-hand and right-hand sides are equal in the~$1/\Lambda$-expansion up to the lowest order in~$\Lambda$ obtained after expanding the right-hand side, but it assumes nothing about the asymptotic behaviour in the large-$\beta_0$ expansion. For completeness we note that the~$\mathcal{O}(\Lambda^0)$ term in~$\mathcal{F}_{\frac{1}{16}\text{near}\frac{1}{8}}^{(1)}$~\footnote{Ignoring logarithmic contributions.} is proportional to
\be\label{eq:LogCorrections}
-\frac{L_1(\varphi_{\widetilde{v}})}{2}\,.
\ee
For later reference we note that the~$\mathcal{O}(\Lambda)$ term in~$\mathcal{F}_{\frac{1}{16}\text{near}\frac{1}{8}}^{(1)}$ is
\be\label{eq:OrderLambda118}
\begin{split}
&N^2\frac{ \pi \text{i} {L_1}(\varphi_{\widetilde{v}})}{2 \beta }+\frac{\pi\text{i}  {L_1}(\varphi_{\widetilde{v}})}{\omega_1}\,.
\end{split}
\ee
~$C$ was defined in~\eqref{eq:C00}. Below we will comment more on it. The term linear in temperature, is essential to compute subleading corrections to the mass gap, not for the leading ones we are looking to compute in this subsection. After extremization with respect to~$\varphi_{\widetilde{v}}\,$, the contribution of this term vanishes exponentially fast as~$\Lambda\to\infty\,$.

The functions~$L_{2}$, and~$L_{1}$ can be recast as combinations of periodic Bernoulli polynomials. Such representation can be straightforwardly recovered by expanding the most general answer~\eqref{eq:FslTotal}, evaluated at~\eqref{eq:LeadingSaddle}) and
\be
\varphi_w\,=\,\omega_1\,-\,\varphi_v\,, \qquad \varphi_v\,=\, \varphi_{\widetilde{v}}\,+\, 2\pi\text{i}(\alpha -\frac{1}{2})\,.
\ee
at large $\Lambda\,$.

Collecting these expansions, we find that the free energy~\eqref{eq:Index2} grows as~$\mathcal{O}(\Lambda^1)$ and (up to corrections of order $\mathcal{O}(\frac{1}{\beta^2})$)
\be\label{eq:F2}
\begin{split}
\mathcal{F}_{\frac{1}{16}\text{near}\frac{1}{8}}\, =\,\mathcal{F}_{\infty}&\,=\, - N^2 \, \frac{1}{\omega_{1}}\,L_2(\varphi_{\widetilde{v}})\\ &\,+\,\pi \text{i} N^2\,C\,\frac{\bigl(\widetilde{\alpha}-\frac{1}{2}\bigr)}{\beta\,\omega_{1}}\,\Bigl( 1\,+\,\mathcal{O}\Bigl(\frac{\omega_1}{\beta},\frac{\omega_2}{\beta}\Bigr)\Bigr)\,L_2(\varphi_{\widetilde{v}})\,+\,\mathcal{O}(\Lambda^0)\,,
\end{split}
\ee
where~$C$
has been defined in~\eqref{eq:C00}. A more complete expression of~\eqref{eq:F2} up-to quadratic order in $\widetilde{\alpha}-\frac{1}{2}$ and with~$C$ expanded as indicated in~\eqref{eq:C00} is given in~\eqref{eq:RefinemenF116}.

Note, that the first asymptotic correction in temperature to the free energy~$\mathcal{F}_{\frac{1}{16}\text{near}\frac{1}{8}}$ is
\be\label{eq:SchwarzianSmallBH}
\begin{split}
&\,\pi \text{i} N^2\,C\frac{\bigl(\widetilde{\alpha}-\frac{1}{2}\bigr)}{\beta\,\omega_{1}}\,\Bigl( 1\,+\,\mathcal{O}\Bigl(\frac{\omega_1}{\beta},\frac{\omega_2}{\beta}\Bigr)\Bigr)\,L_2(\varphi_{\widetilde{v}})\,, 
\end{split}
\ee
which is, essentially, a Schwarzian contribution in grand-canonical ensemble. Let us show this.

In the mixed ensemble defined by taking the Legendre transform, i.e., after extremizing
\be\label{eq:Legendretransform}
-\mathcal{F}_{\frac{1}{16}\text{near}\frac{1}{8}} \,+\, (\text{Some of the source terms in Table~\ref{table:expanded_elements1}})
\ee
with respect to the chemical potentials~$\varphi_{\widetilde{v}}\,$ under the condition that the charge obtained after such transform
\be
R_{\widetilde{v}}\,:=\,-R_1-R_2= N^2 \, \frac{\delta}{\omega_1} \,, \qquad \delta\,\approx\, 0\,, \quad \delta\,\neq\,0
\ee
remains \underline{independent of the chemical potentials~$\{\widetilde{\alpha}\,,\,\beta\}$}, we obtain
\be\label{eq:VarphiSaddle}
\begin{split}
{\varphi_{\widetilde{v}}}&\,\sim\,\varphi^\star_{\widetilde{v}}\,\left(1+\frac{\left(\alpha -\frac{1}{2}\right) (\delta C_{0,0})}{\beta}\right)+O\left(\delta ^1\right)\\
\frac{\varphi^\star_{\widetilde{v}}}{2\pi\text{i}}&\,=\,\frac{1}{2}\,\text{mod}\,1\,,
\end{split}
\ee
where we assume
\be
C_{0,0}\,=:\,\frac{\chi_{0}}{\delta}\,,
\ee
with~$\chi_0$ a constant independent of~$\delta$.

As a result of the intermediate extremization just mentioned we obtain (keeping only leading terms in the large-$\Lambda$ and small-$\delta$ expansion)
\be
\begin{split}
-\mathcal{F}_{eff}&\,:=\,-\mathcal{F}_0\, - \frac{ 2 (C_{0,0}\delta)\,\text{i} \left( \widetilde{\alpha}-\frac{1}{2}\right)}{\pi} \mathcal{F}_0\, - \frac{ C_{0,0} \pi \text{i} \left(\widetilde{\alpha}\, -\frac{1}{2}   \right)}{ \beta } \mathcal{F}_0\,- 
\frac{C_{1,0} \,\pi \text{i} \left(\widetilde{\alpha} -\frac{1}{2} \right)^2}{\beta}\mathcal{F}_0\,
\end{split}
\ee
where
\be
\mathcal{F}_0\,:=\,N^2\,\frac{\pi^2}{2 \omega_1}\,.
\ee
Then we proceed to compute the mixed-ensemble free energy (up to order~$\mathcal{O}(\Lambda^0)$)~\cite{Boruch:2022tno}
\be\label{eq:S18MixedEnsemble}
\begin{split}
-I_{\text{ME}}(\beta,J_+,J_-,\widetilde{\alpha})\, & \,=\, S_{_{\frac{1}{16}\text{near}\frac{1}{8}}}\\
&\,:=\, \underset{\varphi_{\widetilde{v}},{\omega}_{1},{\omega}_2}{\text{ext}}\Biggl( -\mathcal{F}_{\frac{1}{16}\text{near}\frac{1}{8}} \,+\,{\omega}_1 \sqrt{2} J_- \,+\,{\omega}_2 \sqrt{2}J_{+}\, +\,\varphi_{\widetilde{v}} R_{\widetilde{v}}\Biggr)\\
&\,=\, \quad\underset{{\omega}_{1},{\omega}_2}{\text{ext}}\Biggl( -\mathcal{F}_{\text{eff}} \,+\,{\omega}_1 \sqrt{2} J_- \,+\,{\omega}_2 \sqrt{2}J_{+}\,\Biggr)\,+\,\mathcal{O}(\delta^1)\,.
\end{split}
\ee
At leading order, this functional is independent of~$\omega_2\,$. This imposes a further constraint on charges -- on top of the ones declared in equation~\eqref{eq:DomainOFcharges} --
\be\label{eq:18BPSNew}
\begin{split}
\sqrt{2}J_{+}\,:=\,{J}^3_1+{J}^3_2+ \frac{R_1+R_3}{2}\,+\frac{\Delta}{3}\,=\,\frac{\Delta_{+}}{2}\,-\,\frac{\Delta}{6}\,=\,\,-\,\frac{\Delta}{6}\,=\,0\,+\,\mathcal{O}(\frac{1}{\beta^2})\,
\end{split}
\ee
The next-to-last equation follows because by definition~$\mathcal{I}_2$ only counts states with~$\Delta_+=0\,$. The first corrections in~$\omega_2$ come at the order we have reincorporated in the right-hand side. 

Next, we enforce the physical charge
\be
\begin{split}
\sqrt{2}J_{-}\,:=\,{J}^3_1-{J}^3_2+ \frac{R_1+R_3}{2}\,+\frac{\Delta}{3}\,=\,\frac{\Delta_{-}}{2}\,-\,\frac{\Delta}{6}\,=\, \frac{\Delta_-}{2}\,+\,\mathcal{O}\bigl(\frac{1}{\beta^2}\bigr)
\end{split}
\ee
to be
\begin{equation}\label{eq:SourcesTermsZeroModes}
\begin{split}
\sqrt{2} J_{-} &\,=\, \sqrt{2}J_{-}^\star\,+\, \mathcal{O}(\frac{1}{\beta^2}),
\end{split}
\end{equation}
with~$J^\star_{-}$ being a fixed value independent of~$\{\alpha,\beta\}\,$. Under this constraint, the extremization procedure~\eqref{eq:S18MixedEnsemble} fixes
\be
\omega_1\,=\,\omega^\star_{1} \left(1\,-\,\frac{i \pi  \left(\alpha -\frac{1}{2}\right) C_{0,0} }{2 \beta
   }+\frac{i \left(\alpha -\frac{1}{2}\right) {C_{0,0}} \delta }{\pi }\,+\,\mathcal{O}\Bigl((\widetilde{\alpha}-\frac{1}{2})^2\Bigr)\right)\,+\,\mathcal{O}(\delta^1)\,,
\ee
where~$\omega^\star_1$ is a function of the extremal charges~$J^\star_{\pm}$ fixed by the auxiliary extremization problem
\be\label{eq:Legendretransform1}
\begin{split}
&\underset{{\omega}^\star_{1}}{\text{ext}}\Biggl( -\mathcal{F}_{0}[{\omega}^\star_{1}] \,+\,{\omega}^\star_1 \sqrt{2} J^\star_- \,\Biggr)\,=\,\pm \sqrt{2}\pi\sqrt{-\sqrt{2}J^\star_-}\,=\,-\frac{\pi^2}{\omega^\star_1}\,.
\end{split}
\ee
Collecting results we obtain (up to order $\mathcal{O}(\Lambda^0)$)
\be\label{eq:IntemediateS116}
S_{\frac{1}{16}{near}\frac{1}{8}} \,=\,
S_0\,+\,2\pi\text{i}\widetilde{\alpha}R_0 \,-\, \frac{8\pi^2}{M}\frac{\bigl(\widetilde{\alpha}-\frac{1}{2}\bigr)\,+\,\mathcal{O}\Bigl((\widetilde{\alpha}-\frac{1}{2})^2\Bigr)}{\beta}\,+\,\mathcal{O}\bigl(\frac{1}{\beta^2}\bigr)\,,
\ee
with
\be
\begin{split}
\frac{S_0}{N^2} &\,=\,\frac{2 \pi^2}{2 \omega^\star_1} - \frac{i\pi C_{0,0}  \delta}{2 \omega^\star_1}, \\
\frac{R_0}{N^2} &\,=\, -\frac{C_{0,0} \delta}{2 \omega^\star_1}, \\
\frac{1}{M N^2} &\,=\, \frac{C_{0,0} \pi}{16 i \omega^\star_1}\,.
\end{split}
\ee
Now we move on to impose the reality conditions that select the \emph{physical} near BPS-limit 
\be
\text{Im}(R_{\widetilde{v}})=\text{Im}(J_-)=\text{Im}(S_0)=\text{Im}(R_0)=\text{Im}(M)\,=\,0\,.
\ee
The semi-positivity condition
\be
\Delta_-\,\geq\,0\,,
\ee
implies (omiting~$\mathcal{O}(\frac{1}{\beta_0^2})$)
\be\label{eq:SemipositivityEntropy}
-\sqrt{2} J_{-}\,=\,-\sqrt{2} J^\star_{-}\,=\,-\,\Delta_-\,\leq\, 0\, \implies \text{Re}\Bigl(\frac{1}{\omega_1^\star}\Bigr)\,=\, 0\,.
\ee
Together with this identity,
\be
\text{Im}(R_{\widetilde{v}})=0\,\implies\, \text{Re}(\delta)\,=\,0\,.
\ee
Together with the previous conclusions 
\be
\text{Im}(R_0)=\text{Im}(M)\,=\,0 \implies \text{Im}(C_{00})=\text{Re}(\chi_0)\,=\,0\,.
\ee
Together with the previous conclusions
\be
\text{Im}(S_0)\,=\,0 \,\implies\, \chi_0 \,=\,-2\pi\text{i}\,.
\ee
Collecting conclusions, we obtain (at leading order in the large-$\Lambda$, large~$\beta_0$ expansion)
\be\label{eq:Pars18}
\begin{split}
S_0&\,=\,0\,,\\
R_0&\,= \,\mp\, N \sqrt{2\Delta_{-}}\,,\\
\frac{1}{M}&\,=\,\pm \Bigl(\frac{\pi}{4 \sqrt{2}\text{i}\delta}\Bigr)\,N   \sqrt{\Delta_-}\, =\,{ N^2}\,\frac{4 R_{\widetilde{v}}}{J_-}\,.
\end{split}
\ee
The signs in the second and third line are correlated~\footnote{Top with top, and bottom with bottom.}. If~$\text{i}\delta\,>\,0$ (resp.~$<0$) we have picked up the~$+$ (resp.~$-$) in the third line, in such a way~$M\,>\,0\,$, but the other choice is a priori equally relevant, as it will become evident in the more general case that will be analyzed in the following section. The two sign choices in~\eqref{eq:Pars18} correspond to the two saddle points of~\eqref{eq:Legendretransform1}.

At last, we update~\eqref{eq:S18MixedEnsemble} with~$S_0=0\,$. Then substituting~$\widetilde{\alpha}\to\alpha$ we obtain (up to order-$\Lambda^0$)
\be\label{eq:NearEVHEquation}
\begin{split}
S_{\frac{1}{16}{near}\frac{1}{8}} &=\,2\pi\text{i}{\alpha}R_0 \,-\, \frac{8\pi^2}{M}\frac{\bigl({\alpha}-\frac{1}{2}\bigr)\,+\,({\alpha}-\frac{1}{2})^2}{\beta}\,+\,\mathcal{O}\bigl(\frac{1}{\beta^2}\bigr)\,.
\end{split}
\ee
This is the effective low-temperature infrared action consistent with reality of charges and BPS entropy. The latter being constrained to vanish in this case.

In~\eqref{eq:NearEVHEquation} we have reinstated the canonical~$\mathcal{O}\bigl(({\alpha}-\frac{1}{2})^2\bigr)$ contribution to~$S_{_{\frac{1}{16}\text{near}\frac{1}{8}}}$, up to order~$\frac{1}{\beta}$, which is subleading at large-$\Lambda$, i.e.~$\mathcal{O}(\Lambda^0)\,$, and comes from a repetition of the procedure above reported considering~$C_{1,0}\neq 0\,$.

\paragraph{Partial remarks}
\begin{table}[h]
\centering
\begin{tabular}{c|c|c}
\hline
Chemical potential & Dual charge & Source term \\ \hline
$\beta$& $\Delta$ & $+\beta \Delta$\\$\omega_{1}$ & $J^{3}_{1} - J^{3}_{2} + \frac{R_1}{2} + R_2 + \frac{R_3}{2} + \frac{\Delta}{3}$ & $+{\omega}_1 \cdot \left(J^{3}_{1} - J^{3}_{2} + \frac{R_1}{2} + R_2 + \frac{R_3}{2} + \frac{\Delta}{3}\right)$ \\
$ \omega_2$ & $J^{3}_{1} + J^{3}_{2} + \frac{R_1}{2} + \frac{R_3}{2} + \frac{\Delta}{3}$ & $+\widetilde{\omega}_2 \cdot \left(J^{3}_{1} + J^{3}_{2} + \frac{R_1}{2} + \frac{R_3}{2} + \frac{\Delta}{3}\right)$ \\
$\varphi_{\widetilde{v}}$ & $-R_1 - R_2$ & $+\varphi_{\widetilde{v}} \cdot (-R_1 - R_2)$ \\
$\omega_u=-2\pi\text{i}\alpha$ & $-R_3$ & $+\omega_u \cdot (-R_3)$ \\ \hline
\end{tabular}
\caption{The chemical potentials, charges and source terms that can be potentially added to~$-\mathcal{F}_{\frac{1}{16}\text{near}\frac{1}{8}}$.}
\label{table:expanded_elements1}
\end{table}

~\eqref{eq:SemipositivityEntropy} is predicting that in the region of charges~\eqref{eq:RegionCharges} near-$\frac{1}{8}$-BPS black holes within the family of~\cite{Wu:2011gq}, have near vanishing horizon and entropy. This result is consistent with recent expectations~\cite{Chang:2023ywj}. Our analysis, however, only covers the case~\footnote{At~$\delta=0$ this covers the case of two equal R-charges~$Q_1=Q_3$ and a different third one~$Q_2$, and the case of three equal R-charges~$Q_1=Q_2=Q_3\,$ (using the definitions given in equation~\eqref{eq:TranslationCharges}). }
\be\label{eq:RegionCharges}
\frac{R_1}{R_2}\,\to\, -1\,.
\ee
when two out of the three independent R-charges (the ones to be indentified with electric charges of the dual gravitational solutions)~$Q_1$,~$Q_2\,$, and~$Q_3$ (defined in~\eqref{eq:TranslationCharges}) are equal. We do not see any complication though in repeating our analysis in the more general region of charges, but we leave doing so for future work.~Also, being fair, our analysis does not exclude the existence of~$\frac{1}{8}$-BPS black holes which can not be continuously recovered from the~$\frac{1}{16}$-BPS ones of~\cite{Chong:2005hr,Wu:2011gq}.

We should note also that the mass-gap of the Schwarzian mode goes to infinity in the limit
\be\label{eq:MassGap18}
{M}\,=\,\frac{J_-}{4 N^2 R_{\widetilde{v}}}\,\to\,\infty\,,\qquad R_{\widetilde{v}} \,\to\, 0\,.
\ee
Thus the Schwarzian becomes irrelevant signaling the vanishing of the horizon.

It would be interesting to test~\eqref{eq:MassGap18} in supergravity. Relatedly, it would be also interesting to visualize what happens to the horizon of the dual gravitational solutions in~\cite{Wu:2011gq} when one approaches the vicinity
\be\label{eq:}
\Delta\,,\, \Delta_+\,,\,R_1+R_2\,=\,0
\ee
of their moduli space. Note that the dual gravitational solutions relevant for this analysis can not be embedded in minimally gauged supergravity in five dimensions because they have two different electric charges. 

As an unsurprising consistency check, in appendix~\ref{sec:3p2} we recover the free energy~\eqref{eq:F2} starting from the more general near-$\frac{1}{16}$-BPS computation, which we move on to study next. Such a match confirms the selection of the gauge saddle point~\eqref{eq:GaugeSaddlePoint}.~\footnote{From these results ond should be able to understand whether the feature of vanishing mass gap continues to hold if one approaches the continuum around $\frac{1}{8}$ BPS locus in generic ways, not only along the index~$\mathcal{I}_2\,$. We postpone such an analysis for future work and move on to study the generic near-BPS expansion with our proposal. Notice also that in contradisctinction with the more general case that will be analyzed next, reality conditions are enough to fix the mass-gap parameter of the Schwarzian.}

\section{The near-{1}/{16}-BPS Schwarzian mass gap }\label{sec:5}
Let us test the proposed RG-flow procedure against known predictions in gravity.
Let us start from the maximally refined partition function~\eqref{eq:PartitionFunction2}
\be\label{eq:MostGeneralExample}
Z[x\,,\,u=e^{2\pi\i\alpha}\,,\,v\,,\,w\,,\,t,\,y]\,=e^{-\mathcal{F}}\,.
\ee
At~$\alpha=\frac{1}{2}$ this partition function reduces to the $\frac{1}{16}$-BPS index~$\mathcal{I}_1\,$. To compute corrections of~$\mathcal{F}$, again, we focus on the asymptotic behaviour of its Taylor coefficients around the point of supersymmetric cancellations~$\alpha=\frac{1}{2}$ 
\be
\mathcal{F}^{(p)}\,:=\,\frac{1}{p!}\partial^{p}_{\alpha} \mathcal{F} \bigg|_{\alpha=\frac{1}{2}}\,,\qquad p\,=\,0\,,\,1\,,\,2\,.
\ee
~\footnote{The other relevant coefficients~$p=3,4$ contribute only at order $\mathcal{O}(\Lambda^0)$ and~$\mathcal{O}(\Lambda^{-1})\,$ and can be removed by a trivial redefinition of limits.}
In the naive zero-temperature limit~${\beta}\,\to\,\infty$ all the~$\mathcal{F}^{(p\geq 1)}$ vanish exponentially fast. In the holographic low-temperature expansion obtained after implementing~\eqref{eq:ExpansionChemicalPotentialsLargeCharge}, or more precisely after following the Steps 1 to 4. in~\ref{Steps14}, one finds~\footnote{These expansions can be computed directly by expanding~\eqref{eq:FslTotal}. We have chosen to explain how they can be derived from scratch. They can be also derived with the shared Mathematica file.}  
\be\label{eq:Fgeneral}
\begin{split}
\mathcal{F}^{(0)}&\,=\, -\,N^2 \, \frac{\Lambda^2}{\omega_{1,0}\omega_{2,0}}\,L_3(\varphi_v,\varphi_w)
+\,\mathcal{O}(\Lambda)\,\sim\, -\,N^2 \, \frac{L_3(\varphi_v,\varphi_w)}{\omega_{1}\omega_2}\,
\\
\mathcal{F}^{(1)}&\,=\,\pi \text{i} N^2 \, \frac{\Lambda^3}{{\beta}_0\omega_{1,0}\omega_{2,0}}\,\Bigl( 1\,+\,\mathcal{O}\Bigl(\frac{\omega_{1,0}}{{\beta}_0},\frac{\omega_{2,0}}{{\beta}_0}\Bigr)\Bigr)L_3(\varphi_v,\varphi_w) \,+\, \mathcal{O}(\Lambda^2) \\
&\,\sim\,\pi \text{i} N^2\,\frac{1}{{\beta}\,\omega_{1}\omega_2}\,\Bigl( 1\,+\,\mathcal{O}\Bigl(\frac{\omega_1}{{\beta}},\frac{\omega_2}{{\beta}}\Bigr)\Bigr)\,L_3(\varphi_v,\varphi_w)\,,\\
\mathcal{F}^{(2)}&\,=\, \pi^2 N^2\frac{\Lambda^3}{{\beta}_0\omega_{1,0}\omega_{2,0}}\Bigl(1+\mathcal{O}\Bigl(\frac{\omega_{1,0}}{{\beta}_0}, \frac{\omega_{2,0}}{{\beta}_0}\Bigr)\Bigr)\,L_{2,0}(\varphi_v,\varphi_w)\,+\, \mathcal{O}(\Lambda^2) \\
&\,\sim\,\pi^2 N^2\frac{1}{{\beta}\omega_{1}\omega_{2}}\Bigl(1+\mathcal{O}\Bigl(\frac{\omega_{1}}{{\beta}}, \frac{\omega_{2}}{{\beta}}\Bigr)\Bigr)\,L_{2,0}(\varphi_v,\varphi_w)\,,
\end{split}
\ee
where
\be
\begin{split} 
L_{2,0}(\varphi_v,\varphi_w)&:=\,+2 \text{Li}^{\Lambda}_2(1)\,+\text{Li}^{\Lambda}_2\left(e^{-\varphi_v}\right)+\text{Li}^{\Lambda}_2\left(e^{\varphi_v}\right)+\text{Li}^{\Lambda}_2\left(e^{-\varphi_w}\right)+\text{Li}^{\Lambda}_2\left(e^{\varphi_w}\right) \\&\quad \,-3 \text{Li}^{\Lambda}_2\left(e^{\varphi_v+\varphi_w}\right)-3 \text{Li}^{\Lambda}_2\left(e^{-\varphi_v-\varphi_w}\right)
\\
L_3(\varphi_v,\varphi_w)&:=\text{Li}^{\Lambda}_3\left(e^{-\varphi_v}\right)-\text{Li}^{\Lambda}_3\left(e^{\varphi_v}\right)+\text{Li}^{\Lambda}_3\left(e^{{-\varphi_w}}\right)-\text{Li}^{\Lambda}_3\left(e^{\varphi_w}\right)\\ &\quad +\text{Li}^{\Lambda}_3\left(e^{\varphi_v+\varphi_w}\right)-\text{Li}^{\Lambda}_3\left(e^{-\varphi_v-\varphi_w}\right)\,.
\end{split}
\ee
The~$\mathcal{O}(\Lambda)$ term in~$\mathcal{F}^{(0)}$ is
\be\label{eq:ORcontribution1/16}
\,\sim\,N^2 \,\frac{1}{2}\, \frac{\omega_1\,+\,\omega_2}{\omega_{1}\omega_2}\,L_{2,1}(\varphi_v,\varphi_w)
\ee
and the~$\mathcal{O}(\Lambda^2)$ term in~$\mathcal{F}^{(1)}$ is
\be\label{eq:SubleadingSchwarzianC}
 \begin{split}
&\,\sim \,-\,{\pi\text{i}} N^2 \frac{1}{ \omega_1 \omega_2}  L_{2,0}(\varphi_{v},\varphi_{w})\,-\,\pi \text{i} N^2\,\frac{\omega_1+\omega_2}{2\beta\omega_1 \omega_2}\, L_{2,1}(\varphi_v,\varphi_w)\,+\,\mathcal{O}\bigl(\frac{1}{\beta^2}\bigr)
\end{split}
\ee
where
\be\begin{split}
L_{2,1}(\varphi_v,\varphi_w)&\,:=\,2  \,\text{Li}^{\Lambda}_2(1)\,-\,\text{Li}^{\Lambda}_2\left(e^{-\varphi_v}\right)-\text{Li}^{\Lambda}_2\left(e^{\varphi_v}\right)-\text{Li}^{\Lambda}_2\left(e^{-\varphi_w}\right)-\text{Li}^{\Lambda}_2\left(e^{\varphi_w}\right)\\&\,+\,\text{Li}^{\Lambda}_2\left(e^{-\varphi_v-\varphi_w}\right)+\text{Li}^{\Lambda}_2\left(e^{\varphi_v+\varphi_w}\right)\,.
\end{split}
\ee
The functions~$L_{3}$,~$L_{2,0}$ and~$L_{2,1}$ can be recast as combinations of periodic Bernoulli polynomials. Such representation can be straightforwardly recovered by expanding the most general answer~\eqref{eq:FslTotal} (evaluated at~\eqref{eq:LeadingSaddle}) at large $\Lambda\,$.

With these expansions we find
\be\label{eq:F}
\mathcal{F}\, \sim\, - N^2 \, \frac{1}{\omega_{1}\omega_2}\,L_3(\varphi_v,\varphi_w)\,+\,\pi \text{i} N^2\,C\,\frac{\bigl(\widetilde{\alpha}-\frac{1}{2}\bigr)}{{\beta}\,\omega_{1}\omega_2}\,\Bigl( 1\,+\,\mathcal{O}\Bigl(\frac{\omega_1}{{\beta}},\frac{\omega_2}{{\beta}}\Bigr)\Bigr)\,L_3(\varphi_v,\varphi_w)\,,
\ee
after substituting~$\alpha$ by the re-parameterization choice~\eqref{eq:ReparametrizationAlpha}.
Subleading corrections in $\frac{1}{\Lambda}$ expansion to~$\mathcal{F}$ can be recovered reinstating the contribution coming from~$\mathcal{F}^{(2)}$ in~\eqref{eq:116F18}, and the order~$\mathcal{O}(\Lambda^2)$ contribution in~$\mathcal{F}^{(1)}$ in~\eqref{eq:Fgeneral}. Or equivalently, considering only leading corrections and then applying the substitution rule
\be\label{eq:SubstitutionRule}
L_3(\varphi_{{v}},\varphi_{{w}})\,\rightarrow\, L_3(\varphi_{{v}},\varphi_{{w}})\,-\, \frac{\omega_1\,+\,\omega_2}{2}\,L_{2,1}(\varphi_{v},\varphi_{w})\,+\,O(\Lambda^{-2})\,
\ee
on~\eqref{eq:F}. Ignoring logarithmic contributions and spurious c-numbers, the missing~$\mathcal{O}(\Lambda^0)$ to the free energy, or the missing~$\mathcal{O}(\Lambda^{-2})$ contributions in~\eqref{eq:SubstitutionRule}, have origin in
\be\label{eq:AmbigiousTerms}
\begin{split}
\mathcal{F}^{(0)} &:\qquad N^2\frac{\pi \text{i}  \left(\omega _1^2+\omega _2^2\right)}{12 \omega _1 \omega _2} \,+\,\ldots,  \\ \mathcal{F}^{(1)} &: \qquad -\,N^2\frac{\pi ^2 \beta }{3 \omega _1 \omega _2} \,-\,N^2\frac{\pi  \left(\omega _1+\omega _2\right) \left(8 \pi -3 i \left(\varphi _v+\varphi _w\right)\right)}{3 \omega _1 \omega _2}\\&\qquad\,\,\,-\,N^2\frac{\pi ^2 \left(\omega _1^2+3 \omega
   _2 \omega _1+\omega _2^2\right)}{12 \beta \omega _1 \omega _2}\,+\,\ldots.
\end{split}
\ee
where the $\ldots$ denote contributions coming from the subleading analytic corrections to the gauge saddle point values~$u^\star_i\,$ that are unknown to us at the moment. However, these contributions to the onshell action~$\mathcal{F}_{\infty}$, as well as its corrections in the coupling $\lambda$, can be always understood as a convenient choice of representative in the following family of redefinitions of limits to the continuum around the BPS locus $\alpha=\frac{1}{2}\,$, which for obvious reasons we feel inclined to call a~\emph{choice of counterterms} and that include the redefinitions~\footnote{Notice for instance that it may be used to remove a Casimir energy-like term like the one explicited in~\eqref{eq:AmbigiousTerms}.  }
\be\label{eq:Choice of counterterms}
\begin{split}
\omega_{1,2}&\to \omega_{1,2} \left(1\,+\,\frac{\gamma_1 \omega_1 \omega_2}{2}\,+\,\frac{\gamma_2 (\alpha-\frac{1}{2}) \omega_1 \omega_2}{2\beta}\,+\,\frac{\gamma_3 \bigl(\alpha-\frac{1}{2}\bigr) \beta }{2}+\gamma_4 \omega_{2,1}^2+\frac{\gamma_5\bigl(\alpha-\frac{1}{2}\bigr)\omega_{2,1}^2}{\beta }\right.\\ &\qquad\qquad \qquad \left.\,+\gamma_6\bigl(\alpha-\frac{1}{2}\bigr) \omega_{2,1}+\gamma_7\bigl(\alpha-\frac{1}{2}\bigr)
   \omega_{2,1} \left(\varphi_v+\varphi_w\right)\,+\, \gamma_8 \,\omega_{2,1}^3\,+\,\gamma_9\, \omega_{2,1}^2\omega_{1,2}\right. \\&\qquad \qquad \qquad \left. +\gamma_{10} (\alpha-\frac{1}{2}) \omega_{1,2} + \gamma_{11} (\alpha-\frac{1}{2}) \omega_{1} \omega_{2}  \right)\,.
\end{split}
\ee
This reparameterization of the infrared limit only generates changes in the asymptotic form of the free energy at order~$\mathcal{O}(\Lambda^0)$ and below and it will be used to fix a convenient reference form for the~$\mathcal{O}(\Lambda^0)$ and~$\mathcal{O}(\Lambda^{-1})$ contributions to~$\mathcal{F}$ (up to order $\mathcal{O}(\alpha-\frac{1}{2})$)~\footnote{... which we will fix below by using the condition~\eqref{eq:RefinedRelation}.}
\be
 \frac{\eta_{1}(\omega_1^2+\omega_2^2)\,+\,\eta_{2} \omega_1 \omega_2 +\eta_{3} (\omega_1^3+\omega_2^3)+\eta_{4}(\omega_1^2\omega_2+\omega_1 \omega_2^2) }{\omega_1\omega_2}\,\Bigl(1\,-\, \pi \text{i}\,C_{0,0} \frac{\bigl(\widetilde{\alpha}-\frac{1}{2}\bigr)}{\beta}\Bigr).
\ee
Coming back to~\eqref{eq:F}. Note that the first correction in temperature to the asymptotic expansion of~$\mathcal{F}$ \eqref{eq:F} is
\be\label{eq:SchwarzianLargeBH}
\begin{split}
&\,\pi \text{i} N^2\,C\,\frac{\bigl(\widetilde{\alpha}-\frac{1}{2}\bigr)}{{\beta}\,\omega_{1}\omega_2}\,\Bigl( 1\,+\,\mathcal{O}\Bigl(\frac{\omega_1}{{\beta}},\frac{\omega_2}{{\beta}}\Bigr)\Bigr)\,L_3(\varphi_v,\varphi_w)\,,
\end{split}
\ee
which is, essentially, a Schwarzian contribution in grand-canonical ensemble. Let us show this.

\begin{table}[h]
\centering
\begin{tabular}{c|c|c}
\hline
Chemical potential & Dual charge & Source term \\ \hline
$\beta$& $\Delta$ & $+\beta \Delta$\\
$\omega_1$ & $J_{1}^{3} - J_{2}^{3} + \frac{R_1}{2} + \frac{R_3}{2} + \frac{\Delta}{3}$ & $+\omega_1 \cdot \left(J_{1}^{3} - J_{2}^{3} + \frac{R_1}{2} + \frac{R_3}{2} + \frac{\Delta}{3}\right)$ \\
$\omega_2$ & $J_{1}^{3} + J_{2}^{3} + \frac{R_1}{2} + \frac{R_3}{2} + \frac{\Delta}{3}$ & $+\omega_2 \cdot \left(J_{1}^{3} + J_{2}^{3} + \frac{R_1}{2} + \frac{R_3}{2} + \frac{\Delta}{3}\right)$ \\
$\varphi_v$ & $-R_1 $ & $+\varphi_v \cdot (-R_1 )$ \\
$\varphi_w$ & $-R_2$ & $+\varphi_w \cdot (-R_2)$ \\
$\omega_u\,=\,-2 \pi \text{i}\alpha$& $-R_1 - R_3$ & $+\omega_u \cdot (-R_1 - R_3)$ \\ \hline
\end{tabular}
\caption{The chemical potentials, charges and source terms to be added to~$-\mathcal{F}$ before performing extremization.}
\label{table:updated_elements3}
\end{table}

In the mixed ensemble defined by extremizing 
\be\label{eq:LegendretransformLarge}
-\mathcal{F} \,+\, (\text{Some of the source terms in Table~\ref{table:updated_elements3}})
\ee
with respect to the chemical potentials~$(\varphi_{{v}},\varphi_{w})\,$ under the condition~
\be
R_{1}\,=\,R_{2}\,=\,0
\ee
we obtain
\be\label{eq:VarphiSaddleLarge}
\begin{split}
{\varphi_{{v}}}={\varphi_{{w}}}&\,\sim\,\pm\frac{2  \pi \text{i} }{3}\,+\,\frac{1}{3}
   \left(\omega _1+\omega _2\right)\,+\,\mathcal{O}\Bigl((\widetilde{\alpha}-\frac{1}{2})\Bigr)\,.
\end{split}
\ee
The precise form of the $\mathcal{O}\Bigl((\widetilde{\alpha}-\frac{1}{2})\Bigr)$ corrections is reported in the Mathematica file "Section 5-1 ...." together with its derivation.

As a result of such intermediate extremization and fixing the renormalization ambiguities as follows~\footnote{... this choice amounts to fixing~\eqref{eq:RefinedRelation}. This will be shown below.}
\be
\begin{split}
\eta_1&=\mp\frac{\pi\text{i}}{9}\,,\,\eta_2\,=\,\mp\frac{2 \pi\text{i}}{9} \,,\,\eta_3\,=\,-\frac{1}{54}\,,\,\eta_4\,=\,-\frac{1}{18}\,,
   \end{split}
\ee
we obtain 
\be\label{eq:MinusFinfinity}
\begin{split}
-\mathcal{F}_{\infty}&\,=\,-\mathcal{F}_0\, - {  C\, \left( \widetilde{\alpha}-\frac{1}{2}\right)} \mathcal{F}_{0,1}\, + \,\pi\text{i}C\frac{ \left(\widetilde{\alpha}\, -\,\frac{1}{2}   \right)}{ \beta } \mathcal{F}_{0}\,+\,\mathcal{O}\Bigl( (\widetilde{\alpha} -\frac{1}{2})^2\Bigr)\,
\end{split}
\ee
where
\be\label{eq:F0PM}
\mathcal{F}_{0}\,=\,-\,N^2\frac{ (\pm2 \pi\text{i} + \omega_1 + \omega_2)^3}{54\, \omega_1 \omega_2}\,,\qquad  \mathcal{F}_{0,1}\,=\, -N^2\frac{4  \pi ^3 \text{i} }{9 \omega _1 \omega _2}\,
\ee
and the renormalization ambiguity~$C$ was defined in~\eqref{eq:C00}.
Then we proceed to evaluate~\cite{Boruch:2022tno}
\be\label{eq:SGeneric}
\begin{split}
-I_{\text{ME}}(\beta,J_-,J_+,\widetilde{\alpha})\, & \,=\, S\\
&\,:= \underset{\varphi_{{v}},\varphi_{{w}},{\omega}_{1},{\omega}_2}{\text{ext}}\Biggl( -\mathcal{F} \,+\,{\omega}_1 \sqrt{2} J_- \,+\,{\omega_2} \sqrt{2}J_{+}\,\Biggr)\,\\
&\,=\, \quad\,\,\underset{{\omega}_{1},{\omega}_2}{\text{ext}}\,\,\Biggl( -\mathcal{F}_{\infty} \,+\,{\omega}_1 \sqrt{2} J_- \,+\,{\omega}_2 \sqrt{2}J_{+}\,\Biggr)\,,
\end{split}
\ee
by enforcing~$J_{\pm}$
to be
\begin{equation}
\begin{split}
\sqrt{2} J_{-} &\,=\, \sqrt{2}J_{-}^\star\,+\,\mathcal{O}(({\alpha-\frac{1}{2}})^2)\,+\, \mathcal{O}(\frac{1}{\beta^2})\,,\\
\sqrt{2} J_{+} &\,=\, \sqrt{2}J_{+}^\star\,+\,\mathcal{O}(({\alpha-\frac{1}{2}})^2)\,+\, \mathcal{O}(\frac{1}{\beta^2})\,,
\end{split}
\end{equation}
where~$J^\star_{\pm}$ are fixed values, independent of~$\{\alpha,\beta\}\,$, as it was required in the near BPS limit flow used in~\cite{Boruch:2022tno} (see appendix D in that reference). Under this constraint, the extremization procedure~\eqref{eq:SGeneric} fixes (with the choice~$\pm$ in~\eqref{eq:F0PM})
\be
\begin{split}
\omega_1&\,=\,\omega ^*_1\,\mp\,C_{0,0}\,\bigl(\widetilde{\alpha} -\frac{1}{2}\bigr) \omega ^*_1\,+\,\\ &\qquad \qquad\,+\,\frac{\left(\widetilde{\alpha}
   -\frac{1}{2}\right) \omega ^*_1 \left(\pm\frac{1}{6}C_{0,0} \left(\omega ^*_1-2 \omega
   ^*_2\mp2 \pi \text{i} \right)+\frac{ \text{i} C_{1,0} \left(\omega^*_1-2 \omega
   ^*_2\,\pm\,\pi \text{i}\right)}{\pi }\right)}{6 \beta }\\ &\qquad \qquad\,+\,\mathcal{O}(\Lambda^{-2})\,,\\
\omega_2&\,=\,\omega ^*_2 \,\mp\,C_{0,0}\bigl(\widetilde{\alpha}-\frac{1}{2}\bigr)  \omega ^*_2\\&\qquad \qquad +\frac{\omega ^*_2\left(\widetilde{\alpha}
   -\frac{1}{2}\right) \left(\pm\frac{1}{6} C_{0,0}  \left(-2 \omega
   ^*_1+\omega ^*_2\mp2  \pi\text{i} \right)+\frac{\text{i}C_{1,0} \left(-2 \omega ^*_1+
   \omega ^*_2\pm\pi \text{i} \right)}{\pi }\right)}{\beta }\\
   &\qquad\qquad+\mathcal{O}(\Lambda^{-2})  \,,
\end{split}
\ee
where the~$\omega^\star_1$ and~$\omega^\star_2$ are functions of the extremal charges~$J^\star_{\pm}$. These functions are fixed by the auxiliary \emph{minimal supergravity-like extremization problem} (not to confuse with~\eqref{eq:SGeneric})~\cite{Hosseini2017}
\be\label{eq:AuxiliaryExtremization}
\begin{split}
&\underset{{\omega}^\star_{1},{\omega}^\star_2}{\text{ext}}\Biggl( -\mathcal{F}_{0}[{\omega}^\star_{1},{\omega}^\star_2] \,+\,{\omega}^\star_1 \sqrt{2} J^\star_- \,+\,{\omega^\star_2} \sqrt{2}J^\star_{+}\,\Biggr)\,
\\
&=\underset{{\omega^\star}_{1},{\omega^\star}_2}{\text{ext}}\Biggl( 
\frac{N^2 (\pm 2 \pi\text{i} + \omega^\star_1 + \omega^\star_2)^3}{54 \omega^\star_1 \omega^\star_2}
 \,+\,{\omega}^\star_1 \sqrt{2} J^\star_- \,+\,{\omega_2} \sqrt{2}J^\star_{+}\,\Biggr)\,.
 \end{split}
\ee
Collecting the results so far we write down the onshell value of the physical functional~\eqref{eq:SGeneric} in Schwarzian-like form
\be\label{eq:TheSchwarzianForm}
S\,=\, S_0\,+\,2\pi\text{i}\widetilde{\alpha}R_0 \,-\, \frac{8\pi^2}{M} \frac{\bigl(\widetilde{\alpha}-\frac{1}{2}\bigr)\,+\,\mathcal{O}\Bigl((\widetilde{\alpha}-\frac{1}{2})^2\Bigr)}{{\beta}}\,+\,\mathcal{O}\Bigl(\frac{1}{{\beta}^2}\Bigr)\,,
\ee
with
\be
\begin{split}
S_0&\,=\,\mp\left(\frac{\text{i} N^2 \pi (2 \pi \mp \text{i} (\omega^{\star}_1 + \omega^{\star}_2))^2}{9 \omega^{\star}_1 \omega^{\star}_2}\right) - \left(\frac{2 \text{i} N^2 \pi^3}{9 \omega^{\star}_1 \omega^{\star}_2}\right) C_{0,0}\\
R_0&\,=\,+\left(\frac{2 N^2 \pi^2}{9 \omega^{\star}_1 \omega^{\star}_2}\right)  C_{0,0}\\
\frac{1}{M} \,&=\pm
\frac{ N^2 (2 \pi \mp\text{i}( \omega^\star_1 + \omega^\star_2))^3}{432 \pi\omega^\star_1 \omega^\star_2} C_{0,0} \, -\,\left(\frac{\text{i} N^2 \pi }{18 \omega^\star_1 \omega^\star_2}\right) C_{1,0}\,.
\end{split}
\ee
The solution for the~$\omega^\star$'s as functions of~$J_\pm$ can be written in the implicit form~\cite{Cabo-Bizet:2018ehj, Boruch:2022tno}
\begin{equation}\label{eq:OmegaLocus}
\begin{split}
\omega^\star_1 &= \frac{2 (1 - a^\star) (b^\star \mp \text{i} \sqrt{a^\star + b^\star + a^\star b^\star}) \pi}{2 (1 + a^\star + b^\star) \sqrt{a^\star + b^\star + a^\star b^\star} \mp 2 \text{i} (a^\star + b^\star + a^\star b^\star)}\,, \\
\omega^\star_2 &= \frac{2 (1 - b^\star) (a^\star \mp \text{i} \sqrt{a^\star + b^\star + a^\star b^\star}) \pi}{2 (1 + a^\star + b^\star) \sqrt{a^\star + b^\star + a^\star b^\star} \mp 2 \text{i} (a^\star + b^\star + a^\star b^\star)}\,,
\end{split}
\end{equation}
for charges parameterized as follows~\cite{Chong:2005hr}
\begin{equation}\label{eq:ChargesAB}
\begin{split}
\sqrt{2}J^\star_- &\,=\,-\,N^2\,\frac{(1 + a^\star) (1 + b^\star) (a^\star + b^\star) }{2 (-1 + a^\star)^2 (-1 + b^\star)}\,, \\
\sqrt{2}J^\star_+ &\,=\, -N^2\,\frac{(1 + a^\star) (1 + b^\star) (a^\star + b^\star) }{2 (-1 + a^\star) (-1 + b^\star)^2}\,,
\end{split}
\end{equation}
with the parameters~$a^\star=a^\star(\Lambda)$ and~$b^\star=\mathcal{O}(\Lambda)$, being smooth real functions of~$\Lambda$ such that~$0\,\leq\,a^\star\,,\,b^\star\,<\,1$ and~$a^\star-1=\mathcal{O}(\frac{1}{\Lambda})\,$, $b^\star-1=\mathcal{O}(\frac{1}{\Lambda})\,$ at large~$\Lambda$.

Next, we proceed to impose the three physical (reality) conditions
\be\label{RealityConditions}
\text{Im}(S_0)\,=\,\text{Im}(R_0)\,=\,\text{Im}(M)\,=\,0\,.
\ee
The first two conditions fix~$C_{0,0}\,=\,|C_{0,0}| e^{\text{i}\eta_{0,0}}$ with
\be\label{eq:C00Expansions}
\begin{split}
|C_{0,0}|&\,=\,\frac{9 \sqrt{(1 + a^\star) (1 + b^\star)} (a^\star + b^\star)^2}{2 (a^\star + b^\star + a^\star b^\star) \left(1 + a^{\star2} + 3a^\star(1 + b^\star) + b^\star(3 + b^\star)\right)}\,=\,1\,+\,\mathcal{O}(\frac{1}{\Lambda})\,,\\
\eta_{0,0}&\,=\,\arccos{\Biggl(\mp\left(\frac{-1 + a^\star + b^\star + b^{\star2} + a^{\star2} (1 + 2b^\star) + a^\star b^\star (5 + 2b^\star)}{\sqrt{(1 + a^\star) (1 + b^\star)} \left(1 + a^{\star 2} + 3a^\star(1 + b^\star) + b^\star(3 + b^\star)\right)}\right)\Biggr)}\\&\,=\,\frac{\pi}{3}\,+\,\mathcal{O}\Bigl(\frac{1}{\Lambda}\Bigr)\,.
\end{split}
\ee
Note that the leading asymptotic behaviour of $C_{00}$ is a pure phase $e^{\frac{\pi\text{i}}{3}}\,$. 

The last condition fixes a linear relation between the real ($Y_0$) and imaginary ($Y_1$) parts of~$C_{1,0}=Y_0+\text{i} Y_1\,$ that leads to an isomorphism relation between~$M=M(Y_1)\,$ and~$Y_1\,$. In the language of appendix D of~\cite{Boruch:2022tno} we are in a path to the BPS locus defined by
\be\label{eq:VariationGravity}
\epsilon_q\,=\,\epsilon_a \,=\,\epsilon_b =0
\ee
This means that $Y_1$ controls the relation between the variation created by a differential change~$\epsilon_r  \propto \frac{1}{\beta_{g}}$ upon the gravitational charges~$\{R_{g},\ldots\}$ (as given in equation (3.25)-(3.26) in~\cite{Boruch:2022tno}), and the variation created by our differential~$\frac{1}{\beta}\,$ upon the field theory charges~$\{R_{3},\ldots\}$~\footnote{This will be eventually identified with the Bekenstein-Hawking temperature~$\frac{1}{\beta_g}$ but for the moment it is the field theory temperature.}. At this point in the analysis, and up to order $\mathcal{O}(\frac{1}{\beta}=T)\,$, the only non trivial variation of charges remaining to identify is
\be
\delta R_3\, :=\, R_3 + \Bigl( \frac{\partial \widetilde{\alpha}}{\partial \alpha}\Bigr)\,R_0 \:=- \frac{1}{2\pi\text{i}}\,\Bigl( \frac{\partial \widetilde{\alpha}}{\partial \alpha}\Bigr)\,\frac{\partial^2 S }{\partial\widetilde{\alpha} \partial T}\Bigl|_{T=0} T\,=\, +\,\Bigl( \frac{\partial \widetilde{\alpha}}{\partial \alpha}\Bigr)\,\frac{4 \pi \text{i}+\mathcal{O}\Bigl((\widetilde{\alpha}-\frac{1}{2})\Bigr)}{M(Y_1)}\,T\,.
\ee
In the holographic dual side, the analogous quantity (created by a variation of type~\eqref{eq:VariationGravity}) is most easily computed by using equation~(3.73) in~\cite{Boruch:2022tno}
\be
-\delta R_{g} \,:=\,R_g^\star -R_g \,=\, \frac{1}{2\pi\text{i}}\, \frac{\partial^2 I_{\text{ME},g} }{\partial{\alpha_g} \partial T_{g}}\Bigl|_{T_g=0} T_{g}\,=\,-\frac{4 \pi \text{i}+\mathcal{O}\Bigl(({\alpha_g}-\frac{1}{2})\Bigr)}{M_{SU(1,1|1)}}\,T_{g}\,.
\ee
Then identifying our field theory chemical potentials~\footnote{Here we use the active transformation trick once more.}
\be
\begin{split}
(\widetilde{\alpha}\to -\alpha+{1},\beta) \,, 
\end{split}
\ee
$(\alpha,\beta)$ with the gravitational chemical potentials~$(-{\alpha}_g,\beta_g)\,$. More generally,
\be
(\alpha,\beta,a^\star,b^\star,\omega_{1,2})\,\overset{\cdot}{=}\,(-{\alpha}_g,\beta_g,,a_g^\star,b_g^\star,-\omega_{1,2;g})\,
\ee
and
\be
\begin{split}
S&\,\overset{\cdot}{=}\,-\,I_{ME,g}\,,\\
R_0 &\,\overset{\cdot}{=}\, -\Bigl( \frac{\partial {\alpha}}{\partial \widetilde{\alpha}}\Bigr)\,R^\star\,,
\end{split}
\ee
\be
\delta R_3 \,\overset{\cdot}{=}\,\delta R_{g} \implies R_3\,\,\overset{\cdot}{=}\,\, R_g\,,
\ee
fixes~$(Y_0,Y_1)$ to large expressions in terms of $a^\star$ and $b^\star$ that will not be reported here. They are reported in the attached Mathematica file titled "Section 5-2 ...". More importantly, this procedure fixes
\be\label{eq:MassGap}
-\frac{1}{M} \,=\, \frac{N^2(a^\star + b^\star)^2 (3 + a^\star + b^\star - a^\star b^\star) }{8 (-1 + a^\star) (-1 + b^\star) (1 + a^{^\star 2} + 3 b^\star + b^{^\star 2} + 3 a^\star (1 + b^\star))}\, 
\,\overset{\cdot}{=}\,-
\,\Bigl(\frac{\partial \alpha}{\partial \widetilde{\alpha}}\Bigr)\,\frac{1}{M_{SU(1,1|1)}}\,,
\ee
and
\be\label{eq:EntropyRchargePar}
\begin{split}
S_0&\,=\,\frac{\pi  {N}^2 \left(a^*+b^*\right) \sqrt{a^* b^*+a^*+b^*}}{\left(a^*-1\right)
   \left(b^*-1\right)}\,\overset{\cdot}{=}S^\star\,,\\
-\Bigl(\frac{\partial \widetilde{\alpha}}{\partial {\alpha}}\Bigr)\,R_0   &\,=\,\frac{N^2 (a^*+b^*)}{(a^*-1) (b^*-1)}\,\overset{\cdot}{=}\,R^\star\,.
   \end{split}
\ee
Summarizing, this means that
\be\label{eq:TheSchwarzianFormFinal}
S\,=\, S_0\,+\,\Bigl(\frac{\partial\widetilde{\alpha}}{\partial\alpha}\Bigr)\,2\pi\text{i}{\alpha}R_0 \,-\,\Bigl(\frac{\partial\widetilde{\alpha}}{\partial\alpha}\Bigr) \frac{8\pi^2}{M} \frac{\bigl({\alpha}-\frac{1}{2}\bigr)\,-\,\Bigl(\frac{\partial\widetilde{\alpha}}{\partial\alpha}\Bigr)({\alpha}-\frac{1}{2})^2}{{\beta}}\,+\,\mathcal{O}\Bigl(\frac{1}{{\beta}^2}\Bigr)\,,
\ee
where
\be\label{eq:CHoiceALphaALphaTilde}
\begin{split}
\Bigl(\frac{\partial\widetilde{\alpha}}{\partial\alpha}\Bigr)&\,=\,-1 \,. 
\end{split}
\ee
~$S$ in~\eqref{eq:TheSchwarzianFormFinal} exactly matches the supergravity answer~$-I_{ME}$ in~\cite{Boruch:2022tno} upon the identification of chemical potentials and charges summarized in table~\ref{table:Identification} below. In~\eqref{eq:TheSchwarzianFormFinal} we have reinstated the canonical~$\mathcal{O}\bigl(({\alpha}-\frac{1}{2})^2\bigr)$ contribution to~$S$, up to order~$\frac{1}{\beta}$, which comes from a repetition of the procedure above reported considering~$C_{1,1}\neq 0\,$.

\begin{table}[h!]
\centering
\begin{tabular}{c|c|c}
\hline
Quantities & Minimally gauged gravity& Field-Theory \\
\hline  Angular velocities & $-\omega_{1,2;g}$  as in (3.14),(3.15) of~\cite{Cabo-Bizet:2018ehj}& $\omega_{1,2}\,$\\ $U(1)$ potential & $-\alpha$ of~\cite{Boruch:2022tno}& $\alpha\,$\\Charge dual to~$\omega_1$ &$\mathfrak{j}_{1}$ of \cite{Boruch:2022tno} & $\sqrt{2}J^\star_{-}:=\sqrt{2}\,J_-\Big|_{\Delta\,=\,0\,,\, R_{1,2}=0}$ \\
Charge dual to~$\omega_2$ &$\mathfrak{j}_{2}$ of \cite{Boruch:2022tno} & $\sqrt{2}J^\star_{+}:=\sqrt{2}\,J_+\Big|_{\Delta\,=\,0\,,\, R_{1,2}=0}$  \\
$U(1)$ charge &$R^\star  \,(\text{resp.} R)\,$ of \cite{Boruch:2022tno} & $R_0\, (\text{resp.} R_3)$\\
Large BPS entropy & $S^*$ of \cite{Boruch:2022tno} & $S_0$\\Mixed ensemble & $-I_{\text{ME}}$ of \cite{Boruch:2022tno} & $S$  \\
\hline
\end{tabular}
\caption{The identifications $\overset{\large\cdot}{=}$ between gravity and field theory. Working in minimally gauged supergravity corresponds to the choice of~$\varphi_v$ and~$\varphi_w$ reported in~\eqref{eq:VarphiSaddleLarge}. }
\label{table:Identification}
\end{table}

The details of these previous derivations are reported in the attached Mathematica files with section 5 in the title.

\paragraph{Identifying charges in the boundary and the bulk fixes the holographic match among dual Schwarzian mass-gap functions $|M|$} \label{sec:BoundaryBulkIdentification}

Let us summarize the previous analysis in more physical terms. We have defined a large-$\Lambda$
near-BPS expansion such that the relation between physical thermodynamic charges of the gauge theory (as in gravity) in the mixed ensemble defined at fixed variables $\{J_{\pm}, \alpha, T\}$ take the following form
\begin{equation}\label{eq:ChargesExpansion}
\begin{split}
J_{\pm} \left[ J_+^*, J_+^*, T, \alpha \right] &= J_{\pm}^* + O\left( \left( \alpha - \frac{1}{2} \right)^2 \right) + \mathcal{O}(T^2) \\
\Delta \left[ J_+^*, J_+^*, T, \alpha \right] &= \mathcal{O}(T^2) \\
R_3 \left[ J_+^*, J_-^*, T, \alpha \right] &= R^* \left[ J_{\pm}^* \right] + \left( \chi^* \left[ J_{\pm}^* \right] \right) T + \mathcal{O}\left( \alpha - \frac{1}{2} \right)\,+\,\mathcal{O}(T^2)
\end{split}
\end{equation}
in terms of real BPS charges $J_{\pm}^*\,$ above introduced. $R^*=R[J^\star_{\pm}]$ is the non-linear constraint among charges $R_3$ and $J_{\pm}$ at the BPS locus~\cite{Cabo-Bizet:2018ehj}.~\footnote{This non-linear constraint can be substituted by the first condition~$\text{Im}(S_0)=0$ in~\eqref{RealityConditions}.} The susceptibility about the $T=0$ locus of the thermodynamic charge $R_3$ is defined as
\be\label{eq:DefSuscept}
\chi[J_{\pm}^*,\alpha] \,:=\,\left. \frac{\partial R_3}{\partial T}  \right|_{T=0}\,=\,\left. -\frac{1}{2\pi \text{i}}\frac{\partial S}{\partial T \, \partial \alpha} \right|_{T=0}
\ee
and it can be expanded around the supersymmetric locus $\alpha=\frac{1}{2}$ (i.e. around the BPS locus)
\be\nonumber
\chi[J_{\pm}^*,\alpha]\,=\,\chi^*[J_{\pm}^*] \,+\, O\left(\alpha - \frac{1}{2}\right)\,.
\ee
Its BPS value is proportional to the inverse Schwarzian mass gap (by the last equality in~\eqref{eq:DefSuscept})
\be\label{eq:MassGapSusceptibility}
\chi^*[J_{\pm}^*]\,=\, \frac{ 4 \pi \text{i}}{M}
\ee
and pure imaginary. This unphysical feature is consistent with the previous observation~\cite{Cabo-Bizet:2018ehj} that at the supersymmetric locus $\alpha=\frac{1}{2}$ it only makes physical sense to consider $T=0$ if one requires all physical charges to be real.~\footnote{Considering $T$ as a linear differential for simplicity.} This is, $\text{Im}(R_3)=0 \implies T=0\,$.

Moreover, the action 
\eqref{eq:TheSchwarzianFormFinal} takes the form
\be\label{eq:SInCharges}
S = -S^*\,-\,\pi \text{i}R^*\,-\, 2\pi \text{i} (\alpha-\frac{1}{2}) R_3 \,+\, \mathcal{O}\bigl((\alpha-\frac{1}{2})^2\bigr) \,+\,\mathcal{O}(T^2)\,.
\ee
In particular the corrections to $S$ that define the value of the mass gap come from the value of the thermodynamic charge $R_3\,$ (in the mixed ensemble defined by fixing $J_{\pm}$, $\alpha$ and $T$). Thus, equating the boundary charge $R_3$ to the gravitational charge $R_3$ implies matching of the mass gap:
\be
R_{3,bdy}\,\overset{\cdot}{=}\,R_{3,g}\,\implies\,
\chi^*_{bdy}\,\overset{\cdot}{=}\,\chi^*_{g}\,\implies\,
M_{bdy}\,\overset{\cdot}{=}\,M_{g}\,
\ee
in virtue of~\eqref{eq:ChargesExpansion}

Thus we conclude that the holographic dictionary kinematically fixes the matching between the value of the Schwarzian mass gap functions~$|M|$ in both sides of the duality even if the computation of $M$ started at $\lambda=0$ in field theory and the computation of $M$ in supergravity corresponds to $\lambda=\infty$ in the field theory side.

\paragraph{Corrections in $\lambda\,$, are they also kinematically fixed by the holographic dictionary?}
Yes, they are. For example, from the results of~\cite{Boruch:2022tno} it follows that the $1/\lambda$-corrections there computed are encoded in the expectation value of the thermodynamic charge~$R_{3,g}$, specifically in the susceptibility~$\chi^*_{g}\,$
\be
\chi^*_{g,\lambda}\,=\,\chi^*_{g,\lambda=\infty}\Bigl(1 + \frac{1}{\lambda^{3/2}} \frac{3\pi^3 \zeta(3) (1 + a^*)^3 (1 - a^*)^{3/2}}{2a^{*7/2}(3 - a^*)}+\ldots
\Bigr)\,.
\ee
Thus, the holographic dictionary
\be
\chi^*_{bdy}\,\overset{\cdot}{=}\,\chi^*_{g,\lambda}
\ee
fixes kinematically
\be\label{eq:MbdyMLambdaInf}
\begin{split}
M_{bdy,\lambda}&\,=\, M_{\lambda=\infty}\Bigl(1 - \frac{1}{\lambda^{3/2}} \frac{3\pi^3 \zeta(3) (1 + a^*)^3 (1 - a^*)^{3/2}}{2a^{*7/2}(3 - a^*)}+\ldots
\Bigr)\\
&\,=\,M_{\lambda=0}\Bigl(1 - \frac{1}{\lambda^{3/2}} \frac{3\pi^3 \zeta(3) (1 + a^*)^3 (1 - a^*)^{3/2}}{2a^{*7/2}(3 - a^*)}+\ldots\Bigr)\,.
\end{split}
\ee
where
\be
M_{\lambda=\infty}=M_{\lambda=0}= M\,.
\ee
From now on we comeback to our working example $\lambda=\infty$ and drop all subindices $g$, $bdy$, $\lambda=0\,$ and~$\lambda=\infty\,$.

It would be interesting to compute small Yang-Mills ('t Hooft) coupling corrections (See~\ref{sec:ArgumentFreeTheoryStrongCoupling}).~\footnote{Relevant coupling corrections have been recently studied in a series of papers~\cite{Chang_2023,Choi:2022caq,Choi:2023znd,Chang:2023zqk,Budzik:2023vtr,Chang:2024zqi}.} to $\mathcal{F}_\infty$ and explicitly check that those can be absorbed in renormalizations of the coupling. It would be also interesting to study whether S-duality of $\mathcal{N}=4$ SYM relates such perturbative contributions around $\lambda=0$ to the perturbative contributions around $\lambda=\infty$ in~\eqref{eq:MbdyMLambdaInf}, those predicted by supergravity. We leave such analysis for future work.

\paragraph{Recovering perturbations in the large-${\Lambda}$ expansion} 
\label{subsec:RealityConditions}
Once {reality conditions} are imposed on the infrared theory up to order~$\mathcal{O}(\frac{1}{\beta})$ and its Schwarzian physical form has been recovered,~\footnote{By this we mean the reality conditions on the expectation values of charges~$\sqrt{2}J_\pm$, which are proportional to derivatives of the infrared free energy with respected to their dual chemical potentials~$\omega_{1,2}\,$ up to, and including~$\mathcal{O}(\frac{1}{\beta})\,$. These reality conditions correspond to restricting the initial 2-complex plane~$(\omega_{1},\omega_{2})$ to the middle-dimensional complex contour~\eqref{eq:OmegaLocus} spanned by the real values of parameters~$a$ and~$b$ ranging between~$0$ and $1\,$, including $0\,$. Once reality conditions are imposed on~$\sqrt{2}{J}_{\pm}$ the only remaining freedom in reparameterization that is consistent with them is the set of real reparameterizations of the latter complex curve, which corresponds to the infinitely many choices of smooth real functions~\eqref{eq:realReparametrizations}. These residual reparameterizations remain unbroken in the presence of the Schwarzian.} there remains ambiguity in the choice of implicit dependence on $\Lambda$ within the auxiliary chemical potentials
\be
\omega_{1,0}\,=\,\Lambda \,\omega_{1}(a^\star,b^\star),\qquad \omega_{2,0}\,=\,\Lambda \,\omega_{2}(a^\star,b^\star)\,.
\ee
Up to order $\mathcal{O}(\frac{1}{\beta})$ this ambiguity is parameterized by the possible choices of smooth real functions~$0\,\leq\, a^\star_0\, <\, 1$ and~$0\,\leq\, b^\star_0 \,<\, 1$
\be\label{eq:realReparametrizations}
a^\star\,=\,{a^\star_0(\Lambda)}\,, \qquad b^\star\,=\,{b^\star_0(\Lambda)}\,, 
\ee
which at large-$\Lambda$ respect the boundary conditions 
\be
a^\star(\Lambda)\,-\,1\,=\, \,\mathcal{O}(\Lambda^{-1})\,\qquad b^\star(\Lambda)\,-
\,1\,=\,\mathcal{O}(\Lambda^{-1})\,.
\ee
Inverting the parametric expressions of the physical charges~\eqref{eq:ChargesAB} in terms of~$a^\star$ and~$b^\star$ and substituting the result in the parametric representation of BPS entropy and BPS R-charge~\eqref{eq:EntropyRchargePar}, and mass gap~\eqref{eq:MassGap}, one obtains their unambiguous relations to physical charges and entropy, respectively. Here we only show the explicit relation at leading order at large-$\Lambda$ and first order in the low-temperature expansion
\be\label{eq:1/16MicrocanonicalEntropy}
\begin{split}
S_0& \,=\,{\sqrt{3}} \pi (N^2{J}_+ {J}_-)^{
\frac{1}{3}}\,, 
\qquad  R_0\,=\, (N^2{J}_+ {J}_-)^{
\frac{1}{3}}\,,
\qquad -\frac{1}{M}\,=\,\frac{1}{M_{SU(1,1|1)}}\,=\,\frac{S_0}{12 \sqrt{3} \pi}\,.
\end{split}
\ee
This result could have been obtained directly using only leading expressions for the free energy at large~$\Lambda$ (and low temperature), indeed we have performed that simpler computation independently, as a check.

For completeness we note that the hierarchy of charges that this near-BPS RG-flow probes is
\be\label{eq:HierarchyCharges}
\begin{split}
\sqrt{2}J_{\mp}&\,=\,\mathcal{O}(1)N^2\Lambda^3\,,
\\
\Delta &\,=\,\mathcal{O}(1) N^2\frac{|\alpha_0|}{\beta_0^2}\,\Lambda^3\,=\,\mathcal{O}(1)N^2\frac{\bigl(|\alpha-\frac{1}{2}|\bigr)}{\beta^2}\Lambda^2\,.
\end{split}
\ee

Four comments before concluding.

Recall that
\be
J_{\mp}\,=\,\frac{1}{\sqrt{2}}\,\Bigl(\,\frac{3\Delta_{\mp}-\Delta}{6}\Bigr)\,=\, \Bigl(\frac{\Delta_{\mp}}{2\sqrt{2}}\,+\,\mathcal{O}(\frac{1}{\beta_0^2})\Bigr)\,\underset{\beta_0\to\infty}{=}\,\frac{\Delta_{\mp}}{2\sqrt{2}}\,\geq\,0\,,
\ee
Thus, in the limit~$\Delta_+\to 0$ the~$S_0$ vanishes. This is consistent with the findings in the near-$\frac{1}{8}$-BPS expansion, which by definition is located at~$\Delta=\Delta_+=0\,$, and has~$S_0=0\,$,~\eqref{eq:Pars18}.~\footnote{The~$\frac{1}{8}$-BPS reference point used in the previous subsection only intersects with the $\frac{1}{16}$-BPS case in this section at~$R_1=R_2=0\,$.}

One natural extension of the analysis in this section corresponds to relaxing the minimal gauged supergravity constraint
\be\label{eq:IsotropicCharges}
{R_{1}} \,=\, R_2 \,=\,0\, \quad \longleftrightarrow \quad Q_1=Q_2=Q_3=\frac{R_3}{2}\,.
\ee
A gravitational mass gap has not been computed yet in such regime (as far as we understand). This would require a generalization  of the analysis in~\cite{Boruch:2022tno} to non-minimally gauged supergravities.

{Without loss of generality, we have chosen to focus on the Schwarzian corresponding to the winding sector~$n\,=\,j\,=\,0\,$ (in the notation of~\cite{Boruch:2022tno}). Our discussion trivially generalizes to other sectors~$n\,\neq\,0\,$, by expanding~$\alpha$ around other integer shifts of~$\alpha\,
=\,\pm\frac{1}{2}$, i.e., around $\alpha\,=\,\frac{1}{2}-n$ for some integer~$n\,$. Namely the result of those is obtained from~\eqref{eq:TheSchwarzianFormFinal} by substituting~$\alpha\to \pm\alpha+n\,$}. Analogously, the result obtained with the more general orbifold sectors mentioned in section~\ref{par:OrbifoldRGflows} can be obtained following analogous steps to the ones before summarized. The analysis of these more general cases is left for future work. 

\paragraph{A large-$\Lambda$ expansion to the complete quantum Schwarzian.}\label{sec:QUantumSchwarzian}

It should be noted that the ambiguity in selecting large-$\Lambda$ expansion can be used to fix a precise Schwarzian action starting from $\mathcal{F}_\infty$. Namely, there exists a large-$\Lambda$ expansion to the continuum such that, after truncating corrections of order $\mathcal{O}(\frac{1}{\beta^2})$ or higher one obtains an entropy of the form (at fixed $J_{\pm}$, $\alpha$ and $\beta$)
\be\label{eq:Sn}
S_n = S_{\text{in eq. }\eqref{eq:TheSchwarzianFormFinal}}\,+\, \log \left[ \frac{2 \cos[\pi( \alpha+n)]}{\pi (1 - 4 (\alpha+n)^2)} \right]\,+\, 2\pi \text{i} n R^*
\ee
around $\alpha = \frac{1}{2}-n\,$. The last term in the right-hand side exponentiates to $1$ in the UV theory but in the continuum theory the dependence on $n$ remains as an ambiguous choice, which can be justified as coming from the choice of large-$\Lambda$ expansion to the continuum. 

The extra dependence on $\alpha$ in~\eqref{eq:Sn}, which is an infinite Taylor series when expanded around $\alpha=\frac{1}{2}-n\,$, could only come from redefinitions of the large-$\Lambda$ expansion. This is because there exists one-specific large-$\Lambda$ expansion for which the mixed entropy $S$ is polynomial (not a series) in $\alpha-\frac{1}{2}+n\,$. The latter is the naive large-$\Lambda$ expansion that leads to equation~\eqref{eq:FInfinity} in grandcanonical ensemble.

As mentioned before, these large-$\Lambda$ limits pick up a series of equally contributing leading saddle points of the partition function labelled by an integer number $n\,$. Thus, after adding such contributions one obtains,
\be
Z[\beta, J_{\pm},\alpha]\,\underset{\Lambda\to\infty}{\sim}\,e^{-\,2\pi \text{i} \alpha R^*}\,Z_{\text{eff}}[\beta, J_{\pm},\alpha]
\ee~\footnote{The symbol $\sim$ in this equation means up to exponentially suppressed, logarithmic, and/or any other possible non-analytic corrections in $\omega_{1,2}=\omega_{1,2}[J_{\pm}]\,$.}
where~$\Lambda\to \infty$ denotes the specific large-$\Lambda$ expansion that lands on the effective partition function
\be
Z_{\text{eff}}[\beta, J_{\pm},\alpha]\,:=\, e^{S^*}\, \sum_{n\,\in\, \mathbb{Z}}\, \frac{2 \cos[\pi( \alpha+n)]}{\pi (1 - 4 (\alpha+n)^2)} e^{- \frac{2 \pi^2}{M {\beta}} \left( 1 - 4 \alpha^2 \right)
} 
\ee
is the $\mathcal{N}=2$ Schwarzian partition function~\cite{Boruch:2022tno}. Laplace transforming in $\beta$ one encounters the density of states
\be
\rho(\Delta,\alpha)\,=\,\sum_{n\,\in\,\mathbb{Z}}\rho_0(\Delta,\alpha+n)\,
\ee
with
\be
\rho_0(\Delta,\alpha)\,:=\, e^{-\mathcal{F}_0(\alpha)}\delta(\Delta) + e^{-\mathcal{F}_0(\alpha)}\frac{\sqrt{\mathcal{F}_1(\alpha)} I_1\left(2\sqrt{\mathcal{F}_1(\alpha) \Delta}\right)}{\sqrt{\Delta}}\,.
\ee
The function $I_1$ is the modified Bessel function of first kind and
\be\begin{split}
e^{-\mathcal{F}_0(\alpha)} \,:=\, e^{S^*} \, \frac{2 \cos (\pi \alpha)}{\pi (1 - 4\alpha^2)}\,, \qquad \mathcal{F}_1(\alpha) \,:=\, \frac{2\pi^2}{M_{SU(1,1|1)}} \left(1 - 4\alpha^2\right)
\,,
\end{split}
\ee
with~$M_{SU(1,1|1)}=-M$. It should be noted that~\cite{Stanford:2017thb}
\be\label{eq:IdentitiesSch}
\begin{split}
e^{-S^*}\,\sum_{n\,\in\, \mathbb{Z}}\,e^{-\mathcal{F}_0(\alpha+n)} &\,=\,1\,,\\
\sum_{n\,\in\, \mathbb{Z}}\,e^{-\mathcal{F}_0(\alpha+n)}{\sqrt{\mathcal{F}_1(\alpha+n)} I_1\left(2\sqrt{\mathcal{F}_1(\alpha+n) \Delta}\right)}&\,=\,D(\Delta,\alpha) \,\Theta\Bigl(\Delta-\frac{M_{SU(1,1|1)}}{32}\Bigr)
\end{split}
\ee
where the periodic function in $\alpha$, $D(\Delta, \alpha)=D(\Delta, \alpha+1)\,$, which we will not try to find analytically here, obeys the following conditions
\be\label{eq:Props2}
D(\frac{M_{SU(1,1|1)}}{32},\alpha)=D(\Delta,\frac{1}{2}^+)=0\,,
\ee
~\footnote{We define the value of $D[\Delta,\alpha]$ at $\alpha=\frac{1}{2}$ by its lateral limit.} and has positive integer Fourier coefficients as well. Here, equations~\eqref{eq:IdentitiesSch} have been numerically checked in $0\leq\alpha<1$~\footnote{We define the infinite sums over the integer $n\,$,~$\sum_{n\,\in\, \mathbb{Z}}$ by the principal value definition~$\underset{\Gamma\to\infty}{\text{lim}}\sum_{n=-\Gamma}^{\Gamma}\,$.} not analytically proven. At last, from~\eqref{eq:IdentitiesSch} there follows the state-density function~\cite{Stanford:2017thb}
\be
\begin{split}
\rho(\Delta,\alpha)&\,=\,e^{S^*}\delta(\Delta)\,+\,\rho_{cont}(\Delta,\alpha) \\
\end{split}
\ee
where the continuum part
\be
\rho_{cont}(\Delta,\alpha) = \frac{D(\Delta,\alpha)}{\sqrt{\Delta}} \,\Theta\Bigl(\Delta-\frac{M_{SU(1,1|1)}}{32}\Bigr)
\ee
is separated by the gap $\Delta=\frac{M_{SU(1,1|1)}}{32}$
from the BPS sector.

It is easy to check, numerically, that a generic deformation of the limit to the continuum, e.g. a generic deformation of the function $\mathcal{F}_0(\alpha)\,$, spoils the cardinal properties~\eqref{eq:IdentitiesSch},~\eqref{eq:Props2}, the positiveness and integralness of the Fourier coefficients of $D(\Delta,\alpha)\,$, and the existence of the infinitesimal mass gap $\frac{M_{SU(1,1|1)}}{32}\,$. In particular, as advanced early in the Introduction, this gap cannot be identified with the mass gap of~$N=4$ SYM above the BPS sector $\Delta=0\,$. This is because the latter is $\Lambda$-independent while the former it is not.~\footnote{Please read around the emphasized comment below equation~\eqref{eq:EmphComment}.}

\section{Final remarks}
\label{sec:6}
Holographic low-temperature expansions were recovered from the free energy of four dimensional maximally supersymmetric Yang-Mills theory. The analytic part of the free energy associated to the infrared effective theory,~\eqref{eq:FslTotal}, was computed. Assuming supergravity predictions are correctly capturing strong-coupling results in field theory, and in virtue of analyticity, the computed infrared free energy is bound to encode the Gibbons-Hawking free energy of the dual gravitational solutions even well beyond their BPS locus.  The formula allows to identify limits to the continuum (of $\mathcal{N}=4$ SYM) where Schwarzian contributions with a small mass gap~$M$ scaling with a negative power of~$\Lambda\,$ emerges. The value of the gap was fixed by identifying chemical potentials and thermodynamic charges of the infrared theory and the gravitational solution.

These low-temperature expansions localize around complexified values of chemical potentials at which supersymmetric cancellations happen and the physical partition function reduces to a superconformal index~$\mathcal{I}\,$. 

If the index~$\mathcal{I}$ counts states preserving four supercharges (i.e $\frac{1}{8}$-BPS states), then the corresponding Schwarzian deformation is protected against corrections in~$g_{YM}^2 N\,$. {In that case, the index $\mathcal{I}$ can be understood as a 1/16 BPS index, e.g. the index $\mathcal{I}_1$ defined in~\eqref{eq:TraceIndex4m} upon imposition of a second BPS constraints~\eqref{eq:EnhancementSUSY}.} In that case,~$\mathcal{I}$ is the 1/8 BPS index defined in equation~\eqref{eq:TraceIndex4m2p}. If the higher small-temperature deformations on top of this index happen to be appropriately directed along the complexified space of chemical potentials~\footnote{By this we mean that the derivatives with respect to~$\alpha\,$, at~$\alpha=\frac{1}{2}\,$, equal traces over the Hilbert space that receive contributions only from certain supersymmetric states. For examples, this would correspond to the constraint~$\gamma=1$ in~\eqref{eq:PhiVsGamma}.} then all of its terms, not just the Schwarzian contribution are exactly computed by another 1/16 superconformal index,~$\mathcal{I}_2\,\neq \mathcal{I}\,$ and $\mathcal{I}_2\neq \mathcal{I}_1\,$. In an abstract sense the 1/8 BPS index $\mathcal{I}$ is at the \emph{intersection} between the 1/16 BPS indices $\mathcal{I}_1$ and $\mathcal{I}_2\,$.

The \emph{transversal} 1/16 BPS index $\mathcal{I}_2\,$, like $\mathcal{I}_1\,$, counts states preserving only two supercharges out of the four ones preserved by the states counted by~$\mathcal{I}\,$. One difference with $\mathcal{I}_1$ --aside from the fact that they count different states, is that the latter is independent of temperature, while the former does depend on temperature. Other choices of temperature-dependent indices are possible in $\mathcal{N}=4$ SYM. For our purposes it was enough to focus on $\mathcal{I}_2\,$. 

Near-$\frac{1}{8}$-BPS black holes with charges $\Delta\approx 0\,$,~$\Delta_+\approx0\,$, $R_1+R_2\approx0$ and~$\Delta_- \neq 0\,$, were found to have near-vanishing horizon, and thus vanishing BPS entropy~$S_0\,$, due to the BPS contraint~$\Delta_-\geq0\,$. In these cases, the low-temperature corrections to the near-$\frac{1}{8}$-BPS mixed-ensemble free energy~$S$ was reduced to the protected Schwarzian contribution identified in subsection~\ref{subsec:ProtectedExpansion}. This result suggests that near-$\frac{1}{8}$-BPS gravitational solutions within the family found in~\cite{Wu:2011gq} have a near-vanishing near extremal horizon. It also suggests that their horizon vanishes in the strict \underline{and smooth} $\frac{1}{8}$-BPS limit.~\footnote{It would be  interesting to understand how such near-horizon geometry meets the conclusions of~\cite{Goldstein:2019gpz}. Such reference identified BTZ geometries in the near-horizon region of near-$\frac{1}{8}$-BPS black hole solutions. One may then wonder how the effective one-dimensional Schwarzian entropy relates (if they happen to be related e.g. in the spirit of~\cite{Mertens:2017mtv,Ghosh:2019rcj}) to the conjectured two-dimensional conformal field theory computing the entropy of such BTZ geometries~\cite{Goldstein:2019gpz}.}

More generally, our analysis show how the Schwarzian corrections already identified in the supergravity side of the duality~\cite{Boruch:2022tno} emerge after deforming the discrete UV theory,~$\mathcal{N}=4$ SYM to the continuum.

It should be pointed out that the corresponding Schwarzian mass gap, which scales as $\mathcal{O}(\frac{1}{\Lambda^3 N^2})\,$, should not be equated with the gap beyond $\Delta=0$ in the spectrum of the UV theory $\mathcal{N}=4$ SYM which is by construction $\mathcal{O}(1)$~\footnote{... fixing the radius of $S^3$ to $1$...} . The Schwarzian mass gap is a feature of the low energy effective action that remains in a specific large-$\Lambda$ expansion to the continuum,~\eqref{eq:DoubleExpansion} (See around equation~\eqref{eq:Sn}). The gap of $\mathcal{N}=4$ SYM is the gap before the latter is deformed to a continuum, and it does not depend on the imposed scale $\Lambda\,$.~\footnote{Please read around the emphasized comment below equation~\eqref{eq:EmphComment}.}

Other interesting questions become tangible. For instance, one may try to develop analytic tools to quantitatively understand the emergence of quantum chaos~\cite{Cotler:2016fpe,Saad:2018bqo,Turiaci:2023jfa} in gauge theories. In the case of $\mathcal{N}=4$ SYM one next obvious goal is to identify which is the random matrix theory emerging in the infrared theory. We will address this question in forthcoming work. Relatedly, one could also explore this possibility in simpler quantum systems, possibly even in some systems already known to be realized in nature.

In $\mathcal{N}=4$ SYM {this problem has been studied at~$\beta=\infty$~\cite{Choi:2022asl}}.~{The erratic behaviour there identified in the spectral form factor (even at zero temperature), is closely related to interference effects~\cite{Agarwal:2020zwm,Murthy:2020rbd}\cite{Cabo-Bizet:2020ewf} among complex phases associated to different saddle point contributions of the superconformal index~\cite{Cabo-Bizet:2019osg,ArabiArdehali:2019orz,Cabo-Bizet:2019eaf,Copetti:2020dil,Aharony:2021zkr, Cabo-Bizet:2021plf}}. The contribution of all such saddles is encoded in subleading corrections within the RG-flow procedure this paper focused on. The majority of them are $e^{-N^2}$-suppressed at large $N$, but their contribution is essential at finite~$N\,$. There are also instantonic $e^{-N}$-suppressed contributions~\cite{Aharony:2021zkr,Cabo-Bizet:2019eaf}, which come from D3-brane instantons~\cite{Aharony:2021zkr}. Our result predicts -- and calls for the study and understanding of -- the presence of analogous contributions away from the BPS locus.  {That said, it is still unclear to us whether in order to identify the ramp/plateau feature in the spectral form factor of $\mathcal{N}=4$ SYM it will be necessary to resort to these non-perturbatively suppressed contributions.}

\section*{Acknowledgements}
 It is a pleasure to thank Matteo Beccaria, Nana G. Cabo Bizet, Alejandro Cabo Montes de Oca, Justin R. David, Marina David, Stefano Giusto, Alfredo Gonzalez Lezcano, Camillo Imbimbo, Vishnu Jejjala, Albrecht Klemm, Kimyeong Lee, Gabriel Lopes Cardoso, Robert de Mello Koch, Jose F. Morales, Sameer Murthy, Suresh Nampuri, Leopoldo Pando Zayas and Alejandro Ruiperez for useful conversations or discussions. The author would like to thank the Isaac Newton Institute for Mathematical Sciences, Cambridge, for support and hospitality during the programme Black holes: bridges between number theory and holographic quantum information, where work on this paper was undertaken. This work was supported by EPSRC grant EP/R014604/1, and by the INFN grants GSS and GAST. 

\appendix

\section{The saddle point for gauge potentials}\label{app:Saddle}

Discarding \emph{heavy enough} states, the truncated version of the partition function~\eqref{eq:PartitionFunction2} is
\be\label{eq:TruncatedPartitionFunction}
Z_{\Lambda}[x,u,v,w,t,y]=\int [DU] e^{\sum_{n=1}^{\mathcal{O}(1)\Lambda^{n^\prime}}\frac{1}{n} \bigl(f_{Bos}[x^n,u^n,v^n,\,\ldots] + (-1)^{n+1} f_{Fer}[x^n,u^n,v^n,\,\ldots]\bigr)\,\text{Tr}U^n \text{Tr}U^{\dagger n}}\,,
\ee
where the $f_{Bos}$ and $f_{Fer}$ were defined in~\eqref{eq:FBosFFer}. 

$Z_{\Lambda}$ is as good as~$Z\,$ to compute (weighted) degeneracies at charges up to order~$\mathcal{O}(1)\Lambda^{n^\prime}\,$.

Let~$\{e^{2\pi \text{i} u_i}\}_{i=1,\ldots N.}$ be the eigenvalues of~$U$ then integral~\eqref{eq:TruncatedPartitionFunction} can be written in the form
\be
Z_{\Lambda}[x,u,{v},w,t,y]=\frac{1}{N!}\int_{0}^1 \text{d}u_1\ldots \int_{0}^1\text{d}u_N e^{-F_{\Lambda}({u}_i)}\,.
\ee
Expanding, for instance,
\be
Z_\Lambda[x,u=e^{2\pi\i\alpha},e^{-2\pi\i(\alpha-1/2)}\widetilde{v},w={\widetilde{v}}{t y}\,,\,t,\,y]\,=\,\mathcal{I}_{2\Lambda}[x,e^{2\pi\i\alpha},\widetilde{v},t,y]
\ee
at leading order in the low-temperature expansion~\eqref{eq:ExpansionChemicalPotentialsLargeCharge} with finite gauge potentials~$u_i$'s and recalling that
\be
t\,=\,e^{-\frac{1}{6} (\omega_1+\omega_2)},y\,=\,e^{\frac{{\omega_1}-\omega_2}{2}},\widetilde{v}\,=\,e^{\frac{1}{3} (\omega_1+\omega_2-3 \varphi_{\widetilde{v}})}\,,
\ee
we find that at leading order at large~$\Lambda$
\begin{align}\label{eq:EffAction}
-F_{\Lambda}(u_i) \,=\, &\frac{1}{\omega_1} \sum_{i,j=1}^{N} \left( \text{Li}^{\Lambda}_{2}\left(e^{2 \pi\text{i}  \left(u_{ij}-\frac{i \varphi_{\widetilde{v}}}{2 \pi
   }\right)}\right)+\text{Li}^{\Lambda}_{2}\left(e^{2 \pi\text{i}  \left(u_{ij}+\frac{i
   \varphi_{\widetilde{v}}}{2 \pi }\right)}\right) \,-\,\right. \nonumber \\
&\left. \qquad\qquad-2 \text{Li}^{\Lambda}_{2}\left(e^{2 \pi\text{i} 
   \left(u_{ij}\right)}\right)\right) \times \left(1 + \frac{4\pi i}{\beta} {c}\left(\widetilde{\alpha} - \frac{1}{2}\right) + \mathcal{O}\left(\frac{\omega_{1,0}}{\beta_0^2}\right)\right) 
\end{align}
where~$u_{ij}:=u_i-u_j$ and
\be\label{eq:TruncatedDilog}
\text{Li}^{\Lambda}_{p}(z):=\sum_{j=1}^{\mathcal{O}(1)\Lambda^{n}} \frac{z^j}{j^p}\,.
\ee
One saddle solution of~\eqref{eq:EffAction} is
\be\label{eq:saddlePointGauge}
u_i \,=\,u^\star\,, \qquad i\,=\,1\,,\ldots, N\,.
\ee
We note that even though in the limit~$\Lambda\to \infty$
$S_{\Lambda}(\underline{u})$ develops a cusp at~\eqref{eq:saddlePointGauge}
as it was pointed out recently in~\cite{Chang:2023ywj},~\footnote{ This is because~$\partial_u \text{Li}^{\Lambda}_{2}(e^{2\pi\text{i}u})+\partial_u \text{Li}^{\Lambda}_{2}(e^{-2\pi\text{i}u}){\underset{\Lambda\to \infty}{\to}}4 \pi^2 (\{u\}-\frac{1}{2})$ if~$u\neq 0 \,\text{mod}\,1\,$, and~$0$ if~$u\,=\,0 \,\text{mod}\,1\,$. $\{u\}:=u-\lfloor u\rfloor\,$. Here for simplicity we assumed real~$u$, but the generalization to complex~$u$ of the previous identity can be obtained straightforwardly.} the ansatz~\eqref{eq:saddlePointGauge} is a well-defined saddle point of the truncated partition function~$Z_\Lambda\,$ at any finite value of~$\Lambda\,$. This implies, as it will be shown in the following appendix, around equation~\eqref{eq:COntnuousRefPOint}, that this saddle point is bound to capture any potential growth in the number of-$\frac{1}{8}$-BPS states with charges of order~$N^2$ at leading~$\mathcal{O}(N^2)$ order. 

For example, at this saddle point
\be
\begin{split}
-F_{\Lambda}({u}_i=u^\star)&\,\sim\, \frac{N^2\left(\text{Li}^{\Lambda}_{2}\left(e^{ \varphi_{\widetilde{v}}}\right)+\text{Li}^{\Lambda}_{2}\left(e^{-
   \varphi_{\widetilde{v}}}\right)-2 \text{Li}^{\Lambda}_{2}\left(1\right)\right)}{\omega_1}\\ &\qquad\qquad\times\Bigl(1-{\pi\i C}\frac{\bigl(\widetilde{\alpha}-\frac{1}{2}\bigr)\,}{\beta}\,+\,\mathcal{O}\bigl(\frac{\omega_1}{\beta^2}\bigr) \Bigr) +\mathcal{O}\Bigl((\widetilde{\alpha}-\frac{1}{2})^2\Bigr)\,.
   \end{split}
\ee
which is minus the low-temperature expansion of the free energy~$\mathcal{F}_{\frac{1}{16}\text{near}\frac{1}{8}}$ reported in~\eqref{eq:F2}.

\section{The generic near-{1}/{8}-BPS susceptibility is also protected}\label{sec:3p2}
The generic refinement of the partition function
\be\label{eq:NonSUSYexample}
Z[x,u=e^{2\pi\i\alpha},v,w=e^{-\varphi}{v}{t y}\,,\,t,\,y]\,=e^{-\mathcal{F}_{\text{near}\frac{1}{8}}}\,.
\ee
flows to the~$1/8$-BPS index~$\mathcal{I}^{4,-;2,+}$ in the limit
\be
\alpha\to\frac{1}{2}\,, \qquad \varphi\,\to\, 0\,.
\ee
If~$\alpha =\frac{1}{2}$ and~${\varphi}\,\neq\, 0$ then it reduces instead to the 1/16-BPS index $\mathcal{I}_1\,$.~\footnote{There are other possible restrictions that may have been chosen, but none more general than~\eqref{eq:NonSUSYexample}.} 
For later reference we note that
\be
e^{\varphi} \,=\, e^{-\varphi_w} \frac{v\, y}{t}\,,\qquad  \varphi\,=\,\omega_1-\varphi_{v}-\varphi_w\,.
\ee
Our goal next is to show that the generic Schwarzian contribution (to free energy) about $(\alpha,\varphi)=(\frac{1}{2},0)\,$ is protected against gauge-coupling corrections, for example, if in the expansion~\eqref{eq:ExpansionChemicalPotentialsLargeCharge}
\be\label{eq:Varphi0}
\varphi\,=\,\frac{\varphi_0}{\Lambda}\,,\,
\ee
with fixed~$\varphi_0$ and
\be\label{eq:SupportingExpansion}
\frac{|\varphi_0|}{|\alpha_0|}\,=\,\text{finite and} \quad |\alpha_0| \quad \text{is small.}
\ee
The reason for such protectedness is that the first variation (susceptibility) of
\be\label{eq:FPhiNew}
\mathcal{F}_{\text{near}\frac{1}{8}}\,=\,\mathcal{F}[x,u=e^{2\pi\i\alpha},v,w=e^{-\varphi}{v}{t y}\,,\,t,\,y]
\ee
in the variables~$(\alpha,\varphi)$ at the point~$(\alpha,\varphi)=(\frac{1}{2},0)\,$,
\be\label{eq:LinearDifferential}
\delta^{(1)}\mathcal{F}_{\text{near}\frac{1}{8}}\,:=\,  \Bigl(\mathcal{F}^{(1)}_{\text{near}\frac{1}{8}}\Big|_{\varphi=0}\Bigr) \bigl(\alpha-\frac{1}{2}\bigr)\,+\,\Bigl(\partial_\varphi\mathcal{F}^{(0)}_{\text{near}\frac{1}{8}}\Big|_{\varphi=0}\,\Bigr) \,\varphi\,,
\ee
which is encoded in the derivatives
\be\label{eq:TotalSchwarzian}
\begin{split}
 \mathcal{F}^{(1)}_{\text{near}\frac{1}{8}} &\,=\, -\,2\pi \text{i}\,\text{Tr}_{\mathcal{H}} (R_1 \,+\,R_3) (-1)^F x^{\Delta}v^{-R_1-R_2} t^{2 (H+J^3_1)-R_2} y^{2 J^3_2-R_2}\,,\\
\partial_{\varphi} \mathcal{F}^{(0)}_{\text{near}\frac{1}{8}} \bigg|_{\varphi=0} &\,=\, \text{Tr}_{\mathcal{H}} \,R_2 \,(-1)^F \,x^{\Delta}v^{-R_1-R_2} t^{2 (H+J^3_1)-R_2} y^{2 J^3_2-R_2}\,,
\end{split}
\ee
happens to be a linear combination of two~$\mathcal{Q}_1$-protected
\be
-\text{Tr}_{\mathcal{H}} (R_{1,2}) (-1)^F x^{\Delta}v^{-R_1-R_2} t^{2 (H+J^3_1)-R_2} y^{2 J^3_2-R_2}
\ee
and one~$\mathcal{Q}_2$-protected
\be
-\text{Tr}_{\mathcal{H}} R_3 (-1)^F x^{\Delta}v^{-R_1-R_2} t^{2 (H+J^3_1)-R_2} y^{2 J^3_2-R_2}
\ee
traces. Namely, the linear differential~\eqref{eq:LinearDifferential} is a linear combination of three indices and thus it is protected.

To compute the large-$\Lambda$ expansion of the free energy~\eqref{eq:FPhiNew},~\eqref{eq:TotalSchwarzian}, we start from the matrix integral representation~\eqref{eq:PartitionFunction2} at zero gauge coupling. We extract the leading behaviour of such integral in the expansion~\eqref{eq:ExpansionChemicalPotentialsLargeCharge}, assuming the condition~\eqref{eq:Varphi0}, and keeping fixed~$\varphi_0$ and~$\alpha_0\,$, i.e., without imposing~\eqref{eq:SupportingExpansion}. In such expansion the leading saddle-point for the gauge-singlet condition is~$U\sim e^{2\pi \i u^\star}\,\times\,1_{N\times N}$, and thus, again, we can substitute~$\text{Tr} U^n \text{Tr} U^{-n} \,\to\, N^2$ and obtain
\be\label{eq:FnearExp}
\begin{split}
\mathcal{F}^{(0)}_{\text{near}\frac{1}{8}} \bigg|_{\varphi=0} &\,\sim\, -\,N^2\frac{L_2(\varphi_v)}{\omega_1 }\,=\,\mathcal{O}(\Lambda^1)\,.\\
\mathcal{F}^{(1)}_{\text{near}\frac{1}{8}} &\,\sim\,\pi \text{i} N^2\Bigl(\frac{-2}{\omega_1\omega_2}\,+\,\frac{\varphi+\omega_2}{\beta\,\omega_1 \omega_2 }\,+\, \mathcal{O}\Bigl(\frac{1}{{\beta_0}^2}\Bigr)\Bigr)\,\times\,L_2(\varphi_v)\,=\,\mathcal{O}(\Lambda^2)\,.\\
\partial^{1}_{\varphi} \mathcal{F}^{(0)}_{\text{near}\frac{1}{8}}\bigg|_{\varphi=0} &\,\sim\,- N^2\frac{L_2(\varphi_v)}{\omega_1 \omega_2 }\,=\,\mathcal{O}(\Lambda^2)\,. \\
\mathcal{F}^{(2)}_{\text{near}\frac{1}{8}} \bigg|_{\varphi=0}&\,\sim\,-\,2\pi^2\,N^2\,\frac{L_2(\varphi_v)}{\beta \,\omega_1 \omega_2}\,=\,\mathcal{O}(\Lambda^3)\,.
\end{split}
\ee
Again,~$\sim$ means that the objects to the right-hand and left-hand sides have identical leading asymptotic bevarior in the large-$\Lambda$ expansion. 

Using~\eqref{eq:FnearExp} we obtain the leading asymptotic behavior of the free energy~\eqref{eq:Index2} (at order~$\mathcal{O}(\Lambda^1)$)
\be\label{eq:FPhi}
\begin{split}
\mathcal{F}_{\text{near}\frac{1}{8}}\, \sim\,& - N^2 \, \frac{\varphi+\omega_2+2\pi \i (\alpha-\frac{1}{2})}{\omega_{1}\omega_2}\,L_2(\varphi_v)\,\\&+ \pi\text{i} N^2\,\frac{\bigl(\varphi+\omega_2+2\pi \i (\alpha-\frac{1}{2})\bigr)\bigl({\alpha}-\frac{1}{2}\bigr)}{{\beta}\,\omega_{1}\omega_2}\,\Bigl( 1\,+\,\mathcal{O}\Bigl(\frac{\omega_1}{{\beta}},\frac{\omega_2}{{\beta}}\Bigr)\Bigr)\,L_2(\varphi_v)\,.
\end{split}
\ee
In the right hand-side of this asymptotic relation only the quadratic contributions in $(\varphi,\alpha-\frac{1}{2})$
\be
\pi\text{i} N^2\,\frac{\bigl(\varphi+2\pi \i (\alpha-\frac{1}{2})\bigr)\bigl({\alpha}-\frac{1}{2}\bigr)}{{\beta}\,\omega_{1}\omega_2}\,\Bigl( 1\,+\,\mathcal{O}\Bigl(\frac{\omega_1}{{\beta}},\frac{\omega_2}{{\beta}}\Bigr)\Bigr)\,L_2(\varphi_v)\,
\ee
may receive corrections in the gauge coupling. This follows from the protectedness argument above. If on the right-hand side of~\eqref{eq:FPhi} we assume~$\alpha_0$ to be small and fix
\be\label{eq:PhiVsGamma}
\varphi\,=\, -2\pi\i \gamma (\alpha-\frac{1}{2})\,=\,\mathcal{O}\bigl(\frac{1}{\Lambda}\bigr)\,,
\ee
with~$\gamma$ being a finite constant -- which is equivalent to assuming~\eqref{eq:SupportingExpansion} --\,, then we obtain
\be\label{eq:AsymptitoticExpansionFVarphi}
\begin{split}
\mathcal{F}_{\text{near}\frac{1}{8}}&\, \sim\, \mathcal{F}_{\frac{1}{16}\text{near}\frac{1}{8}} \,-\,2\pi\text{i} N^2\frac{(\alpha-\frac{1}{2}) L_2(\varphi_v)}{\omega_1\omega_2} \,-\, N^2\frac{\varphi L_2(\varphi_v)}{\omega_1\omega_2}+\mathcal{O}\bigl((\alpha-\frac{1}{2})\varphi,(\alpha-\frac{1}{2})^2\bigr)\,.\\&\sim\, \mathcal{F}_{\frac{1}{16}\text{near}\frac{1}{8}} \,+\,2\pi\text{i} N^2\frac{\bigl(\gamma-1\bigr)(\alpha-\frac{1}{2}) L_2(\varphi_v)\,}{\omega_1\omega_2} +\mathcal{O}\bigl((\alpha-\frac{1}{2})\varphi,(\alpha-\frac{1}{2})^2\bigr)\,.
\end{split}
\ee
This equation is implicitly saying that the first-order correction in temperature to the free energy~$\mathcal{F}_{\text{near}\frac{1}{8}}$ at order~$\mathcal{O}(\frac{1}{\beta_0})$ is independent of~$\gamma$ (resp.~$\varphi$) and equals the one computed with~$\mathcal{F}_{\text{near}\frac{1}{8}}\,$.

\begin{table}[h]
\centering
\begin{tabular}{c|c|c}
\hline
Chemical potential & Dual charge & Source term \\ \hline
$\beta$& $\Delta$ & $+\beta \Delta$\\
$\omega_1$ & $J_{1}^{3} - J_{2}^{3} + \frac{R_1}{2} + R_2 + \frac{R_3}{2} + \frac{\Delta}{3}$ & $+\omega_1 \cdot \left(J_{1}^{3} - J_{2}^{3} + \frac{R_1}{2} + R_2 + \frac{R_3}{2} + \frac{\Delta}{3}\right)$ \\
$\omega_2$ & $J_{1}^{3} + J_{2}^{3} + \frac{R_1}{2} + \frac{R_3}{2} + \frac{\Delta}{3}$ & $+\omega_2 \cdot \left(J_{1}^{3} + J_{2}^{3} + \frac{R_1}{2} + \frac{R_3}{2} + \frac{\Delta}{3}\right)$ \\
$\varphi_v$ & $-R_1 - R_2$ & $+\varphi_v \cdot (-R_1 - R_2)$ \\
$\varphi$ & $-R_1$ & $+\varphi \cdot (-R_1)$ \\
$\omega_u=-2\pi\text{i}\alpha$ & $-R_1 - R_3$ & $+\omega_u \cdot (-R_1 - R_3)$ \\ \hline
\end{tabular}
\caption{The chemical potentials, charges and source terms to be potentially added to~$-\mathcal{F}_{\text{near}\frac{1}{8}}$ before performing extremization. }
\label{table:updated_elements2}
\end{table}

The extra term, which is protected against coupling corrections because it is included in the protected susceptibility~$\delta^{(1)}\mathcal{F}_{\text{near}\frac{1}{8}}$
\be\label{eq:Tadpole}
 \,+\,2\pi\text{i} N^2\frac{\bigl(\gamma-1\bigr)(\alpha-\frac{1}{2})\,L_2(\varphi_v)}{\omega_1\omega_2} \,\in\,\delta^{(1)}\mathcal{F}_{\text{near}\frac{1}{8}}\,,
\ee
captures the singularity associated to the growth of $\frac{1}{16}$-BPS states,~$\frac{1}{\omega_1\omega_2}\,$, and it is bound to match certain near-$\frac{1}{8}$-BPS (thermodynamic) susceptibility of the~$AdS_5$ black hole solutions found in~\cite{Wu:2011gq,Chong:2005hr}. 

The near-$\frac{1}{8}$-BPS expansion along~$\mathcal{I}_2$~\eqref{eq:F2} is recovered from~\eqref{eq:FPhi} upon the contraint~$\gamma=1$, as it has to be the case, due to consistency with the constraints that reduce~$Z[x,u,v,w,t,y]$ to~$\mathcal{I}_2[x,e^{2\pi\i\alpha},\widetilde{v},t,y]$
\be
\varphi_v\,=\, \varphi_{\widetilde{v}}\,+\, 2\pi\text{i}(\alpha -\frac{1}{2})\,,\qquad \varphi \,=\,-2\pi\text{i}(\alpha-\frac{1}{2})\,.
\ee

\subsection{Confirming the near-{1}/{8}-BPS saddle-point selection}\label{eq:Matching}

It should be noted that the low-temperature expansion of the protected near-$\frac{1}{8}$-BPS susceptibility~\eqref{eq:LinearDifferential}, which is implicit in the first-order differential in the right-hand side of~\eqref{eq:FPhi}, is \emph{continuously} recovered from 
the low-temperature expansion of~$\mathcal{F}\,$,~\eqref{eq:F}, by substituting~\eqref{eq:Varphi0}
\be
\varphi\,=\,\frac{\varphi_0}{\Lambda}\,=\,\mathcal{O}(\frac{1}{\Lambda})\,,
\ee
on the large-$\Lambda$ expansion of the latter. Concretely, by using the relation
\be
\varphi_w\,=\,\omega_1\,-\,\varphi_v\,-\,\varphi
\ee
and the identities
\be
 L_{3}(\varphi_v,-\varphi_v)=0\,,\qquad \,\partial_{\varphi_w} L_{3}(\varphi_v,\varphi_w)\Biggl|_{\,\varphi_w=-\varphi_v}\,=\, -\,L_{2}(\varphi_v)
\ee
on the first-order Taylor expansion of~\eqref{eq:F} around~$\varphi,\,\omega_1=0$, and the identity
\be\label{eq:COntnuousRefPOint}
\begin{split}
 \qquad L_{2,1}(\varphi_v,-\varphi_v)\,=\, -\,2 \,L_{2}(\varphi_v)
\end{split}
\ee
on the zeroeth-order Taylor expansion around~$\varphi,\,\omega_1=0$ of the~$\mathcal{O}(\Lambda)$ (and large-charge subleading) correction~\eqref{eq:ORcontribution1/16}. In particular, this demonstrates that the free energy of the near-$\frac{1}{8}$-BPS phase,~\eqref{eq:F2}, is continuously recovered from the free energy of the~$\frac{1}{16}$-BPS phase,~\eqref{eq:F}. This check also reaffirms the assumption that the saddle-point~\eqref{eq:GaugeSaddlePoint} determines the free energy of the near-$\frac{1}{8}$-BPS sector,~\eqref{eq:FPhi}.

\paragraph{Useful equation}
In this subsection we will use equalities instead of $\sim$ symbols. The equalities should be always understood up to the order we used them in the analysis summarized in the main body of the paper.
\be\label{eq:RefinemenF116}
\begin{split}
\mathcal{F}_{\frac{1}{16}near\frac{1}{8}}&\,=\,- \frac{L_2\bigl(\varphi_{\widetilde{v}}\bigr)+\frac{\omega_1}{2} L_1\bigl(\varphi_{\widetilde{v}}\bigr)}{\omega_1}\,+\,C\frac{  \pi \text{i} (\widetilde{\alpha}-\frac{1}{2}) L_1\bigl(\varphi_{\widetilde{v}}\bigr)}{\omega_1}+\\ & + C\frac{ \pi \text{i}\,(\widetilde{\alpha}-\frac{1}{2}) \Bigl( L_2\bigl(\varphi_{\widetilde{v}}\bigr)+\frac{\omega_1}{2} L_1\bigl(\varphi_{\widetilde{v}}\bigr) \Bigr)}{ \beta \omega_1} + C^2\frac{ \pi^2 (\widetilde{\alpha}-\frac{1}{2})^2 L_1[\varphi_{\widetilde{v}}]}{\beta \omega_1} \\
&=\,- \frac{L_2\bigl(\varphi_{\widetilde{v}}\bigr)}{\omega_1}\,+\,C\frac{  \pi \text{i} (\widetilde{\alpha}-\frac{1}{2}) L_1\bigl(\varphi_{\widetilde{v}}\bigr)}{\omega_1}+ C\frac{ \pi \text{i}\,(\widetilde{\alpha}-\frac{1}{2}) L_2\bigl(\varphi_{\widetilde{v}}\bigr)}{ \beta \omega_1} \\ &  + C^2\frac{ \pi^2 (\widetilde{\alpha}-\frac{1}{2})^2 L_1[\varphi_{\widetilde{v}}]}{\beta \omega_1}\,+\,\ldots\,,
\end{split}
\ee
The dots denote terms that do not contribute to the leading behaviour at large-$\Lambda\,$. The complete expression can obtained from~\eqref{eq:FslTotal} using the relations
\be
\varphi_w\,=\,\omega_1\,-\,\varphi_v\,, \qquad \varphi_v\,=\, \varphi_{\widetilde{v}}\,+\, 2\pi\text{i}(\alpha -\frac{1}{2})\,.
\ee

\section{BPS inequalities and conventions}\label{sec:SemipositivityRels}

In the conventions of charges we have used, the $16$ semi-positivity conditions are --written for instance in~\text{there} --
\be
H_{\text{there}} - \sum_{I=1}^3 s_I Q_{I\text{there}} -  \sum_{i=1}^2 t_i J_{i\text{there}}\,\geq\,0\,,
\ee
where~$s_I=\pm1$, $t_i=\pm1$ and~${s}_1 s_2 s_3 t_1 t_2=1$, are
\begin{align*}
H + 2 J_1^3 + \frac{3 R_1}{2} + R_2 + \frac{R_3}{2} &\geq 0, & 
H - 2 J_2^3 + \frac{R_1}{2} + R_2 + \frac{3 R_3}{2} &\geq 0, \\
H + 2 J_2^3 + \frac{R_1}{2} + R_2 + \frac{3 R_3}{2} &\geq 0, &
H - 2 J_1^3 + \frac{3 R_1}{2} + R_2 + \frac{R_3}{2} &\geq 0, \\
H + 2 J_1^3 - \frac{R_1}{2} + R_2 + \frac{R_3}{2} &\geq 0, &
H - 2 J_2^3 + \frac{R_1}{2} + R_2 - \frac{R_3}{2} &\geq 0, \\
H + 2 J_2^3 + \frac{R_1}{2} + R_2 - \frac{R_3}{2} &\geq 0, &
H - 2 J_1^3 - \frac{R_1}{2} + R_2 + \frac{R_3}{2} &\geq 0, \\
H + 2 J_1^3 - \frac{R_1}{2} - R_2 + \frac{R_3}{2} &\geq 0, &
H - 2 J_2^3 + \frac{R_1}{2} - R_2 - \frac{R_3}{2} &\geq 0, \\
H + 2 J_2^3 + \frac{R_1}{2} - R_2 - \frac{R_3}{2} &\geq 0, &
H - 2 J_1^3 - \frac{R_1}{2} - R_2 + \frac{R_3}{2} &\geq 0, \\
H + 2 J_1^3 - \frac{R_1}{2} - R_2 - \frac{3 R_3}{2} &\geq 0, &
H - 2 J_2^3 - \frac{3 R_1}{2} - R_2 - \frac{R_3}{2} &\geq 0, \\
H + 2 J_2^3 - \frac{3 R_1}{2} - R_2 - \frac{R_3}{2} &\geq 0, &
H - 2 J_1^3 - \frac{R_1}{2} - R_2 - \frac{3 R_3}{2} &\geq 0\,.
\end{align*}
The relations between the charges used in here and the ones used in~\cite{Choi:2018hmj} is
\begin{equation}\label{eq:TranslationCharges}
\begin{split}
H_{\text{there}} &\,=\, H, \quad Q_{1\text{there}} \,=\, \frac{1}{2} (R_1 + 2 R_2 + R_3), \quad  Q_{2\text{there}} \,=\,\frac{1}{2} (R_1 + R_3), \\
Q_{3\text{there}} &\,=\, \frac{1}{2} (-R_1 + R_3)\,,\quad  J_{1\text{there}} \,=\, J_{1}^{3} - J_{2}^{3}\,,\quad  J_{2\text{there}}\, =\, J_{1}^{3} + J_{2}^{3}\,.
\end{split}
\end{equation}

\section{The analytic effective potential for gauge variables}
\label{app:EffPotentialComplete}

The complete \emph{analytic part} of the infrared effective potential of gauge variables is~\footnote{There is a possible $\mathcal{O}(\Lambda^0)$ ambiguity coming from the ambiguity in choice of different branches of the logarithmic contributions, but we can always choose to work in a branch where these contributions vanish. Equivalently, we can always choose to approach the BPS point $\alpha=\frac{1}{2}$ in such a way such contribution vanishes (e.g. the selection of couterterms like~\eqref{eq:Choice of counterterms}).  So, we assume that we work in such a branch. }
\be\label{eq:FslTotal}
\begin{split}
\mathcal{F}^{L R}_{sl}&\,+\,\sum_{j=1}^{LR}\,\sum_{\rho\,\neq\, 0} \frac{e^{2\pi \text{i} j\rho(u)}}{j}\\&\,\underset{\Lambda\to\infty}{\mapsto}\,\mathcal{F}_{\infty,\, sl}\,=\,-\sum_{i\,\leq\,j\,=\,1}^N \sum_{p=-1}^{2}\Bigl( V_{p}(u_{ij})\,+\, V_{p}(u_{ji}) \,+\,\mathcal{O}(\frac{1}{\beta^2})\Bigr)\,,
\end{split}
\ee
where
\be
\begin{split}
V_2(u)&\,:=\,
\frac{4 \pi ^3 \left(\pi  \left(\alpha -\frac{1}{2}\right)+\text{i}
   \beta \right) \left(\overline{B}_3\left[-u-\frac{\text{i} \varphi _v+\text{i} \varphi _w}{2 \pi}\right]+\overline{B}_3\left[-u+\frac{\text{i}
   \varphi _v}{2 \pi }\right]+\overline{B}_3\left[-u+\frac{\text{i} \varphi _w}{2
   \pi }\right]\right)}{3 \beta  \omega _1 \omega _2}\,,\\
V_1(u) &:= \frac{\pi^2}{\beta \omega_1 \omega_2} \left( \pi \left( \alpha -\frac{1}{2}\right) + \text{i} \beta \right) \times \Biggl(2 \pi \left( \alpha -\frac{1}{2}\right) \left(-3 \overline{B}_2\left[-u-\frac{\text{i} (\varphi_v+\varphi_w)}{2 \pi }\right] \right. \\
&\quad\left. + \overline{B}_2\left[-u+\frac{\text{i} \varphi_v}{2 \pi }\right] + \overline{B}_2\left[-u+\frac{\text{i} \varphi_w}{2 \pi }\right] + \overline{B}_2\left[-u\right]\Large\right)\\
&\quad  + \text{i} (\omega_1+\omega_2) \left(\overline{B}_2\left[-u-\frac{\text{i} (\varphi_v+\varphi_w)}{2 \pi }\right] \right. \\
&\quad \left. - \overline{B}_2\left[-u+\frac{\text{i} \varphi_v}{2 \pi }\right] - \overline{B}_2\left[-u+\frac{\text{i} \varphi_w}{2 \pi }\right] + \overline{B}_2[-u]\right)\Biggr)\,,\\
V_0(u)&\,:=\,\frac{8 \pi ^3 \left(\alpha -\frac{1}{2}\right) \left(\pi 
   \left(\alpha -\frac{1}{2}\right)+\text{i} \beta \right)
   \left((\alpha-\frac{1}{2}) -\frac{\text{i} \omega _1+\text{i} \omega _2}{4 \pi
   }\right) \overline{B}_1\left[-\frac{2 \pi  u+\text{i} \varphi_v+i \varphi_w}{2 \pi }\right]}{\beta  \omega _1 \omega _2}\\
&+\left(\frac{\pi  \left(18 \text{i} \pi ^2 \left(\alpha -\frac{1}{2}\right)^2
   \left(2 \beta -\omega _1-\omega _2\right)+24 \pi ^3 ( \alpha
   -\frac{1}{2})^3-3 \text{i} \beta  \left(\omega _1^2+3 \omega _2 \omega
   _1+\omega _2^2\right)\right)}{18 \beta  \omega _1 \omega _2}\right.\\&\qquad\left.-\frac{\pi ^2 \left(\alpha -\frac{1}{2}\right) \left(12 \beta
   ^2-20 \beta  \left(\omega _1+\omega _2\right)+3 \left(\omega
   _1^2+3 \omega _2 \omega _1+\omega _2^2\right)\right)}{18
   \beta  \omega _1 \omega _2}\right)\,\times\,\\& \qquad \times \Biggl(\overline{B}_1\left[-u-\frac{\text{i} \varphi_v}{2 \pi }-\frac{\text{i} \varphi_w}{2
   \pi }\right]+\overline{B}_1\left[-u+\frac{\text{i} \varphi_v}{2 \pi}\right]+\overline{B}_1\left[-u+\frac{\text{i} \varphi_w}{2 \pi }\right]\Biggr)\,,\\
 V_{-1}(u)&:=  -\frac{2 \pi  \left(\alpha -\frac{1}{2}\right) \left(\pi 
   \left(\alpha -\frac{1}{2}\right)+i \beta \right) \left(2 \pi
    \left(\alpha -\frac{1}{2}\right)-i \omega _1\right) \left(2
   \pi  \left(\alpha -\frac{1}{2}\right)-i \omega
   _2\right)}{\beta  \omega _1 \omega _2}\,.
\end{split}
\end{equation}
~$V_p(u)$ is the contribution to the potential at order $\mathcal{O}(\Lambda^p)\,$. We have reported here only the expansion up to order~$\mathcal{O}(\frac{1}{\beta})$ but the complete expansion in terms of rational functions of~$\beta$ is presented in the shared Mathematica file.

\bibliographystyle{JHEP}

\end{document}